\documentclass[a4paper,12pt]{article}
\usepackage{graphicx}

\usepackage{caption}
\newcommand{\Vec}[1]{\mbox{\boldmath$#1$}}
\newcommand{\bfk}{{\Vec k}}
\newcommand{\bk}{{\Vec k}}
\newcommand{\br}{{\Vec r}}
\newcommand{\bfr}{{\Vec r}}

\newcommand{\bq}{{\Vec q}}
\newcommand{\bp}{{\Vec p}}

\newcommand{\ve}{\varepsilon}

\begin{document}

\begin{center}
{\large\textbf{Integer quantum Hall effect}\par
\ \\ 
Hideo Aoki}
\par
\ \\
{\it Department of Physics, University of Tokyo, Hongo,
Tokyo 113-0033, Japan}

\end{center}

{\bf Abstract}  

Integer quantum Hall effect, which is the Hall effect quantized 
into integer times $e^2/h$ ($e$: elementary charge, $h$: Planck's 
constant) observed first in two-dimensional electron gases in strong 
magnetic fields, is reviewed from both theoretical and experimental 
standpoints.  Basic physics underlying the phenomenon 
is explained.   Specifically in this new edition 
we have a fresh look at how the quantum Hall 
effect is captured in a perspective of topological systems, 
since, while the quantum Hall effect is historically the first realization 
of the topological systems, the field has been delved into 
a much wider realm of physics of topological systems.   
We also mention diverse advances such as 
the quantum Hall effect (QHE) in various materials and contexts 
that include graphene, oxides and narrow-gap semiconductors, 
a relation with the fractional quantum Hall effect, and 
the quantum Hall effect as the resistance 
standard and furhter roles in the new SI system.  
We also expound the Floquet topological insulator 
(a light-matter coupled system) as a new paradigm 
in nonequilibrium topological systems, 
where an anomalous quantum 
Hall effect in zero magnetic field is realized as 
theoretically predicted to occur in graphene 
illuminated by a circularly-polarized laser 
and experimentally veryfied recently. 

{\bf Key Points}

\begin{itemize}
\item  Qantum Hall effect (QHE) viewed in a perspective of topological systems,
\item  QHE in graphene systems including the twisted bilayer graphene, 
\item QHE as the resistance standard and furhter roles now 
incorporated in the new SI system,
\item Relation of the integer QHE with the fractional quantum Hall effect,
\item The Floquet topological insulator 
as a prototype of nonequilibrium topological systems.
\end{itemize}

\section{Classification of topological systems by generic symmetries}

\begin{figure}[ht]
\begin{center}
\includegraphics[width=13cm,clip]{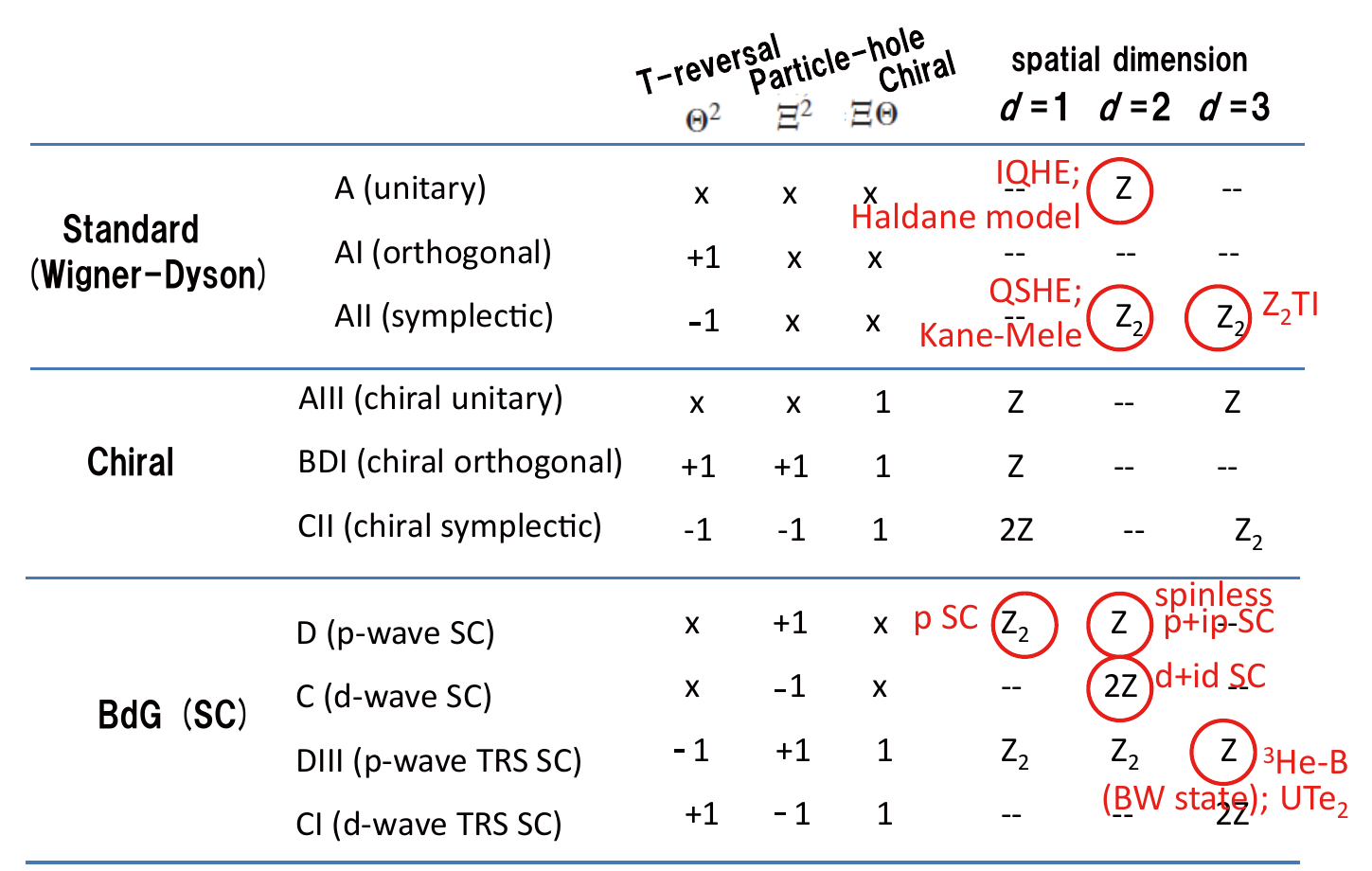}
\caption{Table of topological systems.  For the meaning of 
"x, +1, -1", see text.  For spatial dimensions $d=1,2,3$, 
topological numbers are indicated with symbols 
$Z$: integers, $Z_2$: binary (-1,+1).   After 
A.P. Schnyder, S. Ryu, A. Furusaki, and A.W.W. Ludwig, Phys. Rev. B {\bf 78}, 195125 (2008); 
S. Ryu et al, New J. Phys. {\bf 12}, 065010 (2010). 
Added red inscriptions denote examples of the quantum states and materials.
}
\label{topologicalPeriodicTable}
\end{center}
\end{figure}

Recent years have witnessed a vast 
widening of physics of topological systems, where 
IQHE is still enjoying the status of the very first one 
recognised.  So it will be instructive to 
start the article with a perspective 
of the topological systems 
in terms of the classification scheme of the entire 
topological systems.\cite{classificationTable}  
Let us first show the full classification table in 
Fig.\ref{topologicalPeriodicTable}.  
There are altogether ten universality classes for topological quantum 
states, and examples are indicated in the figure.

The starting point is the essential symmetries in classifying topological 
systems:

\begin{itemize}
\item Time-reversal symmetry (TRS; operator $\equiv \Theta$).
\item Charge-conjugation ($\sim$ particle-hole) symmetry (PHS; operator $\equiv \Xi$).
\end{itemize}

These can be expressed as eigenvalues of (anti-unitary) operators, 
\[
KU
\]
for a given Hamiltonian, where $K$: complex conjugation, $U$: unitary rotation.  For instance, $\Theta= K$ for class AI, $\Xi = -i\sigma_y K$ for AII.  
If each of the symmetries is
\begin{eqnarray}
&{\rm absent} \rightarrow &{\rm "x" in\; the\; table,}\\
&{\rm present} \rightarrow &{\rm "+1"} 
{\rm if\; (operator)}^2 = {\rm identity}\\
& &{\rm "-1"} {\rm if \;(operator)}^2 = -{\rm identity}.
\end{eqnarray}
 
So we have (3 possibilities for TRS)$\otimes$(3 possibilities for PHS), 
i.e., 9 possible cases.  
We can also consider a product TRS$\otimes$PHS (where an operator 
$\Theta \Xi$ represents the chiral symmetry).  
When (TRS, PHS) = (x, x), $\Theta \Xi$ can be either present ($\Theta \Xi=1$) or absent ($\Theta \Xi=$x).  Thus we end up with 10 cases in total.   
The quantum Hall effect belongs to Class A (unitary), with the time-reversal 
symmetry broken by an external magnetic field.  

\begin{figure}[ht]
\begin{center}
\includegraphics[width=9cm,clip]{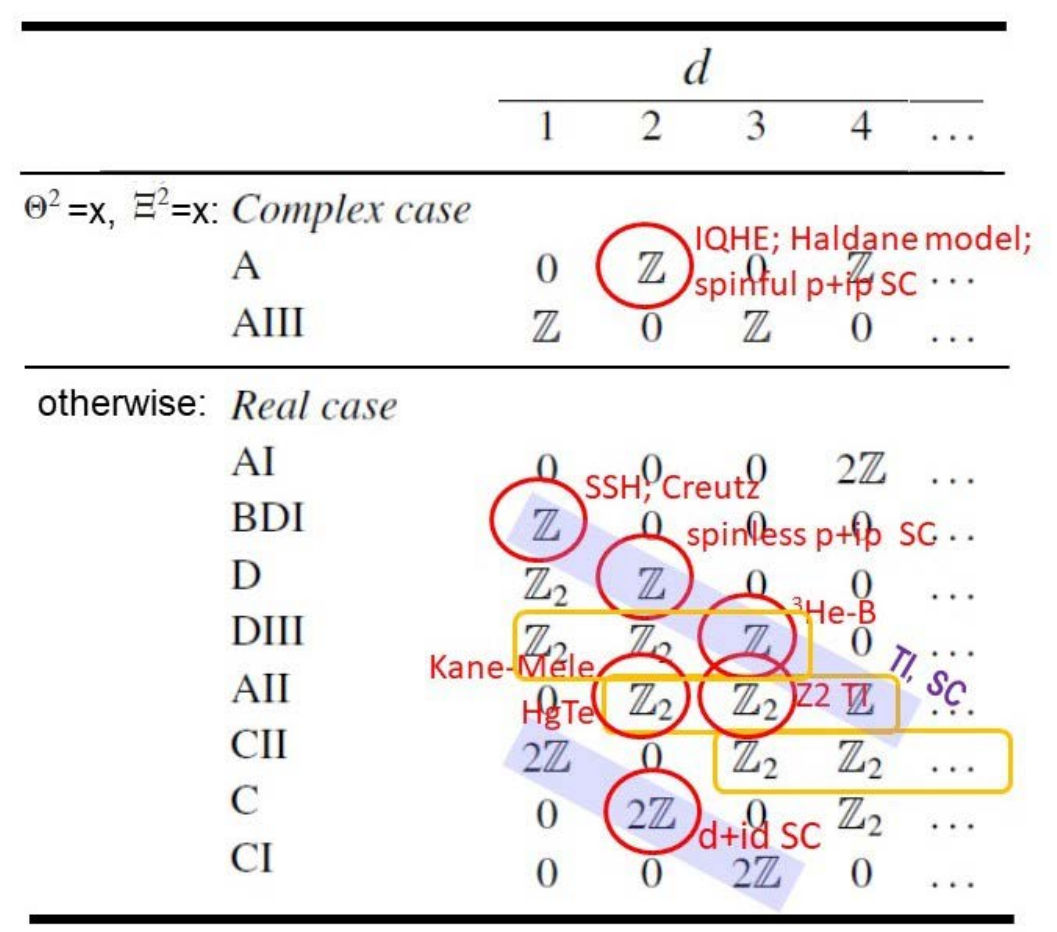}
\caption{Another representation of the 
table of topological systems, after 
A.P. Schnyder, S. Ryu, A. Furusaki, and A.W.W. Ludwig, Phys. Rev. B {\bf 78}, 195125 (2008); 
S. Ryu et al, New J. Phys. {\bf 12}, 065010 (2010).  Purple bands highlight TI (topological insulator) and SC (superconductor) series, 
orange squares $(Z_2,Z_2,Z)$  triplets, while red inscriptions denote examples of the quantum states as in the previous figure.}
\label{topologicalPeriodicTable_CliffordAlgebra}
\end{center}
\end{figure}

The table can be obtained mathematically in terms of Clifford
algebra and dimensional reduction.  Along this line, we can 
rearrange the table as in Fig.\ref{topologicalPeriodicTable_CliffordAlgebra}, where TI and SC states appear in a bunched structure.

\begin{figure}[h]
\begin{center}
\includegraphics[width=12cm,clip]{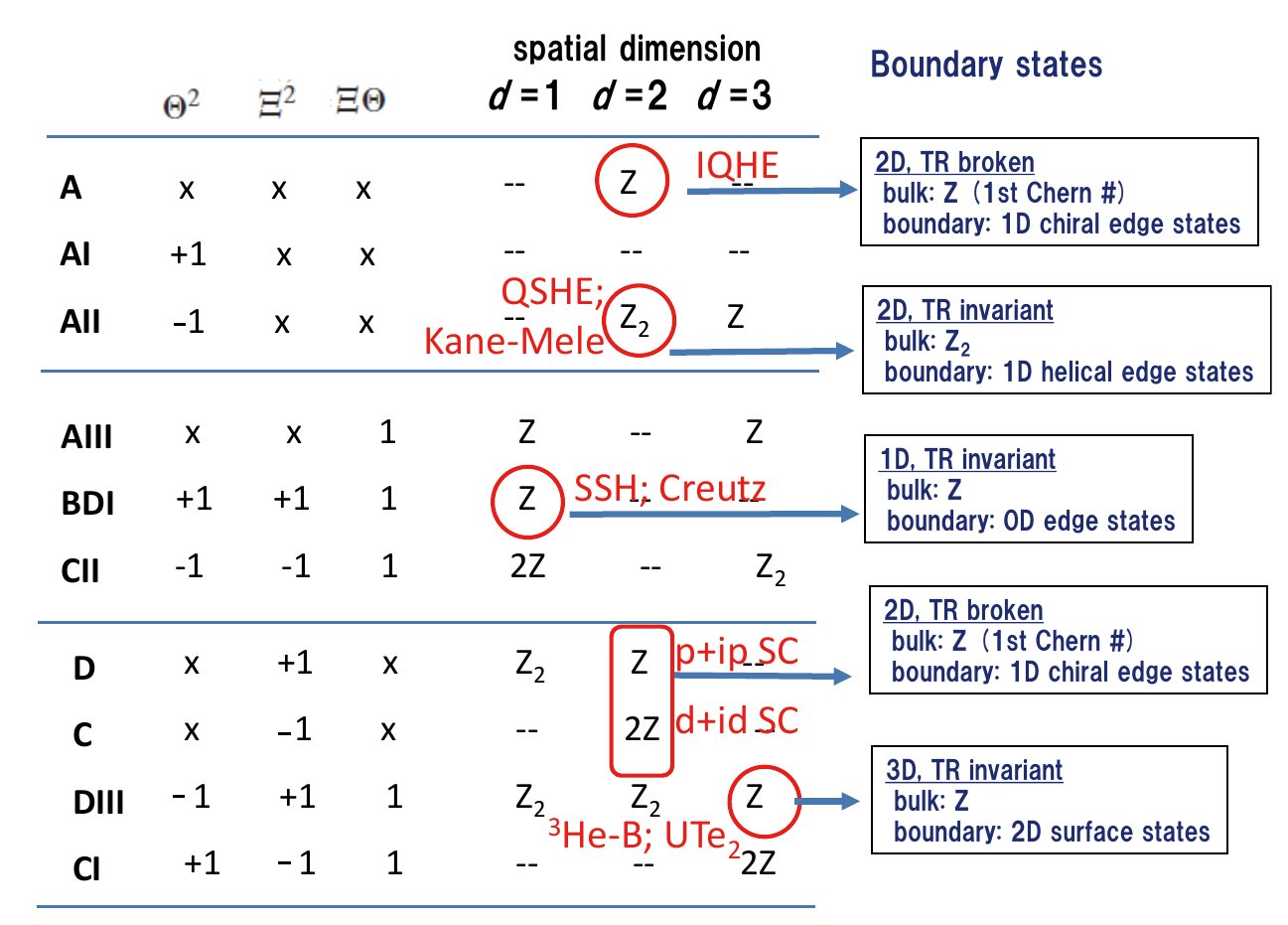}
\caption{
Topological boundary states for various topological states on the table of topological systems.  
TR stands for the time-reversal symmetry.
}
\label{boundaryStates}
\end{center}
\end{figure}

In subsection 'Bulk-edge correspondence' below, we shall describe the 
edge states beginning with the QHE system.  This notion can be 
extended to the general topological states as described above, which is natural since the boundary 
states are generic in the field-theoretic picture.  
Figure \Ref{boundaryStates} summarises, for later references, 
topological boundary states for various topological states on the table of topological systems.

\section{QHE --- Introduction}

Quantum Hall effect (QHE) is undoubtedly one of the most fascinating and 
important phenomena not only in the condensed-matter physics, 
but more generally in a wider scope of physics that encompasses 
cold-atom systems and a multitude of topological systems on one 
hand, and in field theoretic aspects on the other.  
This can be immediately realised if one notes a vast 
spectrum of the quantum Hall 
physics ranging from fundamental physics (topology 
in terms of the quantum field theory) 
down to applicational physics as exemplified by 
the QHE as the resistance standard.  
To start with, QHE originated from quantum mechanical 
physics in two spatial dimensions (2D) as 
opposed to the three-dimensional space in which electrons usually dwell.  
Surprisingly, 2D is a special dimension which accommodates 
phenomena specific to 2D, as 
field theories actually note that even spatial dimensions are 
rather special.  QHE stands out as a most remarkable 
one.   In this sense 2D is definitely not just a reduced dimensionality.  
This accounts for the remarkable width and depth of the physics 
of QHE, 
which has now become as large a field as those for superconductivity/superfluidity.  Interestingly, the QHE physics even has spinoffs into topological 
superconductivity, QHE in 
higher (three) dimensions, etc.  

QHE consists basically of the integer QHE discovered in 1980, which is essentially 
a one-body problem (but see section `Integer vs fractional quantum Hall effects'), and the subsequent fractional QHE 
discovered in 1983, which is a many-body effect.  In this chapter we 
focus on the integer QHE.  Even so, 
the field is so vast that here we shall describe bare essentials.  
(For details, see, e.g., 
\cite{landwehr,aoki1,chakra,prange,Pinczuk,Yoshioka}).

If we just summarise the peculiar properties of the QHE system, 
\begin{enumerate}
\item Energy spectrum: completely discrete, line spectrum (i.e., 
Landau levels) arises 
in the clean limit.   This 
is most unusual, since the system is a bulk.  This gives a starting 
point for the integer QHE.  When an integer number of Landau levels are 
fully filled, then we can regard the system as a giant ``closed shell" 
of electrons.

\item Transport properties: the closed shell is not an ordinary one, since the quantised Hall conductivity (integer times $e^2/h$) is 
given entirely in terms of physical constants 
($e$ : elementary charge, $h$ : Planck's constant). 
\end{enumerate}

One essential feature of the system is that 
the position coordinates, $x$ and $y$, in real space 
become noncommutative operators.  
In other words, an applied magnetic field makes 
the position coordinates $\bfr$ mixed with 
the wavenumber $\bfk$.   On top of this, there is a fortuitous 
coincidence of the Hall conductivity with a topological invariant.

\clearpage

 \subsection{Two-dimensional electron gas}

\begin{figure}[ht]
\begin{center}
\includegraphics[width=5cm]{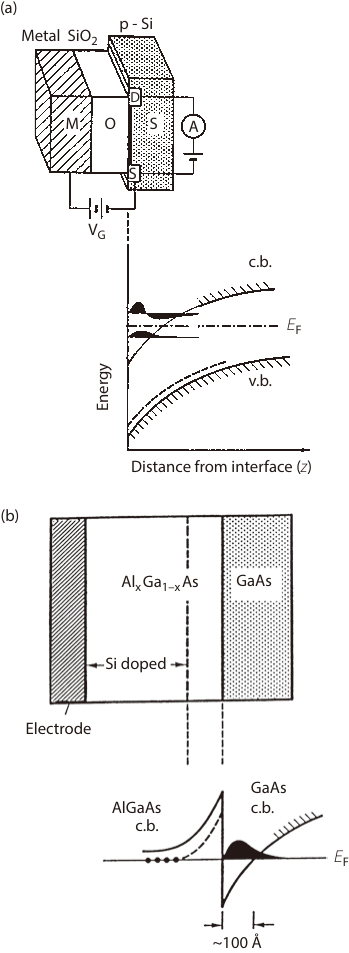}
\caption{
Structures of MOSFET (a) and 
semiconductor heterostructure (b).  In each frame, 
the upper panel shows the sample structure, while the 
lower panel the electronic band structure around the 
interface.  Typical wavefunctions are plotted in black against the 
direction perpecdicular to the interface.}
\label{2DEG}
\end{center}
\end{figure}

As a background we should start with describing the two-dimensional electron gas (2DEG).   The usual electron gas is a system of electrons 
that move more or less freely in a 3D space, as typically 
realised in simple metals.   In semiconductor physics we can realise 2D electron gas  in metal-oxide-semiconductor field-effect 
transistors (MOSFET's; Fig.\ref{2DEG}(a))  mainly before ca 1970's, and subsequently QHE is observed primarily 
in semiconductor heterostructures (Fig.\ref{2DEG}(b))\cite{AndoSternFowler}.

In these structures, electrons are confined to a 2D plane, due to a 
Schottky barrier between the metal and the oxide in an MOSFET, 
or between different semiconductors in a heterostructure.  
There, an electron moves in the 2D interface, with 
the electronic structure of the constituent materials 
entering only through the effective mass in the effective-mass 
approximation.   The wavefunction has a finite thickness in 
the direction perpendicular to the plane, 
but the motion along this direction is quantised, so that 
the band structure comprises 2D electronic bands labelled by 
the quantised levels in the normal direction, which 
are called subbands.  When only the lowest subband is 
occupied by electrons (or, more precisely, if the transitions 
between adjacent subbands can be neglected),  we can regard the motion 
genuinly 2D.  The Schr\"{o}dinger equation reads
\begin{equation}
{\cal H} \psi = \left[\frac{1}{2m^*}p^2 + U(z)\right] \psi = E\psi,
\end{equation}
where $m^*$ is the effective mass of the electron, $\bp$ 
is the momentum in the 2D plane along $\bfr \equiv (x,y)$, 
and $U(z)$ is the confining (Schottky) potential.  If we ignore 
disorder (interface roughness, impurities, etc), the wavefunction is 
expressed as 
\begin{equation}
\psi = {\rm exp}[i(k_x x + k_y y)]f_n(z)
\end{equation}
up to a normalisation constant, where the 2D motion 
is decribed by plane waves with wavenumber $\bfk \equiv (k_x, k_y)$ 
and $f_n$ is the $n$-th quantised wavefunction along $z$.  

In semiconductor heterostructures, typically GaAs/AlGaAs grown 
with the molecular beam epitaxy (MBE), 
the Fermi energy is $E_F \sim 10$ meV, so that 
the electron system is a degenerate Fermi gas at liquid He 
temperature ($\sim 0.4$ meV).

\subsection{2DEG in strong magnetic fields --- Classical mechanics}

\begin{figure}[ht]
\begin{center}
\includegraphics[width=11.7cm,clip]{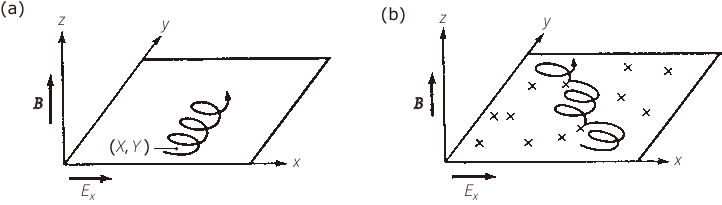}
\caption{
Classical orbits for a charged particle in a magnetic field 
($\parallel z$) in an applied electric field ($\parallel x$) for a 
clean system (a) and in a disordered system with scatterers (crosses) (b).}
\label{classicalorbits}
\end{center}
\end{figure}

If we apply an external magnetic field, ${\Vec B}$,  normal to a 2DEG, classically 
an electron undergoes a circular motion (called Larmor's motion) 
due to the Lorentz force, $-e{\Vec v}\times {\Vec B} \,\,(e$ : elementary 
charge, ${\Vec v}$ : velocity of the 
electron).   When there is an external electric field, ${\Vec E}$, as well, 
the classical orbit is a trochoid (Fig.\ref{classicalorbits}(a)), where the 
centre of the circular motion drifts in a direction perpendicular to 
${\Vec E}$ with a drift velocity $c{\Vec E}\times {\Vec B}/B^2$\,\,
($c$: speed of light).

This is because the electron is accelerated (decelerated) 
when it moves along (against) ${\Vec E}$, while the Lorentz force 
($\propto v$) is always balanced with the centrifugal force ($\propto$ the 
radius of the circular motion), so that the circular trajectory 
is elongated in the accelerated part while 
contracted in the decelerated part.   The resulting drift of the 
centre coordinate accounts for the classical Hall effect.  

When the system is disordered, due to e.g. impurities, 
then an electron is scattered, and the drift velocity 
acquires a component along ${\Vec E}$  (Fig.\ref{classicalorbits}(b)).  
If we define the conductivity, the quantity becomes a tensor in the 
presence of a magnetic field, where the current, ${\Vec j}$, 
and ${\Vec E}$ are related as
\begin{eqnarray}
\left(
\begin{array}{c}
  j_{x}\\
  j_{y}
\end{array}
\right)
&=&
\left( 
\begin{array}{cc}
  \sigma_{xx} & \sigma_{xy} \\
  \sigma_{yx} & \sigma_{yy}
\end{array}
\right)
\left(
\begin{array}{c}
  E_{x} \\
  E_{y}
\end{array}
\right),  \nonumber \\
\left(
\begin{array}{c}
  E_{x}\\
  E_{y}
\end{array}
\right)
&=&
\left( 
\begin{array}{cc}
  \rho_{xx} & \rho_{xy} \\
  \rho_{yx} & \rho_{yy}
\end{array}
\right)
\left(
\begin{array}{c}
  j_{x} \\
  j_{y}
\end{array}
\right).
\end{eqnarray}
Here $\sigma_{\mu\nu}$ is the conductivity tensor and 
$\rho_{\mu\nu}$ the resistivity tensor.   They are inverse 
matrices with each other, so that we have, 
with a symmetry $ \sigma_{xx} = \sigma_{yy}, \sigma_{yx} = -\sigma_{xy}$, 

\begin{eqnarray}
\left(
\begin{array}{cc}
 \rho _{xx} & \rho _{xy} \\
 \rho _{yx} & \rho _{yy}
\end{array}
\right)  &=& \left(
\begin{array}{cc}
 \sigma _{xx} & \sigma _{xy} \\
 -\sigma _{xy} & \sigma _{xx}
\end{array}
\right)^{-1}= \frac{1}{\sigma _{xx}{}^2+\sigma _{xy}{}^2}\left(
\begin{array}{cc}
 \sigma _{xx} & -\sigma _{xy} \\
 \sigma _{xy} & \sigma _{xx}
\end{array}
\right),   \nonumber \\
\left(
\begin{array}{cc}
 \sigma _{xx} & \sigma _{xy} \\
 \sigma _{yx} & \sigma _{yy}
\end{array}
\right) &=& \frac{1}{\rho _{xx}{}^2+\rho _{xy}{}^2}\left(
\begin{array}{cc}
 \rho _{xx} & -\rho _{xy} \\
 \rho _{xy} & \rho _{xx}
\end{array}
\right).
\end{eqnarray}

\begin{figure}[ht]
\begin{center}
\includegraphics[width=11.7cm,clip]{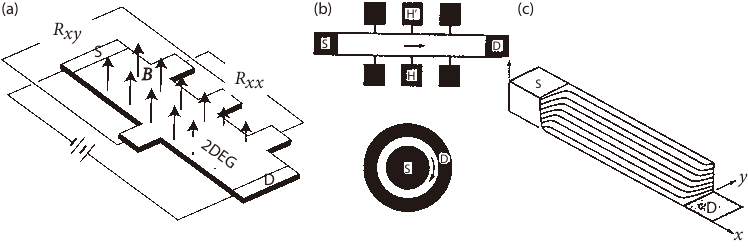}
\caption{
(a) Hall-bar sample geometry for measuring the QHE. 
(b) Top views of a Hall bar and a Corbino sample, with source (S) and 
drain (D) electrodes.  
(c) A birds-eye view of the equipotential lines in the QHE condition.}
\label{samplegeometry}
\end{center}
\end{figure}

If we introduce 
a phenomenological relaxation time, $\tau_0$, in zero 
magnetic field to 
describe the scattering in a classical transport theory 
with the equation of motion given by 
$m^*(d/dt +1/\tau_0){\Vec v} = -e({\Vec E}+{\Vec v}\times {\Vec B})$, 
then we have
\begin{eqnarray}
\sigma_{xx} &=& \frac{\sigma_0}{1 + \omega_c^2\tau_0^2}, \nonumber \\
\sigma_{xy} &=& \frac{\sigma_0\omega_c\tau_0}{1 + \omega_c^2\tau_0^2} 
= -\frac{nec}{B} + \frac{\sigma_{xx}}{\omega_c \tau_0}, \label{sigxytau}
\end{eqnarray}
where $\sigma_0 = ne^2\tau_0/m^*$ is the conductivity in 
zero magnetic field, and 
\[
\omega_c=eB/m^*c
\] 
is the cyclotron frequency.  When $\omega_c \tau_0 \gg 1$ (for 
which an electron can accomplish the Larmor motion 
between scattering events), the 
leading term, $\sigma_{xy} \simeq -nec/B$, on the right-hand side 
of eqn(\ref{sigxytau}) is the main term (the classical Hall conductivity in the clean case), while the second term a small correction.  
Thus we have a voltage (Hall voltage) in the direction perpendicular 
to the electric field when the sample has open boundaries in that 
direction, or we can measure the Hall current if electrodes are 
attached.  
Figure \ref{samplegeometry} depicts the sample geometry.

\subsection{2DEG in strong magnetic fields --- Quantum mechanics}

If we go to quantum mechanics, an electron is still subject to a circular motion 
within the correspondence principle, but now with a 
quantum mechanical uncertainty.  We should start from a Hamiltonian,

\begin{equation}
{\cal H}_{0} = \frac{1}{2m^*} {\Vec  \pi}^2,  \label {eq: ob1}
\end{equation}
where the momentum $\bp$ is now an operator, which is 
replaced, in a  magnetic 
field ${\Vec B} = {\rm rot}{\Vec A}$,  with the 
canonical momentum ${\Vec \pi} = \bp +(e/c){\Vec A}(\br)$ 
with ${\Vec A}$ being the vector potential.  

It is convenient to decompose the cyclotron motion 
into the centre ${\Vec R}\equiv (X,Y)$ of the circular motion (``guiding centre") 
and the relative coordinate ${\Vec \xi}\equiv (\xi, \eta)$ 
to facilitate applying the correspondence principle.   
The relative coordinate is related to the velocity as 
${\Vec v}=\omega_c \hat{{\Vec e}}_z\times {\Vec \xi}$ 
with $\hat{{\Vec e}}_z$ a unit vector 
normal to the 2D plane.  
The presence of a magnetic field thus renders a 
skew (i.e., vector-product) relation between ${\Vec \xi}$ 
and ${\Vec v}$. 

The quantum mechanical expression for the velocity is 
${\Vec v} = (i/\hbar)\,[{\cal H}_0,{\Vec r}] = {\Vec \pi}/m^*$, 
which means that we have
${\Vec \xi}=-(c/eB) \hat{{\Vec e}}_z\times {\Vec \pi}$, 
i.e., 
\begin{equation}
(\xi, \eta) = \frac{\ell ^2}{\hbar}\,(\pi _y, - \pi _x).
\label {eq: cmvel2}
\end{equation}
Here 
\begin{equation}
\ell \equiv \sqrt{\frac{c \hbar}{eB}}
\label{eq:maglength}
\end{equation}
is the length scale of the cyclotron motion, called 
the magnetic length, which does not depend on material parameters 
and has a typical value of  81 \AA$\,$  for the magnetic field 
of 10 T.

Thus the relative coordinate, ($\xi$, $\eta$), is a quantum 
mechanical operator, which in turn implies the centre coordinate, 
$(X, Y) = (x-(\ell ^2/\hbar)\pi _y, \; y+(\ell ^2/\hbar) \pi _x)$, 
is a quantum mechanical operator as well.  
From the standard commutation relation, 
$[x, p_x ] = [y, p_y ] = i \hbar$, 
we obtain commutation relations, 
\begin{equation}
[\xi, \eta]=-i\ell^2, \\
\quad [X, Y]=i\ell^2,
\end{equation}
namely, $x$ coordinate does not commute with 
$y$ coordinate, which implies that 
the relative coordinats have an uncertainty $\sim \ell$.   
The same applies to the centre coordinate.  
This is unusual, since ordinarily it is 
the momentum with which the real-space coordinate 
does not commute.  

Quantum mechanical states can be derived algebraically.  
For this we first note that the commutation 
relations for ($X$, $Y$) and ($\xi$, $\eta$) enable us 
to introduce two sets of harmonic-oscillator (boson) operators, 
\begin{eqnarray}
a &=& \frac{\ell}{\sqrt{2} \hbar}
\left (\pi _x -i \pi _y \right )\ \ =\ \
- \frac{1}{\sqrt{2} \ell}\left (\eta +i \xi \right ),
\nonumber \\
b &=& \frac{1}{\sqrt{2} \ell} \left (X +i Y \right ),
\label {eq: bbozon}
\end{eqnarray}
which have bosonic commutation relations,
\begin{equation}
[a, a^{\dagger}] = [b, b^{\dagger}] = 1.
\end{equation}

Then the one-particle Hamiltonian, ${\cal H}_{0}$ for 
the clean system which is quadratic in ${\Vec \pi}$, 
can be expressed as
\begin{equation}
{\cal H}_{0} = \hbar \omega _c \left (a^{\dagger} a + \frac 12 \right )
\label {eq: aalg}
\end{equation}
in terms of the operator $a$ alone, which is natural 
since $b$ involving ($X$, $Y$) should not appear in a translationally 
invariant system.  
Since the Hamiltonian has the same form as a 
linear harmonic oscillator, the energy eigenvalues are 
\begin{eqnarray}
E_N &=& \hbar \omega _c \left (N+\frac 12 \right ),  \ \ \  (N=0,1,2,...)
\label {eq: Lqz}
\end{eqnarray}
where $N$ is called the Landau index.  

\begin{figure}[ht]
\begin{center}
\includegraphics[width=11.7cm,clip]{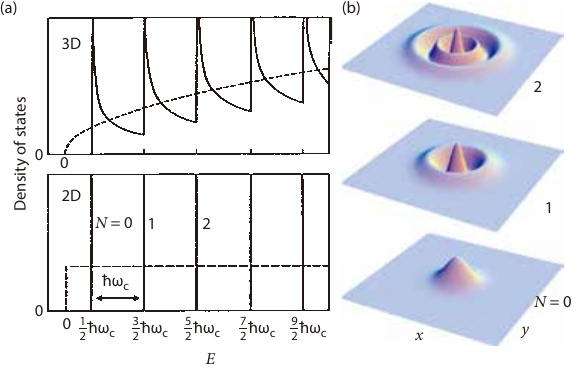}
\caption{
(a) Density of states for a clean system in the absence 
(dashed lines) or in the presence (solid lines) of a magnetic 
field in two-dimensional (lower panel) or in three-dimensional 
(upper) systems.  
(b) Wavefunctions for a 2D system  in the symmetric gauge for various values of the 
Landau index $N$.}
\label{2D3D}
\end{center}
\end{figure}

So we have here a truly abnormal situation where the energy 
spectrum, despite the system being a bulk, is completely 
discrete (Fig.\ref{2D3D}(a)).   
In three-dimensional systems we do have Landau's quantisation, 
but the extra motion along ${\Vec B}$ makes the density of 
states a continuum  (Fig.\ref{2D3D}(a)).
In 2D each level, called the Landau level as 
labelled by Landau index, has then a macroscopic degeneracy.  
The degeneracy, $n_{\phi}$, can be estimated by noting that 
the density of states of a 2DEG, which is a constant, 
$D(E)=m^*/2\pi \hbar^2$ per unit area, integrated over 
an interval $\hbar \omega_c$ should correspond to 
the degeneracy per unit area, 
\begin{equation}
n_{\phi} = \hbar \omega_c D(E) = 1/(2\pi \ell^2).
\end{equation}
This number can be expressed as 
\begin{equation}
n_{\phi} = B/\phi _0,
\end{equation}
where $\phi _0 \equiv c h/e = 4\times10^{-7}\;{\rm G}\cdot {\rm cm}^2$ is the flux quantum, 
so that $n_{\phi}$ amounts to the number of flux quanta 
penetrating the unit area.  Alternatively, we can say that 
the total degeneracy, $S/2\pi \ell^2$, 
is of the order of the number of cyclotron orbits that 
cover the sample area $S$.

We can more precisely realise this in terms of the wavefunctions.  
We first note that the operator $b$ is 
related with the orbit centre as
$X^2 + Y^2 = 2 \ell ^2 \left (b^{\dagger} b + \frac 12 \right )$.
The formulation so far does not depend on the gauge 
(i.e., how we fix the vector potential ${\Vec A}$ which 
has an ambiguity 
related with the gauge transformation). 
The wavefunction does depend on the gauge.  
Let us adopt the symmetric gauge, ${\Vec A} = 
\frac 12 \,{\Vec B} \times {\Vec r}$.   
This amounts to taking, out of the degenerate wavefunctions, 
the set that 
diagonalise, simultaneously with the energy, 
the angular momentum, $\hat{{\Vec L}} = {\Vec r} \times {\Vec p}$, 
which points to $z$ when the motion is within the $(x,y)$ plane.  
We can show that 
\begin{equation}
L_z = [{\Vec r}\times ({\Vec  \pi}-(e/c){\Vec A}]_{z} 
 =  a^{\dagger} a - b^{\dagger} b
\end{equation}
with $\hbar=1$.  
For the lowest $N=0$ Landau level we have a harmonic-oscillator 
form for $\hat L_z =  - b^{\dagger} b$, 
so that the eigenfunction having the eigenvalue of $L_z=-m$ is 
given by
\[
|\,m\,\rangle = \frac {(b^{\dagger})^m}{\sqrt{m !}}
\, |\,0\,\rangle \quad (m=0, 1, 2, \ldots),
\]
where $|\,0\,\rangle$ is the vacuum of boson $b$.  
If we go to the first-quantised form, the wavefunction is expressed, in 
polar coordinates $(r,\theta)$, as
\begin{equation}
\psi_{_{Nm}}({\Vec r}) \propto \exp{(-{\rm i} m \theta-r^2/4\ell^2)} \,
r^{m} \,L_{_N}^{|m|}(r^{2}/2\ell^{2}), 
\label {eq: obwfl}
\end{equation}
where $L_{_N}^{(m)}(z)$ is the associated Laguerre polynomial.  
They are depicted in Fig.\ref{2D3D}(b).  
By restricting the radius $R$ of these wavefunctions 
within a radius of a disk, we recover the 
degeneracy of a Landau level.

In terms of the degeneracy, we can now define 
the Landau level filling factor, i.e., the fraction 
of the occupied states for a given Landau level.  
If we denote the density of electrons by 
$n_e$, the Landau level filling factor $\nu$ is
\begin{equation}
\nu \equiv n_e/n_{\phi} = 2\pi \ell^2 n_e.  \label{nphi}
\end{equation}
In other words, $1/\nu$ is the number of flux quanta 
per electron.  The above is the essence of Landau's quantisation, formulated by 
Landau in 1930, and the discovery of the QHE in 1980 
coincided with its half centenary.

When only the lowest (or lowest few) Landau level(s) 
are occupied (i.e., $\nu <\sim 1$)  for 
strong enough magnetic fields, the situation is 
called the ``quantum limit".  
For GaAs with a typical density of electrons 
$n \sim 10^{11}\,{\rm cm}^{-2}$, we have 
\[
\nu =  \frac{4}{B{\rm [T]}}\times n[10^{11}{\rm cm}^{-2}]
\]
where $B$ is in units of Tesla and $n_e$ in $10^{11} {\rm cm}^{-2}$, 
so that the quantum limit is realised for $B >\sim 4$ T.  

So far we have considered a clean system.  In the presence of 
disorder, such as a random potential $V({\Vec r})$, arising from impurities, 
 interface roughness etc, 
cyclotron orbits that have different centre coordinates $(X,Y)$ are 
no longer degenerate, but are subject to scattering.  
Then the equation of motion for $(X,Y)$
are obtained from its commutator with ${\cal H}$ as
\begin{eqnarray}
\dot{X} &=& \left(\frac{{\rm i}}{\hbar}\right)[V,X] =
\frac{\ell^2}{\hbar}\frac{\partial V}{\partial y}, \nonumber \\
\dot{Y} &=& \left(\frac{{\rm i}}{\hbar}\right)[V,Y] =
-\frac{\ell^2}{\hbar}\frac{\partial V}{\partial x}.
\end{eqnarray}
The fact that 
$\dot{\Vec R} \propto \hat{\Vec e}_z \times \nabla V$ 
implies that 
${\Vec R} \equiv (X,Y)$ moves, classically, along equipotential contours with 
a velocity proportional to $|\nabla V|$.  
Quantum mechanically, however, the Hamiltonian, with the Landau wavefunctions as a basis, reads
\begin{equation}
{\cal H} = \sum_{_{NM}}\left(N+\frac{1}{2}\right)\hbar\omega_c
c^{\dag}_{_{NX}}c_{_{NX}} + \sum_{_{NXN^{\prime}X^{\prime}}}\langle NX
| V | N^{\prime}X^{\prime} \rangle c^{\dag}_{_{NX}}c_{_{N^{\prime}X^{\prime}}},
\end{equation}
where $c^{\dag}_{_{NX}}$ creates an electron in the 
Landau's wavefunction $ |NX\rangle$ (here in the Landau gauge, 
${\Vec A} = (0, Bx)$,  rather 
than in the symmetric gauge), and 
the second term on the right-hand side represents 
the quantum mechanical hopping between cyclotron orbits 
at different positions.  The ``hopping" matrix element $\langle NX
| V | N^{\prime}X^{\prime} \rangle$ becomes large when 
the random potential varies rapidly in space (on the 
magnetic length scale, eqn(\ref {eq:maglength}), 
since each function $|NX\rangle$ 
has a spatial extension $\sim \ell$).  

In the presence of the hopping, each Landau level is broadened from the 
line spectrum.   
Electronic structure and transport properties of the disordered, 
Landau-quantised 
systems in 2D were theoretically 
studied with the self-consistent Born approximation 
in the 1970's by Uemura, Ando and 
coworkers \cite{Ando, AndoSternFowler}.   
In this formalism 
the lifetime of an electronic state due to disorder is taken into 
account in a self-consistent way, which is imperative, 
since the unperturbed system has anomalous, 
delta-function spectra.  
Let us assume that the random potential 
$V({\Vec r}) = \sum_i 2\pi\ell^2V_0\delta(\br-\br_i)$ is expressed as a 
sum of the contributions from short-range (delta-function like) 
impurity potentials at position $\br_i$ with the 
average number of impurities per unit area $n_i$.  
For dense impurities (to be more precise, for the dimensionless 
concentration $c_i \equiv 2\pi\ell^2n_i \gg 1$) 
we can adopt the Born approximation.  There, 
the self-energy due to the impurity 
scattering is given as 
$\Sigma(E)  = c_iV^2_0 G(E)$, where the self-consistency 
demands that Green's function, $G(E)$, contain the effect of $\Sigma(E)$.  
Then each Landau level is broadened with a width, 
$\Gamma = 2\sqrt{c_i}V_0$.  

In other words, the ratio between the Landau-level broadening $\Gamma$ 
and the cyclotron energy is 
\begin{equation}
\Gamma/(\hbar\omega_c) \sim 1/(\omega_c\tau_0)^{1/2}, 
\end{equation}
where $\tau_0 = (n_i V_0^2m^*/h^3)^{-1}$ is the scattering relaxation time due to the disorder 
in zero magnetic field.  Intuitively, the quantity $\omega_c\tau_0$ 
gives a measure of how many times an electron can 
rotate on a cyclotron orbit between scattering events on average.  
For sufficiently large magnetic field and/or small disorder we have 
$\Gamma/\hbar\omega_c < 1$ (i.e., $\omega_c\tau_0 > 1)$, 
for which Landau levels are separated, while Landau levels are 
merged in the opposite condition.  
To give an idea about the magnitudes of relevant quantities, we have 

$\hbar\omega_c = 0.12 B{\rm [T]}(m_0/m^*)$ meV, 

$\Gamma \sim 0.15\sqrt{B{\rm [T]}/\mu}$, 

\noindent where $B$ is measured in units of Tesla, the effective 
mass is set to $m^* = 0.067m_0 \,\, (m_0$: bare mass of an electron) for GaAs 
and $\mu = e\tau_0/m^*$ 
is the carrier mobility here measured in units of $10^4$ cm$^2$/(V$\cdot$s).

\clearpage

\section{Integer quantum Hall effect  --- experiments}

\begin{figure}[ht]
\begin{center}
\includegraphics[width=11.7cm,clip]{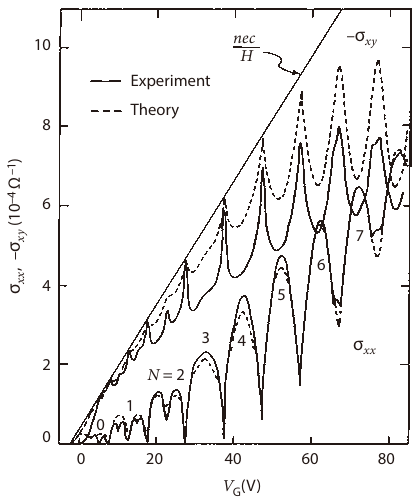}
\caption{
A typical experimental result (solid lines) with a 
theoretical fit (dashed) for $\sigma_{xx}$ and $\sigma_{xy}$, 
against the gate voltage $V_G$  which is roughly $\propto n$ in a 
MOSFET sample [after T. Igarashi, J. Wakabayashi and S. Kawaji, 
Suppl. Progr. Theoret. Phys. No.57, 176 (1975)].}
\label{Kawaji}
\end{center}
\end{figure}

The Landau quantisation in 2DEG described above is 
reflected in various properties, especially in transport 
properties. 
There are prehistories for the 
the QHE physics that are precursors of the discovery.  
As early as in the 1960's Shubnikov de Haas effects 
were observed in Si-MOSFET's by Landwehr's group, and by 
groups in US and in Japan.  
Shubnikov de Haas effect is an 
oscillations (versus an external magnetic field $B$) in 
transport properties, which is general enough but the effect becomes 
peculiar in 2DEG, 
since the oscillations in $\sigma_{xx}$ and $\sigma_{xy}$ 
reflect the line-like Landau levels.  
  In the 1970's more elaborate experimental 
studies did exhibit oscillatory behaviours 
indicative of Landau levels in 2D.  Figure \ref{Kawaji} shows a 
typical example\cite{kawaji}, along 
with a theoretical result.  
It was also recognised that there are regions of vanishing 
$\sigma_{xx}$ and flat $\sigma_{xy}$ between Landau levels, which are 
called ``plateaux".

\begin{figure}[ht]
\begin{center}
\includegraphics[width=11.7cm,clip]{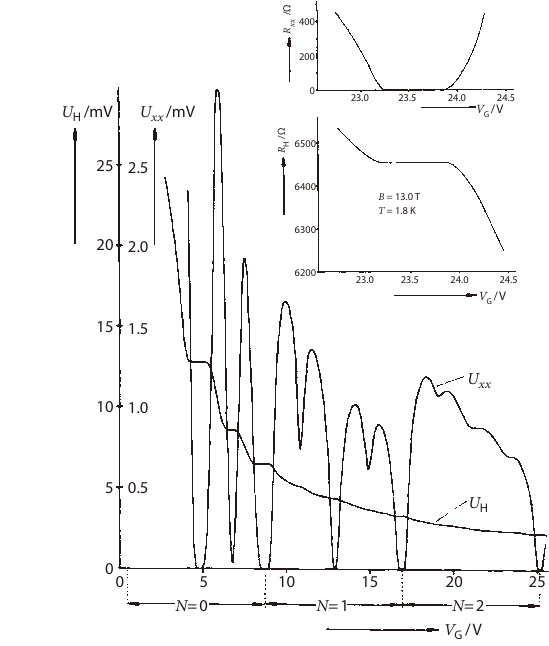}
\caption{The original QHE result for the Hall ($U_{\rm H}$) 
and longitudinal ($U_{xx}$)  voltages against the gate voltage $V_G$ at 
$T=1.5$ K and $B=18$ T.  Inset shows a detail around the fully 
occupied $N=0$ Landau level
[after K. von Klitzing, G. Dorda and M. Pepper, Phys. Rev. Lett.  {\bf 45}, 494 (1980)].
\label{Klitzing}}
\end{center}
\end{figure}

\begin{figure}[ht]
\begin{center}
\includegraphics[width=11cm,clip]{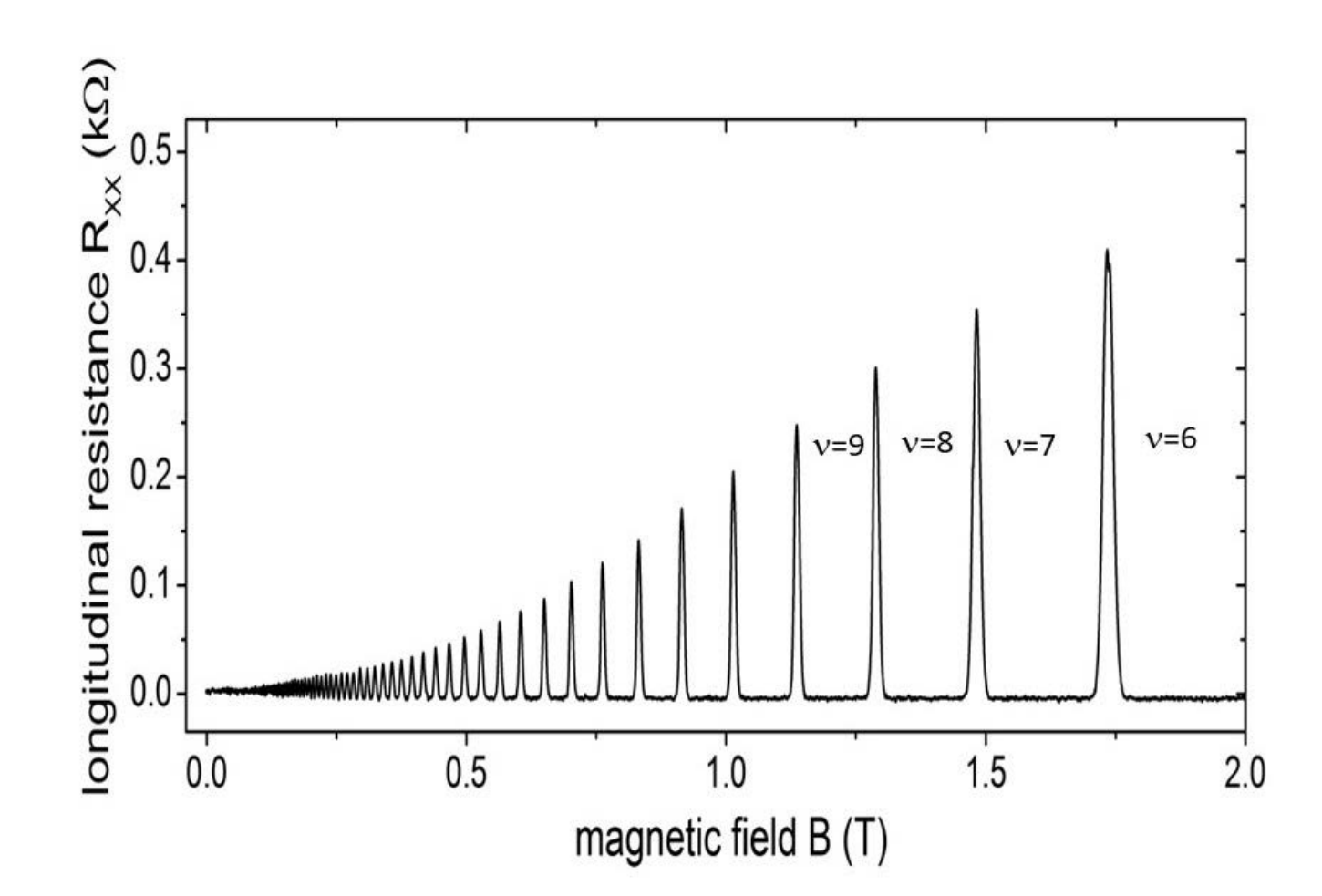}
\caption{A typical result for the longitudinal resistance 
$R_{xx}$ and Hall resistance 
$R_{\rm H}$ in the GaAs-Al$_x$Ga$_{1-x}$As  IQHE system.  
[after B. Friess, PhD thesis (Technische Universit\"{a}t 
M\"{u}nchen, 2014) 
https://mediatum.ub.tum.de/1219842].
\label{Paalanen}}
\end{center}
\end{figure}

In 1980, von Klitzing, Dorda and Pepper 
found an astonishing behaviour 
in the Hall conductivity as shown in 
Fig.\ref{Klitzing}.\cite{klitzingPRL}  In a Si MOSFET in strong magnetic fields 
($\sim 20$ T), 
the heights of the Hall resistance $R_{xy}$ are quantised very 
accurately 
into (integer)$^{-1}$ times $h/e^2 = 25\;813\;\Omega$, or when 
translated to the Hall conductivity $\sigma_{xy}$ as
\begin{equation}
\sigma_{xy}= - N\frac{e^2}{h}\hspace{3em}(N:\mbox{integer}).
\end{equation}
Astonishing, because (a) the quantised value only contains the fundamental physical constants, 
the elementary charge $e$ and Planck's constant $h$ (whereas 
ordinarily transport properties are naturally affected by 
various material parameters, etc),  and 
(b) this occurs in disordered systems.  
This has become known as the quantum Hall effect.  A few years after this 
discovery the fractional 
quantum Hall effect was discovered, so the original effect 
is called 
the integer quantum Hall effect.   Remarkably, 
the accuracy of the quantisation 
is experimentally confirmed to be better than $10^{-7}$. 
Hall effect itself was discovered by Edwin Hall in 1879, so 
it was almost exactly a century later when the quantum 
Hall effect was discovered.   
Subsequently, experimental data were refined, where the 
quantum Hall steps are almost a series of step 
functions, as typically shown in Fig.\ref{Paalanen}, 
which are then interpreted in terms of the localisation, 
as we shall see in the section on localisation.

 \subsection{Materials and sample geometry}

Historically the integer quantum Hall effect was discovered 
in Si-MOSFET's, but subsequently studied for 2DEG's 
in 
semiconductor heterostructures, typically GaAs/AlGaAs 
(Fig.\ref{2DEG}).  
Integer QHE has also been observed in 
heterostructures other than GaAs/AlGaAs, 
which include 2D 
hole (as opposed to electron) gas systems in p-type GaAs, 
type III heterostructures such as GaSb-InAs where 2D electron and 
2D hole gases coexist side by side, and 
Si/Si$_{1-x}$Ge$_x$ 
strained heterostructures with different 
band structure and higher effective masses than 
those in Si MOSFET's.

In measuring Hall resistivity, basically two sample geometries are 
used --- 
Hall bar geometry and Corbino geometry (Fig.\ref{samplegeometry}(b)).  
The former is usually adopted with a multi-terminal 
geometry.  In the latter, where electrodes are attached 
to the inner and outer perimeters of an annular sample, 
an advantage is that we do not have to worry about 
the sample edges and edge transports, nor 
do we need to worry about ``hot spots" (Fig.\ref{samplegeometry}(c)); 
see the section 
on real-space imaging below).  Namely, 
in the QHE condition, where we have 
$\sigma_{xy}$: integer$\times e^2/h, \sigma_{xx} = 0$ 
with the Hall angle [$= {\rm tan}^{-1}(\sigma_{xy}/\sigma_{xx}$)] 
of 90 degrees, 
the Hall current flows around the annulus.  
QHE has also been observed in Corbino geometry.  
In fact some of the early Shubnikov de Haas experiments 
were done for this geometry.  

An electron has a spin 1/2, and the associated Zeeman energy in a 
magnetic field.   The Zeeman energy is much smaller than 
the cyclotron energy in usual experimental situations 
(although the spin splitting is enhanced due to 
exchange interactions to be precise), 
so that every Landau level is almost twofold spin degenerated.   
Later, the effects of the spin splitting on 
QHE have been elaborated with various experimental 
techniques.  In the case of Si-MOSFET, a valley degeneracy also 
exists, since the bulk Si has multiple valleys in the band 
structure. 

\clearpage

\subsection{Optical properties}

QHE systems exhibit characteristic optical properties as well as 
characteristic transport properties.  
Optical measurements have advantages that (i) 
this is a contactless method, and (ii) we can probe 
both luminescence and absorption.  
In the QHE regime, luminescence spectra 
have been obtained for Si MOSFET's and 
semiconductor heterostructures.  The luminescence spectra, 
which measures the radiative recombination of 2D electrons 
with photoexcited holes, 
reflect the Landau level structure.  One feature is  
that the Landau-level width as deduced 
from the width of the luminescence spectra oscillates with 
magnetic field with a maximum 
every time a Landau level is filled.  This is 
associated with the screening which becomes 
effective when the Landau levels are partially filled (i.e., an open shell).  
There are other properties which include spin-dependent relaxations.

\begin{figure}[ht]
\begin{center}
\includegraphics[width=6cm,clip]{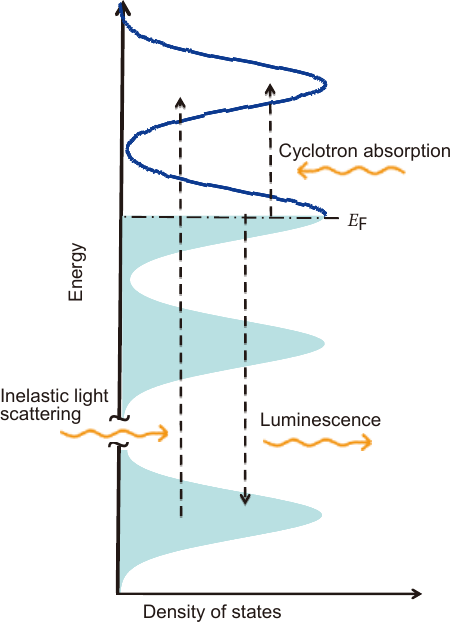}
\caption{Various optical processes are schematically shown.
\label{optical}}
\end{center}
\end{figure}

Cyclotron absorption is another important experimental technique 
in the QHE regime.  The technique can probe lower carrier-density regions than in transport measurements.  Landau level structures 
have been probed, where oscillatory line widths, 
spin splitting, nonparabolicity effects,  etc have been observed.  
Inelastic light scattering is another powerful optical method that 
compliments luminescence.  These methods are schematically 
illustrated in Fig.\ref{optical}.  

More recently, the optical Hall conductivity $\sigma_{xy}(\omega)$ 
is theoretically predicted to have 
a Hall plateau structure in the THz regime 
in quantum Hall systems both in 2DEG and graphene, which we shall 
describe in Section ``Optical properties in graphene" below.

\subsection{Other properties}

\begin{figure}[ht]
\begin{center}
\includegraphics[width=8cm,clip]{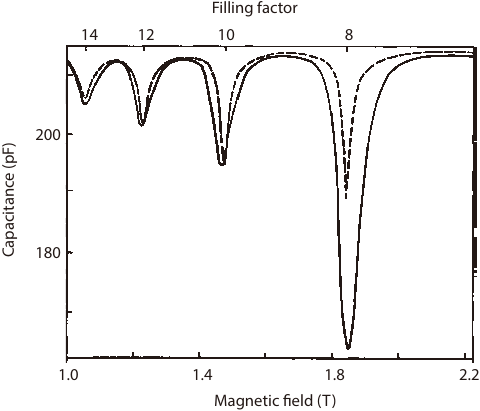}
\caption{ Experimental (solid line) and theoretical (dashed) results 
for the capacitance against magnetic field.  Landau level filling 
is indicated on the upper axis  [after T. P. Smith et al, Phys. Rev. B {\bf 32}, 2696 (1985)].
\label{capacitance}}
\end{center}
\end{figure}

\begin{figure}[ht]
\begin{center}
\includegraphics[width=8cm,clip]{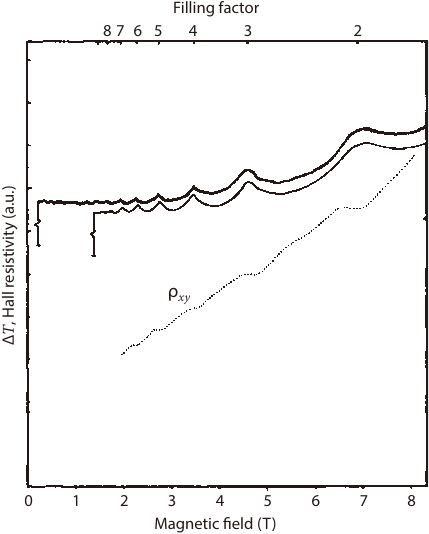}
\caption{Experimental (thick line) and theoretical (thin) results 
for the magnetisation as probed by the temperature change, 
$\Delta T$, in the heat-pulse method, along with $\rho_{xy}$ against magnetic 
field [after E. Gornik et al, Phys. Rev. Lett. {\bf 54}, 1820 (1985)].
\label{magnetisation}}
\end{center}
\end{figure}

There are host of other properties that have been measured.  
One is the electronic specific heat, which probes the Landau quantisation 
through the density of states (that include both localised and 
delocalised states), while transport measurements mainly probe 
the delocalised states. 
Another method to probe the density of states 
is the magnetocapacitance (Fig.\ref{capacitance}).  
Magnetisation has also been measured, with torque magnetometers or 
micromechanical cantilevers,  where jumps 
are observed as $E_F$ traverses the Landau 
levels (Fig.\ref{magnetisation}).

\begin{figure}[ht]
\begin{center}
\includegraphics[width=8cm,clip]{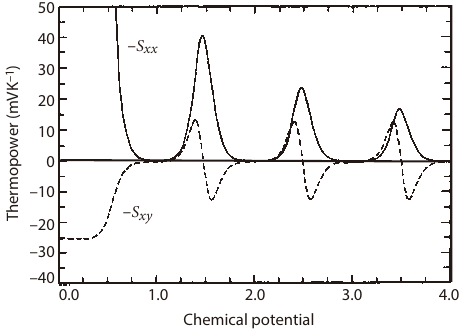}
\caption{A typical theoretical result 
for the diagonal ($S_{xx}$) and off-diagonal ($S_{xy}$) components of 
the thermopower tensor against chemical potential  
[after M. Jonson and S. M. Girvin, Phys. Rev. B {\bf 29}, 1939 (1984)].
\label{thermopower}}
\end{center}
\end{figure}

Various other  properties have been experimentally studied, 
among which is  the thermoelectric effect.  
This probes how the electrons sustain temperature 
gradients in 2DEG's, 
where the dominant mechanism for the thermopower is 
phonon drag.    Another magneto-thermoelectric effect is 
a heat flow (or a temperature gradient) 
that is perpendicular 
to both the electric current and the magnetic field, known as the 
Nernst-Ettingshausen effect.   Namely, the electric 
field ${\Vec E}$ and the temperature gradient $\nabla T$ are 
related as 
${\Vec E} = {\cal S}\nabla T$.   Here ${\cal S}$ is the 
thermopower tensor, where $S_{xx}$ corresponds to the thermopower while 
 $S_{xy}$ to the Nernst-Ettingshausen coefficient 
(Fig.\ref{thermopower}).\cite{fletcher}.  
The effect has been studied both theoretically and experimentally 
in conjunction with the breakdown of QHE

\subsection{QHE in other materials}

\begin{figure}[ht]
\begin{center}
\includegraphics[width=7cm,clip]{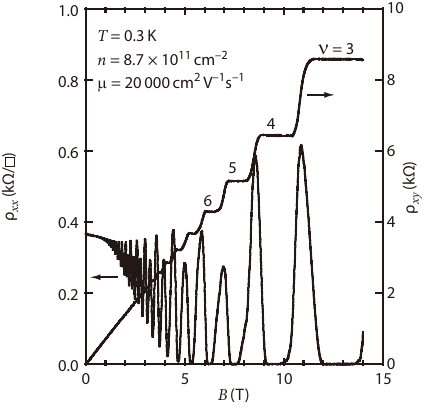}
\caption{QHE in an oxide heterostructure ZnO/Mg$_x$Zn$_{1-x}$O. 
After A. Tsukazaki et al, Phys. Rev. B {\bf 78}, 233308 (2008).
\label{oxide}}
\end{center}
\end{figure}

There are a multitude of developments for the integer QHE in 
wider classes of materials.  
Let us here just mention the realisation of QHE 
in oxide heterostructures  as an example.  
The system is zinc oxide (ZnO), which is an insulator, or 
a wide-gap semiconductor.   When heterostructures such as 
ZnO/Mg$_x$Zn$_{1-x}$O are grown with MBE, MgZnO layer acts as 
a potential barrier for the 2DEG in ZnO layer in realising 
a 2DEG.   The carrier density (typically $\sim 10^{12}\;{\rm cm}^{-2}$) systematically depends on the composition 
$x$ and the growth temperature, where spontaneous and piezoelectric 
polarisation effects work to accumulate carriers at the heterointerface.  
The condition necessary to have QHE (i.e., 
$\omega_c \tau_0 > 1$) is met, and a typical result (Fig.\ref{oxide}) exhibits 
a clear IQHE.\cite{Tsukazaki}

\clearpage

\section{Integer quantum Hall effect --- theories}

\subsection{Localisation in Landau levels}

Let us start with the localisation problem in the QHE system, 
since this has to do with a first essential question about the 
integer QHE --- the presence of plateaux in $\sigma_{xy}$ 
versus the density of electrons $n$ (or vs $B \propto 1/\nu$ for a fixed $n$).  Experimentally the plateau in $\sigma_{xy}$ 
goes hand in hand with a region of vanishing $\sigma_{xx}$.  
This is a curious situation, since, usually $\sigma_{xy}$ should 
be an increasing function of $n$ while $\sigma_{xx}$ as 
a function of $n$ should not 
be zero (as opposed to the quantities as 
functions of energy $E$).  

\begin{figure}[ht]
\begin{center}
\includegraphics[width=11.7cm,clip]{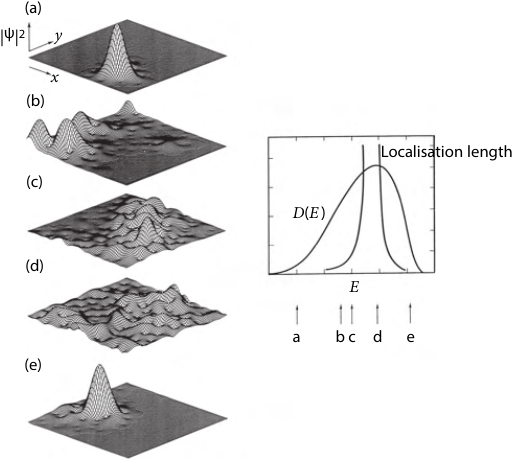}
\caption{Typical wavefunctions in a disordered QHE system obtained 
in a computer simulation for various eigenenergies as indicated on the 
right panel, which shows the density of states and the localisation 
length against energy [after H. Aoki, J. Phys. C {\bf 10}, 2583 (1977)].}
\label{wavefunction}
\end{center}
\end{figure}

\begin{figure}[ht]
\begin{center}
\includegraphics[width=13cm,clip]{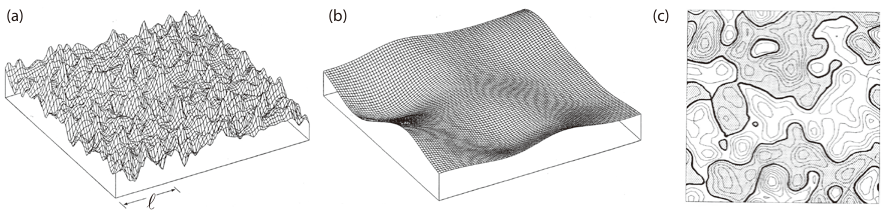}
\caption{A random potential rapidly-varying as compared with 
the magnetic length $\ell$ (left panel), and a 
slowly-varying one (rigth) are schematically shown.  
(c) A typical contour plot with shaded areas 
indicate $E<0$ and thick lines $E=0$ contour 
[after H. Aoki, Rep. Prog. Phys,  {\bf 50}, 655 (1987)].
\label{percolation}}
\end{center}
\end{figure}

This has been explained from the Anderson localisation 
of the wavefunctions in the Landau levels that arises from 
disorder in the system.  The Anderson localisation was 
proposed back in 1958 for general electron systems in 
the presence of disorder, and 
was subsequently culminated as the scaling theory of localisation in 1979.  
In a clean crystal every wavefunction is a Bloch state that extends over the 
entire sample.  By contrast, in a disordered system wavefunctions can be spatially 
localised, whose spatial extension is characterised by the localisation 
length that depends on energy.  
In the presence of magnetic fields, the Anderson localisation was suggested to occur as well for 
wavefunctions in the Landau levels in the 1970's 
(Fig.\ref{wavefunction}).  
For random potentials that vary slowly in space, 
an electron follows, semiclassically, an 
orbit that is an equipotential contour as we have shown in a section above.  
Quantum mechanically, however, 
an electron tunnel between these orbits even for a slowly-varying 
potential, where the quantum mechanical hopping becomes 
more frequent for rapidly-varying (i.e., short-ranged) random potentials 
(see Fig.\ref{percolation}).   
So the question becomes how the localisation arises 
in the presence of the quantum mechanical hopping.   
Then the plateaux in $\sigma_{xy}$ and regions of vanishing $\sigma_{xx}$ 
have been suggested to come from the localisation.\cite{AokiAndo}  
A peculiarity in the localisation in the QHE system is that 
the disorder has a dual role: disorder causes the localisation 
on one hand, but it gives rise to quantum mechanical 
hopping of cyclotron guiding centres to contribute to transport at the 
same time, as pointed out by Aoki and Kamimura\cite{KamimuraAoki}.

 \subsection{Linear-response theory}

The most standard method to treat conductivities microscopically 
is the linear-response 
theory, or the Kubo formula.  Aoki and Ando\cite{AokiAndo} have applied this to the QHE 
problem.  While the plateaux in $\sigma_{xy}$ and vanishing regions 
in $\sigma_{xx}$ are explained by the localisation, the puzzle remains 
as to why the value of the plateaux in 
disordered systems is 
quantised into the universal $e^2/h$, which is originally the value for 
the clean 2DEG when the Landau level filling coincides with an integer.  
In other words, if a disorder makes $\sigma_{xx}$ zero, why 
does the disorder allow $\sigma_{xy}$ to stick to the quantised value.

In the linear-response theory, the conductivity is given by a 
current-current correlation function.  
In the QHE system, the current has to do with the dynamics of the 
cyclotron guiding centre $(X,Y)$, which is subject to quantum mechanical 
hopping.  
The conductivities, both longitudinal and Hall, are given as 
correlation functions of $\dot{X}$ and  $\dot{Y}$ (with a dot 
denoting a time derivative) as
\begin{eqnarray}
\sigma_{xx} &=& e^2\langle\langle\dot{X}\dot{X}\rangle\rangle,\nonumber \\
\sigma_{xy} &=& -\frac{ne}{B} + \Delta\sigma_{xy},\\
\Delta\sigma_{xy} &=& 
\frac{e^2}{2}(\langle\langle\dot{Y}\dot{X}\rangle\rangle -
\langle\langle\dot{X}\dot{Y}\rangle\rangle).
\end{eqnarray}
Here $\langle\langle AB \rangle\rangle \equiv 
\frac{1}{L^2}\int_{0}^{\infty}{\rm
  d}t\,{\rm exp}(-\delta t) \,\int_{0}^{\beta}{\rm d} \lambda \,\langle A(-{\rm i}\hbar\lambda)B(t)\rangle$, where $\langle \rangle$ is the canonical 
ensemble average plus the average over disorder, 
$\beta = 1/k_BT$, $\delta$ a positive infinitesimal, and 
$A(t)$ the Heisenberg representation of an operator $A$.

$\Delta \sigma_{xy}$ is expressed as a combination 
$\langle \langle \dot{Y}\dot{X} \rangle \rangle - \langle \langle \dot{X}\dot{Y} \rangle \rangle$ for the following reason.  
For the conductivity tensor in a magnetic field $B$, 
Onsager's reciprocal theorem dictates that
\[
\sigma_{\mu\nu}(B) = \sigma_{\nu\mu}(-B),
\]
so that we can decompose $\sigma$ into the symmetric part 
$\sigma^s$ and the antisymmetric part $\sigma^a$.  
Then the current ${\Vec j}$ is expressed as 
\begin{equation}
j_{\mu}
= \sum_{\nu}\sigma_{\mu\nu}^{s}E_{\nu}
+ \sigma_{xy}^{a}({\Vec E}\times\hat{{\Vec e}}_{z})_{\mu},
\end{equation}
where the part of the current (i.e., the Hall current) 
that is induced by the magnetic field ($\parallel z$ and 
perpendicular to the applied electric field ${\Vec E}$) 
is related to $\sigma^a$.\cite{LandauLifshitzEM}

The linear-response formula may be written in terms of 
Green's function as 
\begin{equation}
\Delta\sigma_{xy}
=\frac{e^2\hbar}{{\rm i}\pi L^2}
\int_{-\infty}^{\infty} {\rm d} E \, f(E)
\left\langle {\rm Tr}
\left[
\dot{X}\frac{\partial}{\partial E} {\rm Re} G(E) \dot{Y} {\rm Im} G(E)
- (\dot{X}\leftrightarrow\dot{Y})
\right]
\right\rangle,
\end{equation}
where $G(E) = (E-{\cal H} + {\rm i} \delta)^{-1}$
is the Green's function, 
$f(E) = [e^{^{\beta(E-\mu)}+1}]^{-1}$ 
Fermi-Dirac distribution function, and 
$\leftrightarrow$ denotes the term with $\dot{X}$ and $\dot{Y}$ exchanged.  
 In this expression the 
Hall conductivity $\sigma_{xy}$ is contributed by 
all the states below $E_F$, unlike $\sigma_{xx}$ that 
is related to an energy-dissipating process around $E_F$.  
In terms of the eigenstates of the Hamiltonian, the Hall 
conductivity is expressed as
\begin{eqnarray}
\sigma_{xy}(\omega)
&=&
\frac{i\hbar e^2}{ L^2}
\sum_{\alpha\neq \beta}\frac{f(\epsilon_{\beta})-f(\epsilon_{\alpha})}
{\epsilon_{\beta}-\epsilon_{\alpha}}
\left(\frac{j_x^{\alpha \beta}j_y^{\beta \alpha}
-j_y^{\alpha \beta}j_x^{\beta \alpha}}{\epsilon_{\beta}-\epsilon_{\alpha}+i\delta} 
\right) ,
\label{kuboformula}
\end{eqnarray}
where $\epsilon_{\alpha}$ is $\alpha$-th eigenenergy, and 
${\Vec j}^{\alpha \beta}$ the current matrix 
elements between the eigenstates.

\begin{figure}[ht]
\begin{center}
\includegraphics[width=7cm,clip]{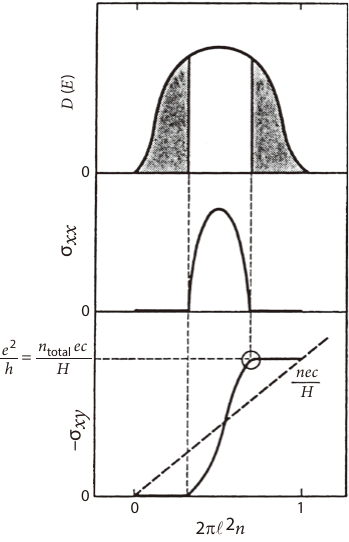}
\caption{Schematic density of states $D(E)$ (with shaded areas indicating 
the localised states), the longitudinal conductivity $\sigma_{xx}$, 
and the  Hall conductivity $\sigma_{xy}$ are schematically 
plotted against the 
Landau level filling $2\pi\ell^2 n$.  In the bottom panel 
the horizontal dashed line indicates how the quantised 
value is achieved despite the presence of localisation.
\label{AokiAndo2}}
\end{center}
\end{figure}

From this formalism, we can show, 
in the presence of localisation arising from disorder, 
that (i) the Hall conductivity $\sigma_{xy}$ should be 
rigorously flat (at $T=0$) as a function of the 
density $n$ of electrons in the region 
where the states are localised.  This accounts for the flatness 
of plateaux in the QHE, as depicted in Fig.\ref{AokiAndo2}.

\begin{figure}[ht]
\begin{center}
\includegraphics[width=7cm,clip]{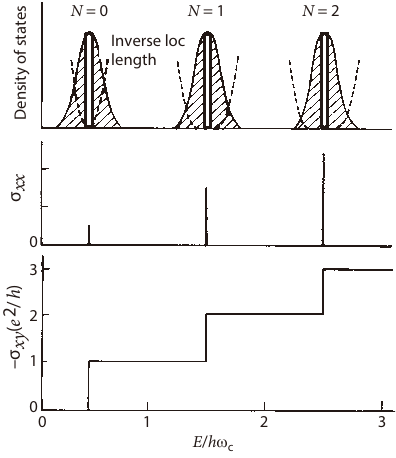}
\caption{Theoretical density of states (with shaded areas indicating 
the localised states) along with the inverse localisation length, $\sigma_{xx}$, and $\sigma_{xy}$ are schematically 
plotted against energy for a series of 
Landau levels.  In the top panel each white region representing the 
delocalised states actually has zero width in the thermodynamic limit 
at $T=0$ 
[after H. Aoki and T. Ando, Solid State Commun. {\bf 38}, 1079 (1981)]. }
\label{AokiAndo}
\end{center}
\end{figure}

We can also show that the quantised 
$\sigma_{xy} = -Ne^2/h$ in 
a plateau should hold in 
the limit of strong magnetic fields where adjacent Landau 
levels are well separated and when $E_F$ is in the gap 
between them.  This in turn implies that, 
(ii) at least in 
the limit of strong magnetic fields all the states cannot be localised, 
otherwise $\sigma_{xy}$ would be zero over the whole region.  
This is important, since, according to the scaling theory of 
localisation all the states must be localised in two dimensions.  
The only way to go around this universal argument is to go to 
another universality class, either to the unitary class 
(e.g., systems in magnetic fields where the time-reversal 
symmetry is broken, see Fig.\ref{topologicalPeriodicTable}) or to the symplectic class (e.g., 
systems that have spin-orbit interactions).  QHE system belongs to 
the unitary class, which is why delocalised states are allowed to exist, 
which in turn makes the QHE to be realised.  
The theoretical picture explained here 
for broadend Landau levels, $\sigma_{xx}$ 
and $\sigma_{xy}$ are summarised in Fig.\ref{AokiAndo}.

As we shall see below, the topological picture 
beautifully explains the quantisation of $\sigma_{xy}$ into 
$e^2/h$ for finite magnetic fields, where the 
required condition is $E_F$ being in a gap in the density of 
states or in a localised regime (mobility gap).  
Even in the treatment above, however, different Landau levels 
are not treated as being completely independent.  
As seen from the fact that the Landau's quantisation 
is formally equivalent to a 1D harmonic oscillator, the 
current operator has matrix elements between $N$ and $N\pm 1$ 
levels, so the mixing between them is implicitly included 
(in fact, the treatment in terms of the guiding centre ($X, Y)$ 
corresponds to taking this mixing to the leading order).  

It may first seem counter-intuitive that 
$\sigma_{xy}$ attains the quantised value 
despite the presence of localised states.  
Physically, we can say the following.  
From the equation of motion for the guiding centre 
the Hall current $j_x$ in an electric field $E_y$
is given as 
\[
j_{x}=-(1/B)\sum_{\alpha^{\prime}}^{\rm delocalised}
(\langle\alpha^{\prime}|\partial V/\partial
  y|\alpha^{\prime}\rangle + eE_{y}).  
\]
There is a kind of sum rule, 
$\sum_{\alpha^{\prime}}^{\mbox{\scriptsize all}}
  \langle\alpha^{\prime}|
  \partial V/\partial y |\alpha^{\prime}\rangle = 0$, 
which means that delocalised states move faster to just 
{\it compensate} the localised states, which 
has been confirmed numerically and field theoretically as well. 

\subsection{St\v{r}eda-Widom formula}

As another formalism for the Hall conductivity, 
St\v{r}eda's formula\cite{widomStreda} is also often evoked.  
He showed that the Kubo formula for the Hall conductivity 
can be decomposed into a form, $\sigma_{xy} = \sigma_{xy}^I + \sigma_{xy}^{II}$, 
where $\sigma_{xy}^I$ is the Drude-like part that tends to 
$\sigma_{xy}^I \rightarrow -\omega_c\tau_0\sigma_{xx}$ in 
the relaxation time picture.  The second term is expressed as
\begin{equation}
\sigma_{xy}^{II} = ec\left[ \frac{\partial N(E)}{\partial B}\right]_{E=E_F},
\end{equation}
where $N(E)$ is the  integrated density of states.  
When the Fermi energy $E_F$ is in an energy gap, $\sigma_{xy}^I$ vanishes, 
while we can show that $\sigma_{xy}^{II} = -(e^2/h)N$ 
when $E_F$ is in a gap between the $N$-th and $(N+1)$-th 
Landau levels.   Thermodynamically, we have 
the electric field, ${\Vec E} = \nabla(\mu/e)$ with $\mu$ being the 
electrochemical potential, so that the 
Hall current ${\Vec j}_H = c\nabla \times {\Vec M}$ 
with ${\Vec M}$ being the magnetisation can be 
expressed as 
${\Vec j}_H = ec {\Vec E} \times (\partial {\Vec M}/\partial\mu)$.  
If we use, following Widom, the thermodynamic Maxwell's relation, 
$\partial {\Vec M}/\partial\mu = \partial N/\partial{\Vec B}$, 
which comes straight from ${\Vec M} = - \partial \Omega/\partial{\Vec B}$ 
with the grand potential $\Omega$, we recover the formula for 
$\sigma_{xy}^{II}$.\cite{widomStreda}  
Note that the St\v{r}eda-Widom formula is 
applicable to the situation where the spectrum has an energy gap.

\clearpage

 \subsection{Laughlin's gauge argument}
 
\begin{figure}[ht]
\begin{center}
\includegraphics[width=7cm,clip]{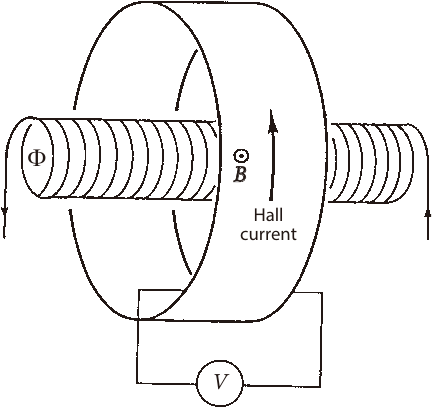}
\caption{Geometry considered by Laughlin.}
\label{Laughlingeometry}
\end{center}
\end{figure}

As a transparent approach to QHE, Laughlin has proposed 
an argument based on a gauge transformation.\cite{laughgauge}  
Consider a Gedankenexperiment, in which a QHE system is 
wound into a cylinder (but the magnetic field is still applied 
perpendicular to every point on the cylinder), 
as in Fig.\ref{Laughlingeometry}.  
To this we add 
a magnetic flux $\Phi$ (due to, say, a solenoid) that pierces 
the cylinder.  Through its vector potential, 
$A_{\theta}=\Phi/2\pi r$ (in cylindrical coordinates), 
the flux exerts an Aharonov-Bohm effect on the electrons.  
We can eliminate the vector potential with a gauge 
transformation at the cost of the boundary condition twisted to 
$\psi(\theta)\rightarrow{\rm exp}[{\rm i}(\Phi/\phi_{0})\theta]$  
on the wavefunction $\psi$, where 
$\phi_{0}=ch/e$ is the flux quantum.

If we increase $\Phi$ with time, 
we recover the original periodic boundary condition 
every time $\Phi$ increases by 
$\phi_0$ in a time interval $\Delta t$.  
When the Fermi energy $E_F$ is in a gap in the density of 
states or in the energy region for localised states (as described 
in the previous section), 
the occupation of electrons must be the same when we make 
$\Phi \rightarrow \Phi +\phi_0$.  The only change, then, should be 
a transfer of an integer ($M$) number of electrons from 
one electrode to another, which we assume to be 
attached to either edge of the cylinder as Hall probes.  
From the Maxwell equation, an electric field 
$LE_y = -(1/c)\partial \Phi/\partial t$ 
is induced along the cylinder of circumference $L$, which should result in a Hall 
current, $j_{x}=-(-e)M/(L\Delta t) = - (Me^{2}/h)E_{y}$, 
and we have $\sigma_{xy}=-Me^2/h$.  
So the derivation applies to arbitrary strength of the 
magnetic field as far as 
the Fermi energy $E_F$ is in a gap in the density of 
states, or in the energy region for localised states.

\clearpage

 \subsection{TKNN's topological arguments}

In Laughlin's argument we have introduced an Aharonov-Bohm magnetic flux, but subsequently it has been shown that we can go even further to express the 
Hall conductivity as a topological invariant, which guarantees the quantisation in a mathematically rigorous manner.   
Namely, Thouless, Kohmoto, Nightingale and den Nijs (TKNN)\cite{TKNN} have considered, 
for a QHE system periodic both in $x$ and $y$ directions 
with  twisted boundary condition 
for each of $x$ and $y$ directions, which corresponds to introducing 
a vector potential, ${\Vec A} = (A_x,A_y)$, describing fictitious fluxes.   
Then the Hall conductivity averaged over $(A_x,A_y)$ reads
\begin{eqnarray}
\frac{\langle\sigma_{xy}\rangle}{e^{2}/h}
&=& \frac{1}{2\pi{\rm i}}\sum_{j}^{\mbox{{\scriptsize occup}}}\int\!\int
\left(
\left\langle
\left.
\frac{\partial u^{j}}{\partial A_{x}}
\right|
\frac{\partial u^{j}}{\partial A_{y}}
\right\rangle
-
\left\langle
\left.
\frac{\partial u^{j}}{\partial A_{y}}
\right|
\frac{\partial u^{j}}{\partial A_{x}}
\right\rangle
\right)
dA_{x} dA_{y} \nonumber \\
&=& -i\frac{e^2}{L^2}\int\frac{d\bk}{(2\pi)^2}\sum_{\alpha}f(\varepsilon_{\alpha\bk})
\left[\nabla_{\bk}\times \langle \alpha\bk| \nabla_{\bk}| \alpha\bk \rangle \right]_z  \nonumber \\
&=& \frac{e^2}{h} \sum_{\alpha} C_{\alpha}.
\end{eqnarray}
This is the celebrated TKNN formula. 
In the first line $u^{j}$ is the $j$th eigenfunction and 
the summation is over the occupied states, 
while in the second line $L$ the sample size, 
$f$ the Fermi distribution function, $\varepsilon_{\alpha\bk}$ the energy 
of the Bloch wavefunction $|\alpha\bk \rangle$ in the $\alpha$th band, 
and $\nabla_{\bk}$ is the gradient with respect to ${\bk}$.  
The expression coincides with a topological invariant $C_{\alpha}$ 
(that always takes 
integer values) known as the first Chern character 
in the differential geometry.  
This is seen if we rewrite the formula as 
\begin{equation}
\langle\sigma_{xy}\rangle \sim \sum_{\alpha} \int d\bk \; \nabla_{\bk} \times 
{\cal A}_\alpha(\bk), 
\label{ChernCharacter}
\end{equation}
where ${\cal A}_\alpha(\bk) \equiv -i\langle \alpha\bk| \nabla_{\bk}| \alpha\bk \rangle$ is a fictitious gauge potential.  
So the Hall conductivity  (in units of $e^2/h$) is just the 
topological invariant.  
The reason why a differential geometry is relevant is that 
the wavefunction in 2DEG in a magnetic field has such a property 
that wavefunction changes in a manner that depends on the 
path along which we 
adiabatically change $(A_x, A_y)$ 
as 
$\rightarrow (A_x+\delta A_x, A_y) \rightarrow (A_x+\delta A_x, A_y+\delta A_y)$ or as 
$\rightarrow (A_x, A_y+\delta A_y) \rightarrow (A_x+\delta A_x, A_y+\delta A_y)$.  The difference is related to the phase of the wavefunction, 
which is a kind of Berry's phase that generally 
arises when a quantum system 
is subject to an adiabatic change.  
When a vector (to which a wavefunction belongs in a Hilbert 
space) ends up with different vectors for different parallel transports, 
we say the space in which the vectors reside is curved.  
The integrand in the expression for the Hall conductivity 
indeed represents the curvature (``Berry's curvature" here), 
and Euler's theorem dictates that the curvature 
integrated over the surface is an integer.  

Alternatively, we can express the linear-response Hall conductivity in terms of Green's function $G$ as
\begin{equation}
\frac{\langle\sigma_{xy}\rangle}{e^{2}/h}
= \frac{1}{8\pi^{2}}\int_{\cal C}{\rm d}z\int\!\int_{0}^{^{\phi_{0}/L}}
{\rm d}A_{x}A_{y}{\rm Tr}
\left[
G\frac{\partial G^{-1}}{\partial A_{x}}
G\frac{\partial G^{-1}}{\partial A_{y}}
G\frac{\partial G^{-1}}{\partial z}
-(x\leftrightarrow y)
\right].
\end{equation}
When $E_F$ is in a localised regime, we can close the contour ${\cal C}$ 
on the plane of complex energy $z$, and the above expression coincides with a 
topological invariant (Pontrjagin number).   This expression reduces, 
for a fixed number of electrons,  to the TKNN formula if we 
note ${\Vec j} = \partial {\cal H}/\partial {\Vec A}$.  

\begin{figure}[ht]
\begin{center}
\includegraphics[width=14cm,clip]{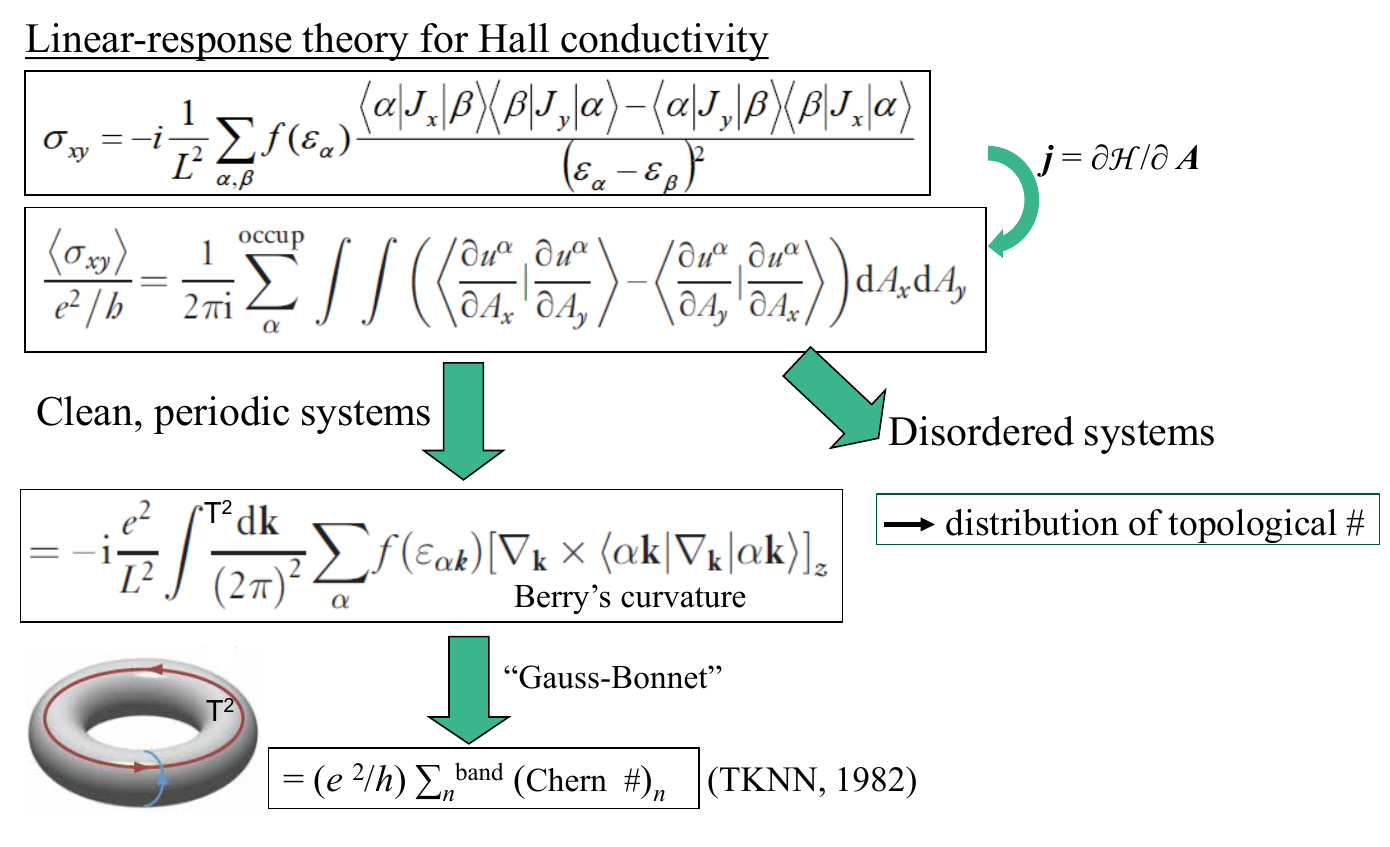}
\caption{Topological expressions for the quantum Hall conductivity via the linear-response 
theory.  For clean systems we arrive at the TKNN formula, while for 
disordered system we can consider the distribution of topological numbers. 
Bottom left inset is a torus representing the 
first Brillouin zone in k-space.
}
\label{LinearResponse_Aoki}
\end{center}
\end{figure}

For clean systems, the topological expression indicates that the Hall conductivity, 
for arbitrary strength of $B$, 
is quantised topologically when $E_F$ is in a gap of the energy 
spectrum.  
For disordered systems, the topological expression indicates that the Hall conductivity is quantised when $E_F$ is 
in the energy region of localised states, since in that case 
${\Vec A}$-averaged $\sigma_{xy}$ is equal to the unaveraged one.  
In practice, $\sigma_{xy}$ is a smooth function (rather than a 
step function) of energy for finite samples or at a finite $T$.  
Namely, we can show that, even though $\sigma_{xy}$ for 
each of disordered samples is individually quantised, the quantity becomes a 
smooth function of energy after the ensemble average over 
randomness (which corresponds to the observable quantity), and 
the latter tends to the $\sigma_{xy}$ (not integrated over ${\Vec A}$) 
in the thermodynamic limit. \cite{AAprl1}  
Figure \ref{LinearResponse_Aoki} shows the outline of the 
topological numbers for clean and random systems.

Field theoretically, we can introduce a topological term 
(as in the $\theta$ term in the Yang-Mills field theory) in the 
Lagrangian in the nonlinear $\sigma$ model for the QHE system\cite{pruisken}.  Topological phenomena abound in condensed-matter physics, 
as classically exemplified by vortices in superfluids and 
fluxoids in superconductors, but QHE 
thus forms an important and distinct class in the topological phenomena, which has 
spinoffs in the physics of  topological insulator, etc.  
As examplified this, QHE is a superb place for exploring field-theoretical 
interpretations.  In addition to the 
Chern number as the topological quantum number,  
there are  other views that deepen our 
understanding of the effect.  One is 
related to the quantum anomaly in the 
field-theoretic language.   
Quantum anomaly, in the context of particle 
physics, refers to the phenomena in which 
a symmetry inherent in a physical system is 
broken due to quantum-mechanical effects.  
In the context of condensed-matter physics, 
quantum anomaly is referred to those phenomena 
that can be described in terms of Feynman diagrams 
resembling those representing the quantum anomaly.  
This applies to QHE, spin Hall effect, 
and especially Haldane's model for anomalous QHE (QHE in 
zero magnetic field), which may be considered as a 
realisation of what is called the parity anomaly.\cite{fradkinEncy}  

\clearpage

\section{Localisation problem}

  \subsection{Scaling theory of localisation in two-dimensional systems}

As we have seen, the integer QHE is intimately related with 
the localisation problem.  We have also mentioned that 
the system falls upon the unitary class in 2D.  Details of the localisation  (which is a non-perturbative problem) has been 
extensively studied with numerical methods and field theoretical 
methods.  As shown schematically in Figs. \ref{wavefunction},\ref{AokiAndo}, 
the localisation becomes stronger towards the edges of a Landau 
level and the weakest at the centre.  
Classically, this corresponds to the fact that 
contours of a random potential, along which the cyclotron guiding centre 
drifts in the semiclassical picture, are valleys (hills) for energies well below (above) 
the level centre, while the contours tend to percolating paths 
for energies close to the centre.  However, the localisation is quantum mechanical 
phenomenon related with interference of wavefunctions.   

If we look more closely at the behaviour of the localisation length 
in a numerical method (finite-size scaling 
for the Thouless number), 
the localisation length, although a 
smooth function of $E$, exhibits a rather singular behaviour, 
in which the inverse localisation length becomes zero 
only at the level centre (i.e., extended states only occur at the 
single point on energy axis).  Namely, 
the delocalised states, 
marginally allowed to appear in the presence of 
magnetic fields in two dimensions, has an energy region coalescing into a single point 
$E_0$ (the centre of each Landau 
level).  
The localisation length, denoted here as $\xi$, behaves around this point as\begin{equation}
\xi \sim 1/|E - E_0|^s ,
\end{equation}
where $s$ (numerically shown to be 
$\sim> 2)$ is the localisation critical exponent.\cite{AAprl2} 
This value may be compared with the exponent $s=7/3$ 
for the percolation problem.  The effect of quantum mechanical 
tunnelling has also been extensively studied in terms of a 
network model due to Chalker and Coddington, 
who have modelled the quantum mechanical tunnelling 
across the paths (see Fig.\ref{percolation}(c)).\cite{Chalker,OhtsukiKramer}

The critical exponent $s$ is shown to depend on the 
Landau level (larger for $N=1$ than for $N=0$).  
A mapping onto the 
nonlinear $\sigma$ model gives an estimate 
$\xi \sim \ell{\rm exp}(\sigma_{\rm SCBA}^2)$ 
with $\sigma_{\rm SCBA} \sim (N+1/2)e^2/h$ being $\sigma_{xx}$ 
in the self-consistent Born approximation\cite{Ando} and $\ell$ 
the magnetic length.   So the localisation length increases with 
$N$, so does the critical exponent.  
The localisation length also depends strongly 
on whether the randomness is rapidly varying or 
slowly varying in real space
 as compared with the magnetic length.  
  
  \subsection{Quantum criticality and 
$\sigma_{xx}-\sigma_{xy}$ diagram}

\begin{figure}[ht]
\begin{center}
\includegraphics[width=15cm,clip]{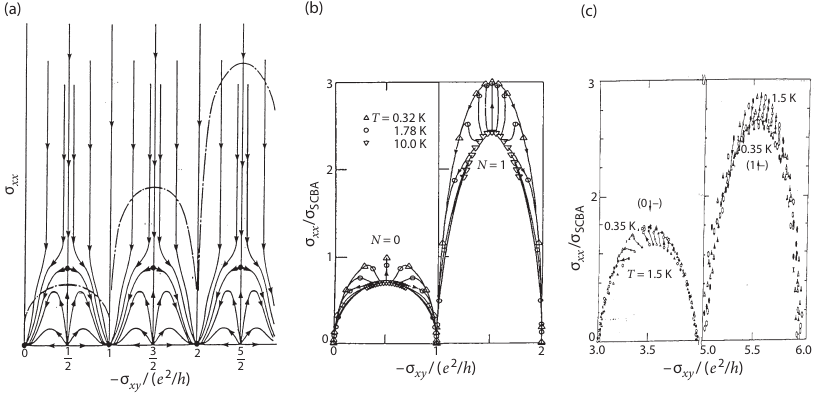}
\caption{Results for the $\sigma_{xx}$-$\sigma_{xy}$ diagram: 
(a) Renormalisation flow lines obtained in the nonlinear $\sigma$ model 
[after A. M. M. Pruisken, Phys. Rev. Lett. {\bf 61}, 1297 (1988)].  (b) A numerical result [after 
H. Aoki and T. Ando, Surf. Sci. {\bf 170}, 249 (1986)].  
(c) An experimental result for 
$(N,$ spin, valley) indices indicated [after M. Yamane et al, J. Phys. Soc. 
Jpn {\bf 58}, 1899 (1989)].
\label{sigmaxxxy}}
\end{center}
\end{figure}

One way to display the critical behaviour is 
the $\sigma_{xx}-\sigma_{xy}$ diagram.  
Namely, while 
at $T=0$ $\sigma_{xy}$ is a step 
function and $\sigma_{xx}$ nonzero at discrete points, 
at finite temperatures not only the Fermi distribution is smeared 
by $k_BT$, but we have also a finite inelastic scattering length $L_{\epsilon}$.  This can be summarised as the 
$(\sigma_{xx},\sigma_{xy})$ flow diagram when 
$T$ is varied for various values of $E_F$, as originally 
evoked in the nonlinear $\sigma$ model (Fig.\ref{sigmaxxxy}(a)).  
Namely, we can map the QHE system onto a field-theoretic model 
called  the nonlinear $\sigma$ model, for which the renormalisation 
into larger sample sizes can be discussed.\cite{hucke}

We can also 
vary the samle size for numerical works on the QHE system to 
look at the renormalisation flow.  Alternatively we can 
change the temperature, which changes the inelastic 
scattering length, hence effectively changes the sample size 
 (Fig.\ref{sigmaxxxy}(b)).  
This implies that experimental $(\sigma_{xx},\sigma_{xy})$ flow lines 
can be obtained by varying the temperature  (Fig.\ref{sigmaxxxy}(c)).

\begin{figure}[ht]
\begin{center}
\includegraphics[width=12cm,clip]{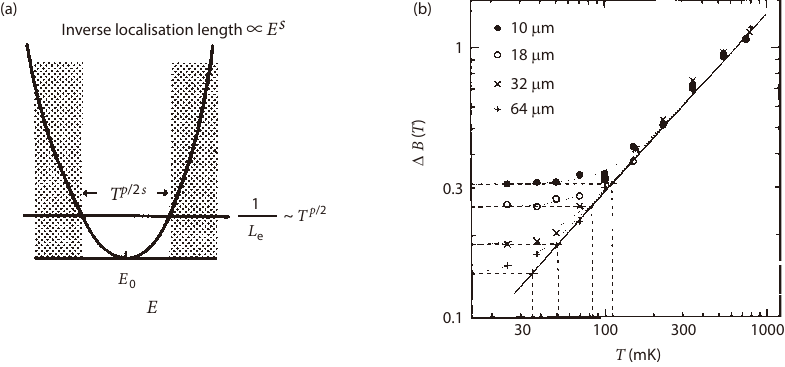}
\caption{(a) In a theoretical plot for the inverse localisation length 
against energy, the effectively extended states (white region) 
is shown for the inelastic scattering length 
$L_{\varepsilon}$ at  a given temperature $T$ 
[after H. Aoki and T. Ando, Surf. Sci. {\bf 170}, 249 (1986)].  
(b) An experimental result for the plateau-to-plateau 
transition width $\Delta B$ on the $B$ axis plotted against $T$ 
for various values of the sample size [after 
S. Koch et al, Phys. Rev. Lett. {\bf 67}, 883 (1991)].}
\label{Tp2s}
\end{center}
\end{figure}

We can also discuss the so-called ''plateau-to-plateau transition", 
which is the energy interval over which one plaleau crosses over to 
another.  In terms of the localisation, the states that satisfy 
$L_{\varepsilon} < \xi (E)$ are effectively delocalised, where 
$L_{\varepsilon}$ is the inelastic scattering length 
and $\xi$ the localisation length (Fig.\ref{Tp2s}(a)).  
If we assume the inelastic scattering 
time behaves like 
$\tau_{\varepsilon} \sim T^{-p}$ with $p=2$ in the diffusive regime,  
the energy region whose width is $\sim T^{p/2s}$ behaves as 
delocalised, 
since the localisation length has a critical behaviour 
$\xi \sim |E-E_0|^{-s}$ while $L_{\varepsilon} \propto 1/T^{p/2}$.  
If the magnetic field is 
varied instead, the plateau transition width against $B$ has 
the width $\Delta B \sim T^{p/2s}$ with the same exponent. 
Figure \ref{Tp2s}(b) displays an experimental result.

\begin{figure}[ht]
\begin{center}
\includegraphics[width=11.7cm,clip]{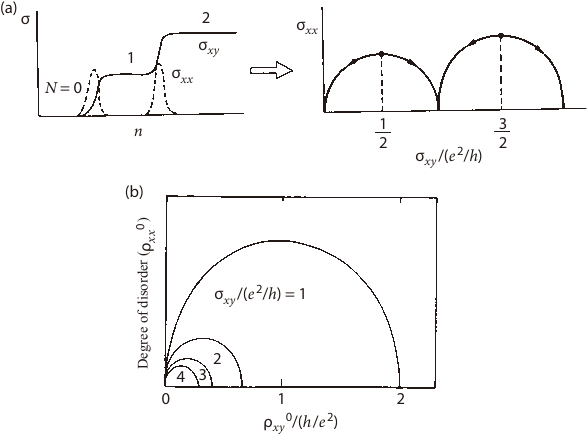}
\caption{
(a) How a $\sigma_{xx}$-$\sigma_{xy}$ diagram is obtained 
from their $n$-dependence is schematically shown, with the 
flow lines bifurcating at $\sigma_{xy}/(e^2/h) = $ half integer indicated.  
(b) The global phase diagram, against the Hall resistivity 
$\rho_{xy}$ and the longitudinal resistivity 
$\rho_{xx}$ (the latter being a measure of the degree of disorder.}
\label{global}
\end{center}
\end{figure}

One important question is how the QHE plateaux vanish 
as the degree of disorder is increased, where 
for a large enough disorder the Landau quantisation, along with 
QHE, should go away.  
Kivelson et al\cite{global}  have considered the problem in 
terms of what is called the global phase diagram on a 
$\rho_{xy}-\rho_{xx}$ plane.   In this diagram 
$\rho_{xx}$ is regarded as a measure of the 
strength of disorder, while $\rho_{xy}$ reflects 
the Landau level filling.  
As shown in Fig.\ref{global}(a), the midpoints, 
$\sigma_{xy} = (N+1/2)e^2/h$, between 
the plateaux in the Hall conductivity 
are bifurcation points in the $\sigma_{xx}-\sigma_{xy}$ 
flow lines, so that 
we can take the points, 
$\sigma_{xy} = 1/(\rho_{xx}^2+\rho_{xy}^2) = (N+1/2)e^2/h$, 
as boundaries 
up to which the QHE effect survives 
when the strength of disorder is increased (Fig.\ref{global}(b)).  

It is also important to consider 
how the QHE in strong magnetic fields crosses over to the situation 
of weak magnetic fields, 
since all the states should 
become localised in the limit of $B \rightarrow 0$, 
so that the fate of the delocalised states is quite nontrivial.  
It has been suggested that the energies corresponding to the 
delocalised states go up in energy, which is called the ``'floating".

Another question of heuristic interest is what is the classical limit 
(i.e., Planck's constant $h \rightarrow 0$).  In this limit the Landau level filling 
$2\pi \ell^2n = (h/eB)n$ vanishes for a fixed $n$.  
The measure of the Landau level mixing, on the other hand, 
is the Landau level broadening divided by the cyclotron 
energy, 
$\Gamma/\hbar\omega_{c} \sim 1/\sqrt{\omega_{c}\tau_{0}}$.  
Since $\Gamma/\hbar \omega_c$ diverges like 
$1/\sqrt{h}$ for $h \rightarrow 0$ with fixed $B$ and $\tau_0$, the classical limit amounts to a regime of dilute filling $\nu (\propto h)$ in weak magnetic fields.

\subsection{Fractal wavefunctions and dynamical scaling}

It is an intriguing question to ask whether 
a transition between localised and 
extended states, in general, is similar to phase transitions in statistical mechanics.  
A suggestion came in 1983 by Aoki to the effect that 
the wavefunction at the Anderson transition, 
where a characteristic 
length scale is absent with the diverging localisation length, 
should be self-similar (fractal),\cite{fractal}   
just as the critical point in a phase transition 
is characterised by scale-invariant states with a diverging correlation 
length.   

\begin{figure}[ht]
\begin{center}
\includegraphics[width=11.7cm,clip]{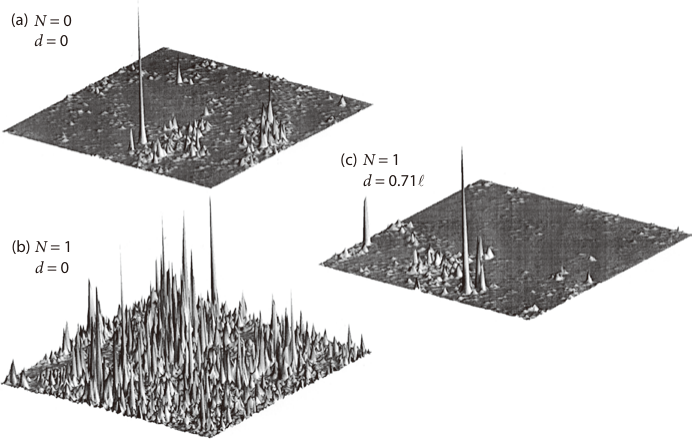}
\caption{Typical wavefunctions (represented as their squared amplitude) 
around the centre of each 
Landau level obtained in a computer simulation 
for a disordered QHE system with a system size of 
$300\ell \times 300\ell$ for (a) $N=0$ and (b) $N=1$ Landau 
levels with short-range scatteres, and for (c) $N=1$ 
with long-range ($d \neq 0$) scatteres 
[after T. Terao et al, Phys. Rev. {\bf 54}, 10350 (1996)].}
\label{wavefunction2}
\end{center}
\end{figure}

For the QHE system in particular,  
the transition point occurs at the centre of each Landau level,  
so that  the delocalised states at the centre of a Landau level 
are just not the usual extended states, but fractal (or sometimes 
called critical) wavefunctions 
$\psi({\Vec r})$ that have a fractal dimensionality, $d^* <2$.  In other words, the density autocorrelation 
decays with a power law, 
$\langle |\psi({\Vec r}) \psi({\Vec r}+{\Vec R})|^2 \rangle \sim 1/R^{2-d^*}$.  
Typical wavefunctions in Fig.\ref{wavefunction2}, which are obtained numerically for larger sample sized than in Fig.\ref{wavefunction}, show that 
the wavefunction is indeed scattered both in amplitude and spatial extent.

To be more precise, the self-similarity in critical wavefunctions 
extends beyond a single scale transformation, 
so the idea has been subsequently developed into 
the multifractal analysis, 
with which we can analyse the delocalised states around 
the Landau level centre.\cite{hucke,terao} 
As described in the section below, real-space imaging such as 
STM can visualise such states.

Among the physical quantities that are affected by the fractality of the wavefunctions is the ``anomalous 
diffusion" in transport.  
Namely, the dc conductivity $\sigma$ obeys, in ordinary systems,  
Einstein's relation, 
$\sigma = {\rm lim}_{\bq, \omega \rightarrow 0}\sigma(\bq, \omega) = e^2D(E_F){\cal D}_0$ 
where ${\cal D}_0$ is the diffusion constant.  
For the 
$\bq, \omega$-dependent conductivity $\sigma(\bq, \omega)$, 
the dynamical scaling ansatz asserts that  
the ac conductivity should depend on $\omega \tau$, where the 
relaxation rate goes to zero like $1/\tau \sim 1/\xi^z$ around 
the transition.   Here $z$ is the dynamical critical exponent, 
which is usually 2 in noninteracting systems.  
If we combine this with the sample size scaling in $L/\xi$, the 
$\bq, \omega$-dependence should take a form, 
\begin{equation}
\sigma(\bq, \omega) \sim e^2D(E_F){\cal D}(q/\omega^{1/z}). 
\end{equation}
At the criticality the fractality of the states is shown to 
lead to ${\cal D}(q/\sqrt{\omega}) \sim {\cal D}_0/(q/\sqrt{\omega})^{\eta}$ with $\eta = 2-d^*$.   
For a dynamical scaling for graphene, see \cite{morimotoAvishai10} 

\clearpage

\section{QHE edge states and edge transport}

\subsection{Bulk-edge correspondence}

\begin{figure}[ht]
\begin{center}
\includegraphics[width=9cm,clip]{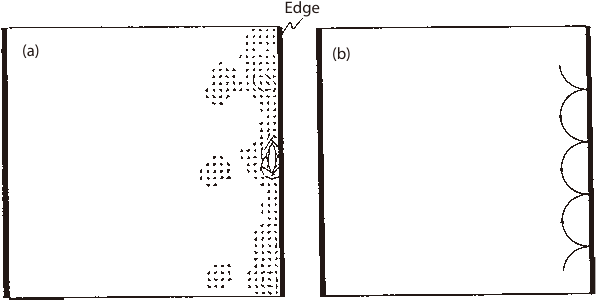}
\caption{Typical edge states in (a) a QHE system in quantum 
mechanics in a simulation for a finite system, here represented by 
the current distribution ${\Vec j}({\Vec r})$  with arrows, and (b) in classical mechanics 
[after H. Aoki, in G. Landwehr (ed.): {\it Application of High Magnetic 
Fields in Semiconductor Physics} (Springer, 1983), p.11].
\label{edgestates}}
\end{center}
\end{figure}

\begin{figure}[ht]
\begin{center}
\includegraphics[width=11.7cm,clip]{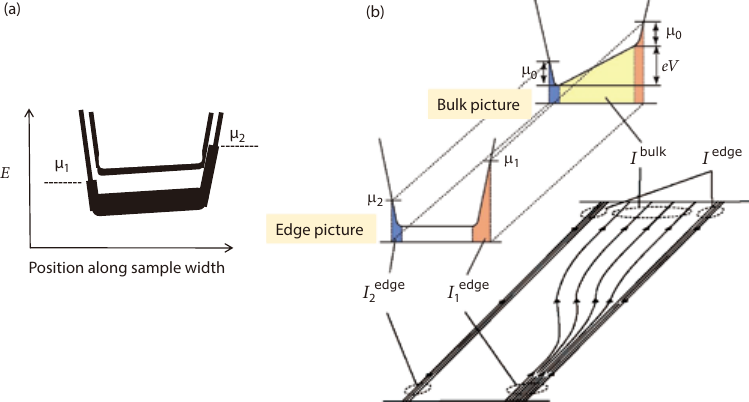}
\caption{(a) Energy spectrum against the position 
along the width is schematically shown for 
a sample with edges with the chemical potential at the left edge ($\mu_1$) 
and at the rigth  ($\mu_2$).  
(b) How the edge picture (in which all the currents are carried by 
edge states) continuously crosses over to the bulk picture 
(where a bulk contribution exists) is schematically shown 
[after M. Koshino, H. Aoki and B.I. Halperin, Phys. Rev. B {\bf 66}, 081301(R) (2002)].}
\label{edge}
\end{center}
\end{figure}

The role of sample edges in the QHE transport has been an 
issue of interest from an early stage when the problem was raised 
by Halperin in the 1980's.\cite{Haledge}  
Classically, there are edge currents that correspond to cyclotron 
motions skipping along the edges (Fig.\ref{edgestates}(b)).  Quantum mechanically, 
edge states also exist  (Fig.\ref{edgestates}(a)).   
The energy diagram plotted against the real-space position 
across the sample width will look like Fig.\ref{edge}(a).  
Edge states (or, in a broader context, boundary states in field theory) are 
ubiquitous 
in quantum mechanics, but the peculiarity in the QHE is that 
the edge currents flow in a definite direction dictated by the 
direction 
of the applied magnetic field (hence they are dubbed chiral 
edge currents), with no backscattering even when the edge 
is e.g. rugged.  
When one takes the electron-electron interaction into account, 
the edge states may be regarded as special, incompressible electronic 
states called the chiral Tomonaga-Luttinger liquid, 
as discussed in the context of the fractional QHE.

In terms of the topological nature of the QHE, 
the appearance of edge states is a prime example of 
a general concept of the ``bulk-edge correspondence" as 
introduced by Hatsugai,\cite{Hatsugaiedge} 
i.e., in topological systems, the nature 
of edge states are dictated by the nature of bulk states, 
as in the boundary states in field theories.  
In a finite QHE sample, the edge states 
appear on the energy spectrum as the edge modes 
that cross from one Landau level to another in a Landau gap.  
These modes have to exist, as sometimes called 
``topologically protected", in QHE systems.  
In a broader context, topological boundary states 
arise as we have displayed on the classification of 
topological systems in Fig.\ref{boundaryStates}.

The topological nature of the QHE edge states 
manifests itself, as shown by Hatsugai, 
in the properties that (i) the edge Hall conductivity is also expressed as 
a topological invariant (Chern number), and (ii) this quantity 
exactly coincides with the  topological Chern number 
for the bulk Hall conductivity.\cite{Hatsugaiedge}  
Alternatively we visualise, in Fig.\ref{edge}(b), that we can continuously change 
the situation from the edge-picture limit (where currents exist 
only at edges) to the bulk-picture  (where a bulk potential 
gradient exists) when the current distribution is deformed, 
but there is always the total current conservation, 
$I^{\rm bulk} = I^{\rm leftedge}-I^{\rm rightedge}$, which implies 
$\sigma_{xy}^{\rm bulk} =  \sigma_{xy}^{\rm edge}$.\cite{koshino}

One way to describe the edge transport is 
B\"{u}ttiker's formula\cite{buttiker}, which is an extension of 
Landauer's formula for transport processes in terms of 
an S matrix for transmission and reflection channels in 
ballistic transports.  
B\"{u}ttiker's formula can explain some of the experimental results, 
including the quantisation of the Hall conduction.  
However, this does not mean that all the QHE currents are carried by 
edges.  In general there exist both bulk and edge Hall currents.  
Bulk and edge states can even be hybridised.  
The details have been studied experimentally and theoretically, 
which largely indicate that there are contributions from both 
bulk and edges, whose proportions depend on the width 
of the sample, $T$-dependent $\sigma_{xx}$, etc.

Physically, a macroscopic system may be regarded as comprising 
subsystems each having a linear dimension of the phase coherence 
length $L_{\phi}$.  The transport is described by the Kubo-formula 
$\sigma_{xx}, \sigma_{xy}$ 
if the sample size is much greater than $L_{\phi}$ where the 
Hall field and Hall current distributed over the sample, 
while the edge transport a la B\"{u}ttiker starts to dominate when the size 
is shrunk to $\sim L_{\phi}$.  

One of the characteristic features in the edge transport is that 
highly nonlocal phenomena can arise 
that extend over length scales far exceeding the 
bulk mean free path.   This can be probed in multi-terminal 
Hall-bar samples by 
injecting currents from some contacts, where 
the conductance can strongly 
depend on which terminals to select for 
its measurment.

The width of edge states in real space 
have experimentally been 
measured with various methods 
including transport (e.g. QHE breakdown), 
magnetocapacitance, and magnetoplasmon 
measurements.  We can also utilise 
the nuclear spins.  Usually, the nuclear spins are 
irrelevant to electron systems, since the nuclear 
Zeeman energy is as small as $1/2000$ of the electronic 
counterpart.  When there are almost degenerate 
electronic states, however, nuclear spins can affect 
the electron system via the hyperfine interaction.  
This occurs in the fractional QHE systems 
close to a spin-polarisation transition, 
but also in the integer QHE where left 
and right edge states 
are degenerate in energy, so that the nuclear spin 
is probed via the scattering between the
left and right edge channels. 

In semiconductor superlattices, which realises a stack of 
2DEG's, an application of a strong magnetic field 
along the growth direction makes the system a stack of 
IQHE systems.  There, the edges states are also stacked 
along the edge surfaces, which is called `sheath currents" 
and have been experimentally observed.  

\clearpage

 \subsection{Real-space imaging}
 
\begin{figure}[ht]
\begin{center}
\includegraphics[width=6cm,clip]{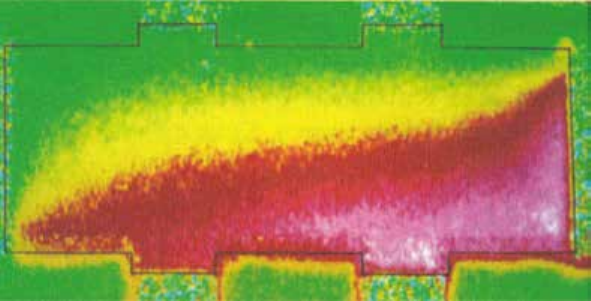}
\caption{A typical real-space image for the potential distribution 
(colour-coded from blue to red with increasing potential) 
obtained with the electro-optic imaging for $B=10.8$ T that 
corresponds to the centre of a plateau.  Black line 
delineates the Hall bar system  
[after R. Knott et al, Semicond. Sci. Tech. 
{\bf 10}, 117 (1995)].
\label{imaging}}
\end{center}
\end{figure}

Experimentally, there have been a body of studies for real-space 
imaging of the QHE systems, including the Hall field.  
Fontein et al\cite{fontein91} have studied a 
potential profile imaging with the electro-optical effect 
(which utilises the birefringence (Pockels effect) where 
the phase difference between different polarisations 
probes the potential) to show that 
an electric field exists over the whole sample although 
the field becomes 
stronger towards the edges 
as shown in Fig. \ref{imaging}.\cite{knott95}  There are various 
other methods as well, including scanning capacitive, force and 
polarisation optical microscopies,\cite{morgenstern} 
the imaging with a metal single-electron transistor (SET)\cite{wei98}, 
and the microwave
impedance microscope\cite{lai11} 
.

\begin{figure}[ht]
\begin{center}
\includegraphics[width=11.7cm,clip]{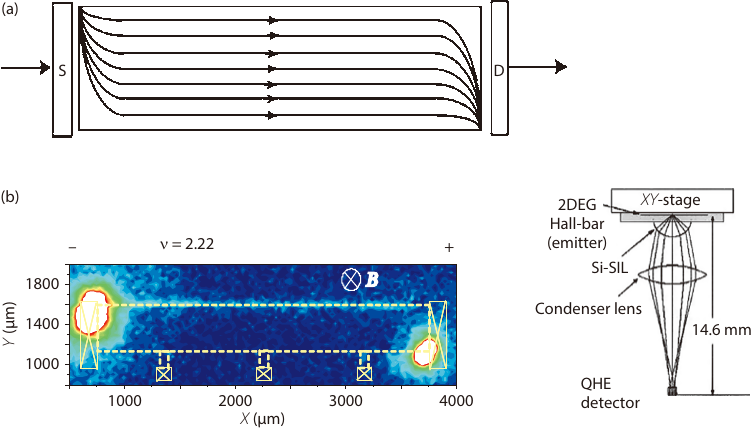}
\caption{(a) The electric lines of force are schematically shown 
for the QHE condition, with source (S) and drain (D) electrodes.  
(b) A real-space imaging for a 
Hall bar sample (delineated by 
yellow lines) obtained with the cyclotron emission as 
detected by the QHE device (inset) [after K. Ikushima et al, Phys. 
Rev. Lett. {\bf 93}, 146804 (2004)].}
\label{hotspot}
\end{center}
\end{figure}

One such probing, that is particularly suited to examine nonequilibrium 
carriers around the ``hot spots", is the imaging of cyclotron 
emission due to Komiyama and coworkers.\cite{ikushima}
In a Hall-bar 
sample in the QHE condition (with $\sigma_{xx}=0, \sigma_{xy}\neq 0$), 
the electric lines of force are forced to be distorted to 
make the rectangular sample geometry compatible 
with the Hall angle of 90 degrees, which results in singularities 
appearing around two corners across the opposite electrode (Fig.\ref{hotspot}(a)).  
These are called the hot spots.  In 
imaging the cyclotron 
emission  (Fig.\ref{hotspot}(b)), the QHE 
detector, which is itself an IQHE system having by nature 
sensitive photoresponses at the cyclotron resonance 
frequency, is used to scan the QHE system. 
 The imaging captuires the hot spots 
clearly, with edge channel also visible for the Landau 
level filling $\neq$ integer.  
The  imaging of cyclotron 
emission has also been used to examine the breakdown of QHE 
(see Section `Breakdown of QHE').

\begin{figure}[ht]
\begin{center}
\includegraphics[width=11.7cm,clip]{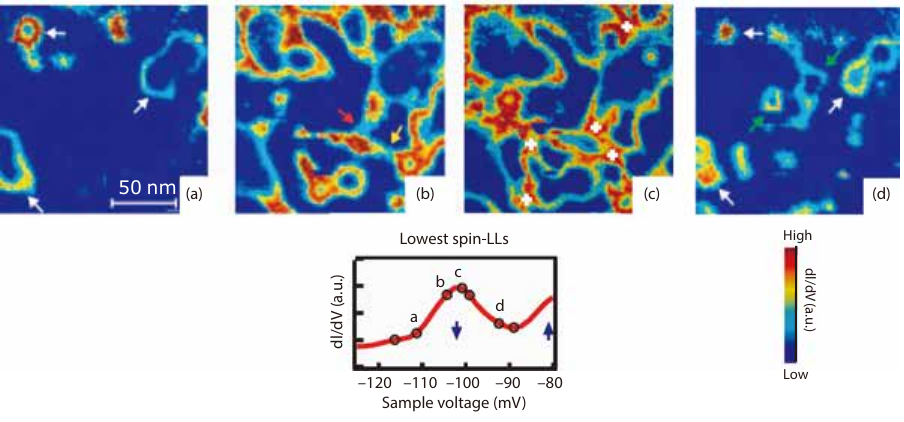}
\caption{A real-space image obtained by STS for 0.01 monolayer of Cs on a cleaved n-InSb(110) at  $T = 0.3$ K for various values of the 
Landau level filling, (b)-(f), as indicated in inset 
[after K. Hashimoto et al, Phys. Rev. Lett. {\bf 101}, 256802 (2008)].
\label{STM}}
\end{center}
\end{figure}

Scanning tunnelling microscopy (STM) and 
scanning tunnelling spectroscopy (STS) are also performed, 
for a special kind of adsorbate-induced 2DEG with e.g. Cs atoms 
deposited on cleaved n-InSb that makes STM/STS feasible.  
The STS imaging can also directly observe 
the quantum Hall transition from one localised regime to another 
via the delocalised states that have a fractal character as 
shown in Fig.\ref{STM}.\cite{hashimoto}

\clearpage

\section{QHE resistance standard, the fine-structure constant and new SI}

After the recognition that the accuracy of the QHE quantisation is experimentally better than 
$10^{-6}$ initially and even refined to $10^{-7}$, it was decided, in 1990 by CIPM (Comit\'{e} International des Poids et Mesures), that the QHE 
be adopted as the electrical 
resistance standard, with which ohm is defined 
via the QHE resistance as 
\begin{equation}
R_{K} (\equiv h/e^2) = 25\;812.807 \,\Omega,
\end{equation} 
where $R_K$ is called the von Klitzing constant.  

\begin{figure}[ht]
\begin{center}
\includegraphics[width=14cm,clip]{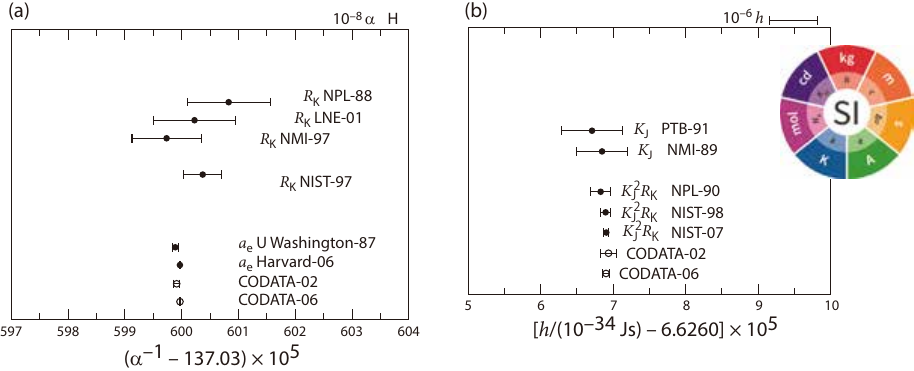}
\caption{(a) Values of the fine-structure constant $\alpha$ 
obtaned with various methods.  Those marked with $R_K$ 
are from the QHE measurements in various institutes, while 
$a_e$ from the electron magnetic moment anomaly.  
(b) A similar plot for Planck's constant $h$.  
CODATA-recommended values are also indicated.
After P. J. Mohr et al, Rev. Mod. Phys. {\bf 80}, 633 (2008).  
The inset is the logo for the new SI (the International System of Units) by BIPM 
(from 
https://www.bipm.org/en/measurement-units).
\label{finestructure}}
\end{center}
\end{figure}

The constant, now adopted as the resistance standard, 
has a profound significance related with a fundamental 
physical constant.  Namely, the fine-structure constant (in cgs), 
\begin{equation}
\alpha = \frac{e^2}{\hbar c} \simeq 1/137.036,
\end{equation}
which is the coupling constant in the QED 
(quantum electrodynamics) \cite{kinoshita} and 
one of the most basic physical constants, 
is directly related with $R_K$ via 
$\alpha = 2\pi/R_K c $.  
Here $c = 2.99792... \times 10^{8}$m/s, the speed of light in 
vacuum, is, in SI, a defined value.  Figure \ref{finestructure}(a) displays 
the measured values of $\alpha$ obtained by various methods 
including that from $R_K$ obtained in various institutes.  
If we combine $R_{K}$ with the Josephson constant, $K_J = 2e/h$, 
we can deduce the value of Planck's constant via 
$h = 4/K_J^2R_K$ (Fig. \ref{finestructure}(b)).   
For a review, see\cite{TaylorRMP}.

Recently, BIPM (Bureau International des Poids et Mesures) decided that 
the International System of Units 
(Syst\`{e}me International d'Unit\'{e}s, abbreviated as SI) 
are defined, from 20 May 2019, in terms of designated fixed numbers as:

\begin{itemize}
\item $\Delta \nu$: The 
ground-state hyperfine transition frequency  of $^{133}$Cs is $9 192 631 770$ Hz.
\item $c$: The speed of light in vacuum is 299 792 458 m/s.
\item $h$: The Planck constant is $6.626 070 15 \times 10^{-34}$ J$\cdot$s.
\item $e$: The elementary charge is $1.602 176 634 \times 10^{-19}$ C.
\item $k$: The Boltzmann constant is $1.380 649 \times 10^{-23}$ J/K.
\item $N_{\rm A}$: The Avogadro constant is $6.022 140 76 \times 10^{23}$ mol$^{-1}$.
\item $K_{\rm cd}$: The luminous efficacy of monochromatic radiation of frequency $540 \times 10^{12}$ 
Hz is $683$ lm/W.
\end{itemize}

Kibble balance (formerly known as Watt balance) 
can make a connection between the Planck constant
and the kilogram, since we have
\begin{itemize}
\item $h/e^2$ from QHE,
\item $h/e$ from Josephson effect,
\item $m_e = 2R_{\infty}h/(c\alpha^2)$: electron mass, where  
$R_{\infty}$: Rydberg constant, $\alpha = e^2/(2\varepsilon_0 hc) 
= \mu_0 ce^2/(2h)$: fine-structure constant 
with a CODATA-recommended valus of $1/[137.035999084(21)]$, and 
$\varepsilon_0$: permitivity of vacuum, $\mu_0 = 1/(\varepsilon_0 c^2)$: 
permeability of vacuum. 
\end{itemize}
The kilogram, for instance, can be expressed as [kg] = [Hz][J $\cdot$ s]/[m/s]$^2$, 
which can be defined in terms of $\Delta \nu, c, h$ defining respectively 
Hz, m/s, and J$\cdot$s.
For the von Klitzing constant, $R_{\rm K} = h/e^2 = \mu_0 c/(2\alpha) 
= 25812.807\; \Omega$, $\mu_0$ (with dimension [N/A$^2$]) 
and $c$ [m/s] have fixed values in the new SI, so that 
a measured uncertainly in $R_{\rm K}$ 
translates to the uncertainly in $\alpha$.  
Experimentally, the universality (relative accuracy in sample and 
material dependence) of the constant has been experimentally 
shown to be as small as $\sim 10^{-10}$. 
The inset of Fig.26 is the logo for the new SI by BIPM. 
See an article, 
K. von Klitzing, Nature Phys. {\bf 13}, 198 (2017).

\clearpage

 \section{Breakdown of QHE}

\begin{figure}[ht]
\begin{center}
\includegraphics[width=7cm,clip]{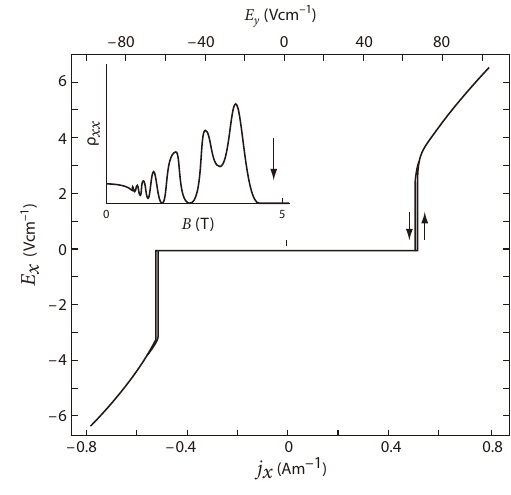}
\caption{A typical current-voltage characteristics at $\nu = 2$ (thick arrow in the inset) 
[after G. Nachtwei, Physica E {\bf 4}, 79 (1999)].}
\label{breakdown}
\end{center}
\end{figure}

When we increase the source-drain 
voltage to increase 
the source-drain current in Hall measurements in the 
Hall-bar geometry, 
the longitudinal resistance $R_{xx}$, 
which is close to zero in the QHE plateau region, 
abruptly increases for sufficiently large voltages, as typically 
shown in Fig.\ref{breakdown}.  This is 
accompanied by disrupted plateau 
structures in  $R_{xy}$.  Thus the Hall current, which is originally dissipationless 
in the QHE condition, becomes dissipative for large enough 
voltage, and 
this phenomenon, known from an early stage of the QHE studies, 
is called the breakdown of QHE.  The breakdown is 
important from both applicational aspects (since it affects the accuracy of the QHE), and fundamental aspects since this is an interesting 
nonequilibrium 
phenomena.  Several factors can be involved, among which are 
tunnelling between different Landau orbits or Landau levels, and/or electron heating effect.   Also relevant is how the breakdown is 
related with the current profile in real space in a Hall bar.  
A review on experimental and theoretical results may be found, e.g., 
in \cite{nachtwei}. 
Sometimes metastable or bistable states are observed around the 
breakdown regime, which is also of interest. 
As for nonequilibrium phenomena, we can mention that 
the microwave-induced magnetoresistance oscillation, 
although this occurs 
in a weak magnetic field regime rather than in the QHE regime.\cite{mani02}

\clearpage
 
\section{Quantum dots and periodically modulated systems in strong magnetic fields}

\subsection{Quantum dots in magnetic fields}

\begin{figure}[ht]
\begin{center}
\includegraphics[width=11.7cm,clip]{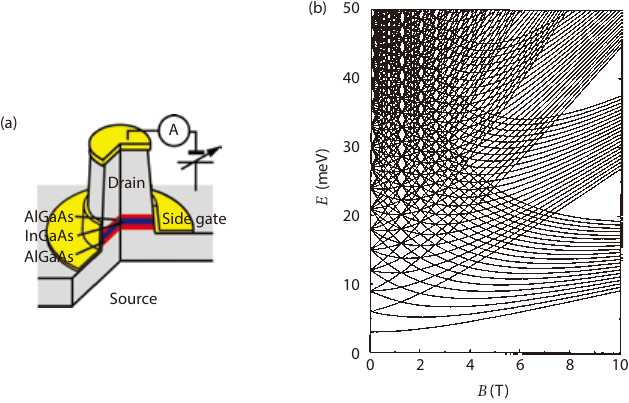}
\caption{(a) A typical mesa-etched quantum dot structure 
[after P. A. Maksym et al, Phys. Rev. B {\bf 79}, 115314 (2009)]. 
(b) Energy levels of the Fock-Darwin states against magnetic 
field for a harmonic confinment 
[after P. A. Maksym et al, J. Phys. Condensed Matter {\bf 12}, R299 (2000)].}  
\label{dot}
\end{center}
\end{figure}

Semiconductor quantum dot is a nanostructure that confines a small 
number of electrons.  The physics of quantum 
dots is a very wide area, so here we only mention its relevance to 
QHE.  Quantum dots are usually fabricated by applying a 
lateral confining potential to a 2DEG system.  
This can be realised either by an electrode that 
exerts an electrostatic confining potential, 
or alternatively we can mesa-etch the system to have a finite system (Fig.\ref{dot}(a)).    
The confining potential is usually cylindrically parabolic 
to a good approximation.

Quantum dots are investigated in strong magnetic fields as well.
In a magnetic field, ${\Vec B} = {\rm rot}{\Vec A}$, applied perpendicular to 
the dot plane the Hamiltonian in 
the one-body problem reads
\begin{equation}
{\cal H} =  \frac {1}{2m^*}
\left[{\Vec p}_i + (e/c){\Vec A}({\Vec r})\right]^2 +
\frac{1}{2} m^* \omega_0^2 r^2,
\end{equation}
where $\hbar \omega_0$ is the confinement energy for a parabolic 
potential, and we have ignored 
the Zeeman energy. 

The exact eigenstates of this Hamiltonian  are known as the Fock-Darwin 
states, since they were first investigated by Fock 
and Darwin in the 1920's.  They are given, up to a normalisation 
constant, as
\begin{equation}
\psi_{nm}({\bf r}) = 
r^{|m|} 
L_n^{|m|}(r^2 / 2\lambda^2)\exp({-r^2 / 4\lambda^2}) e^{-im\theta}
\end{equation}
in 2D polar coordinates with eigenenergies given by
$E_{nm} = (2n + 1 + |m|) \hbar \Omega - m\hbar \omega_c /2.
$
Here $m$ is the angular 
momentum, $n$ a radial quantum 
number, $L_n^{\mid m \mid}$ the associated Laguerre polynomial,
$\Omega^2 = \omega_0^2 + \omega_c^2/4$ with 
the cyclotron frequency $\omega_c = eB/m^*$.  
Thus the wavefunction is almost the usual Landau's wavefunction 
when the confinement potential is parabolic, where the 
difference is in the length parameter, 
$\lambda = \sqrt{\hbar / (2m^*\Omega)}$.  
In Fig.\ref{dot}(b) depicting the energy levels we can indeed see that 
a crossover from the states of a 2D harmonic oscillator in the zero magnetic 
field limit 
to the Landau states in the strong field limit. 
When we consider the electron-electron interaction, a variety of 
states emerge, which include the maximum-density droplet that corresponds to the Landau level filling 
$\nu = 1$, and the electron-molecule states. \cite{Maksym}

So the dots in strong mangetic fields are a kind of confined QHE systems, 
or artificial atoms in magnetic fields. 
Their properties have been extensively studied with various methods.  
For instance, dot wavefunctions have been probed experimentally with techniques 
such as magneto-tunnelling spectroscopy.  Other properties include 
magnetocapacitance, electron addition spectra and transport 
spectra in magnetic fields, from which the ground-state quantum
numbers are deduced.  
Dot arrays have also been studied in the QHE regime, where 
resonant scattering effects etc have been reported.


 \subsection{Hofstadter spectrum}

\begin{figure}[ht]
\begin{center}
\includegraphics[width=8cm,clip]{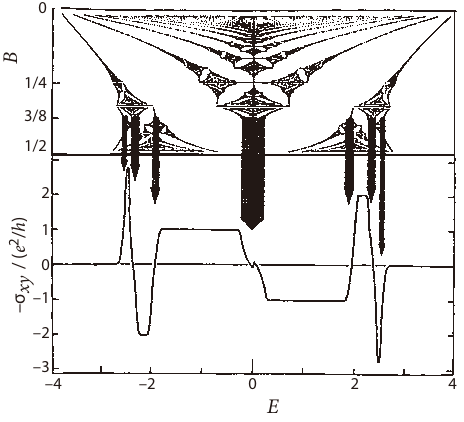}
\caption{Top: Hofstadter butterfly (energy spectrum, against magnetic field 
$B$ measured by the flux penetrating a unit cell) 
for a square lattice.  Bottom: 
QHE for a square lattice 
exemplified for a magnetic 
flux in a unit cell of the periodic system being $3/8$ 
in units of the flux quantum 
[after  H. Aoki, in G. Landwehr (ed.): {\it Application of High Magnetic 
Fields in Semiconductor Physics III} (Springer, 1991), p.17].
\label{Hofstadter}}
\end{center}
\end{figure}

What happens to QHE when we have periodic systems 
rather than a translationally invariant 2DEG?  
Condensed-matter physics tells us that in a periodic system 
we have Bloch's theorem, which dictates that electronic states for 
periodic systems are 
dominated by Bragg's reflection, resulting in band structures 
and Bloch states.  
So, when we apply an external magnetic field, we are talking about Landau's quantisation  in the presence of 
Bragg's reflection.  We can realise that Bragg's reflection 
and Landau's quantisation interfere with each other, since 
the application of a magnetic field, ${\Vec B} = {\rm rot}{\Vec A}$, 
gives rise to, semiclassically, 
the Peierls phase, 
${\rm exp} (
-{\rm i}\frac{e}{\hbar}\int^{\br}{\Vec A}({\br^{\prime}})
\cdot{\rm d}{\br^{\prime}})\psi({\br})$, 
in the wavefunction $\psi$.  
Landau's quantisation in periodic and lattice systems 
was first considered by Wannier and by Hofstadter\cite{Hofs}.  It has been 
shown that 
the energy spectrum plotted against magnetic field is a curious 
fractal (sometimes called Hofstadter's butterfly).  
Fractal, because each Landau level splits into $p$ levels 
when the magnetic flux within a unit cell in units of flux quantum 
equals to a rational number $q/p$.(Fig.\ref{Hofstadter})

When $E_F$ is in one of these gaps, we should have IQHE, for 
which $\sigma_{xy}$ in units of $e^2/h$, the Chern 
number, can be calculated with the topological TKNN formula.  
 
Recent advances in the electron beam lithography has made it 
possible to fabricate 2DEG's with 2D periodic modulations, 
and the butterfly was observed.\cite{geisler}

Hofstadter's problem has been examined not only in condensed-matter 
systems, but also in 
cold-atom systems with greater experimental 
controllability\cite{PhysRevLett.111.185301,PhysRevLett.111.185302}.  
The bulk-edge correspondence (as described in that Section) 
in the Hofstadter problem 
has in fact been confirmed in cold-atom systems.\cite{Goldman}

A recent important addition to the Hofstadter-butterfly 
systems is the superstructures realised as the Moir\'{e} 
pattern in twisted bilayer graphene\cite{dean13}, and 
the butterfly was actually 
observed in 2021 in twisted bilayer graphene at a magic twist angle.\cite{lu21} 
.  
We shall describe this in Section `Twisted bilayer graphene' below.

\clearpage
 
\subsection{QHE in three dimensions}

While usually the QHE is inherent in 
two-dimensional (2D) systems, we can raise a question: 
can we conceive a 
similar effect in three-dimensional(3D) systems, 
and, if so, how?  
If we recall the adiabatic argument 
for topological properties, we essentially exploit the 
presence of inter-Landau-level gaps in the energy spectrum.  
This implies that, if there exist, for some reason, 
energy gaps in 3D systems 
in magnetic fields, we may have a quantisation 
when $E_F$ is in a gap, 
as pointed out already in the 1980s and 1990s 
by Avron et al, and by Halperin et al.\cite{avron}  Note that 
for 3D 
we are talking about 
the Hall conductance $R_{xy}^{-1}$ rather than the Hall 
conductivity, 
since for $d$-dimensional systems 
of size $L$ the conductance $R^{-1}$ and conductivity $\sigma$ are related as 
$R_{xy}^{-1}=L^{d-2}\sigma_{xy}$, so that only in 2D do they 
happen to coincide with each other.  

\begin{figure}[ht]
\begin{center}
\includegraphics[width=13cm,clip]{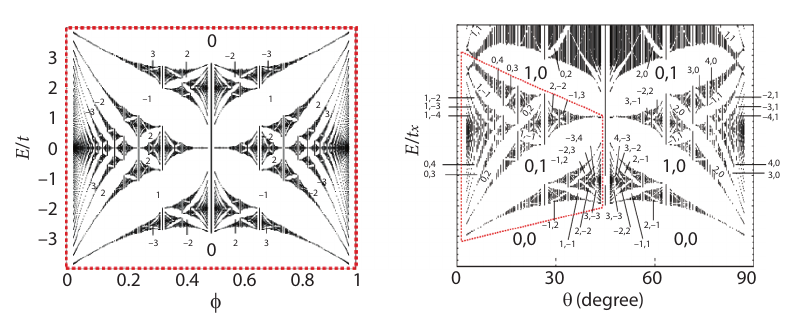}
\caption{ Right: Energy spectrum against the tilting angle, $\theta$, of 
the magnetic filed applied to a three-dimensional, periodic system.  
The QHE values (in units of $e^2/ah$ with $a$: lattice constant) 
for  $(\sigma_{xy}$, $\sigma_{zx})$ 
are indicated by a pair of numbers attached to each gap in the 
spectrum.  The region indicated by a dashed line has a 
one-to-one correspondence to the Hofstadter butterfly 
($E$ vs magnetic flux $\phi$ within a unit cell) in 
2D (left panel) 
[after M. Koshino et al, 
Phys. Rev. Lett. {\bf 86}, 1062 (2001)].
\label{Chernnumber}}
\end{center}
\end{figure}

Usual wisdom, however, is that gaps do not tend to appear
in 3D.  One possibility is to use 3D systems 
that have periodic structures or potentials.  
Koshino et al have shown that 
3D Hofstadter spectra can 
appear in periodically modulated structures 
in the energy spectrum against the tilting angle 
in tilted magnetic fields, where an interference of 
Landau's quantisations due respectively to the 
components $B_y$ and $B_z$ of the magnetic field 
is responsible (as compared with 2D where Hofstadter's butterfly 
comes from an interference between Bragg's reflection and Landau's 
quantisation).\cite{koshino01}   Then each of $\sigma_{xy}$ and 
$\sigma_{zx}$ are quantised when $E_F$ is in each gap 
with the current ${\Vec j} = - {\Vec \sigma} \times {\Vec E}$ 
where ${\Vec \sigma} = (\sigma_{yz}, \sigma_{zx}, \sigma_{xy})$ 
(Fig. \ref{Chernnumber}).   

A kind of Landau quantisation has been known to occur in 
an anisotropic organic conductor (TMTSF in the Bechgaard salt family).  This 
occurs for the field-induced spin-density wave
(SDW)  in strong magnetic
fields with a many-body origin. 
Namely, the Landau quantisation takes place within the pockets 
formed by incompletely-nested Fermi surfaces, which gives
rise to a series of gaps around the main SDW gap, and 
an associated QHE in 3D.\cite{koshino02} 

\clearpage

 \section{Anomalous quantum Hall effect and spin quantum Hall effect}
 
\begin{figure}[ht]
\begin{center}
\includegraphics[width=10cm,clip]{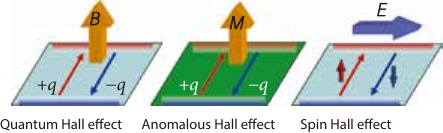}
\caption{Quantum Hall effect, anomalous Hall effect, and spin Hall effect 
are schematically shown.  $\pm q$ stands for the charge of a carrier.
\label{spinHall}}
\end{center}
\end{figure}

QHE has close relatives in the anomalous Hall effect and the spin Hall effect (Fig.\ref{spinHall}), so let us briefly describe them in relation to 
the IQHE.  
While the QHE is a topological phenomenon generic to 2DEG, the anomalous Hall effect (Hall effect 
in ferromagnetic materials) and the spin Hall effect (Hall effect for 
the spin degrees of freedom 
in materials that have strong spin-orbit interactions) arise from 
system's magnetic structure, band structure and interactions, 
so they occur in three-dimensional systems as well.

In the anomalous Hall effect in metallic ferromagnets, 
the Hall resistivity is expressed as 
\begin{equation}
R_H = R_{\rm normal}B + R_{\rm anomalous}M 
\end{equation}
with $R_{\rm normal}$ being the normal Hall resistivity, 
$B$ the external magnetic field, 
$R_{\rm anomalous}$ the anomalous Hall resistivity and $M$ 
the magnetisation of the material.  In this effect, the Hall current flows 
even in zero external magnetic fields.  
Theoretical analysis was  initiated in the 1950's by Karplus and Luttinger 
as 
a (multi-)band effect in the presence of spin-orbit interaction.  
To be more precise, 
two mechanisms have been identified, one (called extrinsic) is the 
skew scattering + side jump by impurities 
when both of spin-orbit interaction and magnetisation are present, 
and the other (called intrinsic) 
is Berry's-curvature contribution to the Hall conductivity as 
in the QHE,  which is 
included in the linear-response formula.  Experimentally, 
the anomalous Hall effect has been observed in various 
materials, typically Nd$_2$Mo$_2$O$_7$ with non-collinear 
spin configurations and 
(Sr,Ca)RuO compounds.  

In the spin Hall effect, which also occurs in zero magnetic field, 
$\uparrow$ spins and $\downarrow$ spins 
flow in the opposite directions in an electric field 
(as opposed to the ordinary Hall effect in which 
opposite charges 
flow in the opposite directions in a magnetic field).  
This effect, which also comes from the spin-orbit 
interaction, is another topological effect.  
The effect, predicted in the 1970's, is later 
experimentally observed  with e.g.
 Kerr rotation microscopy in 
both n- and p-type GaAs-based semiconductor heterostructures.

\begin{figure}[ht]
\begin{center}
\includegraphics[width=7cm,clip]{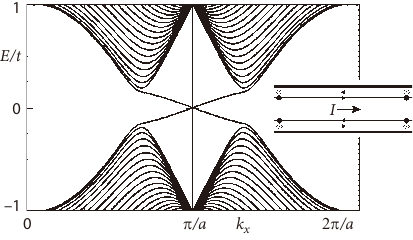}
\caption{Energy spectrum for a system that exhibits 
quantum spin Hall effect.  Bunch of lines represent bulk 
states, while a pair of isolated lines the edge states.  
Inset shows how the spin-up and spin-down electrons flow 
in real space.  [after C. L. Kane and E. J. Mele, Phys. Rev. Lett. {\bf 95}, 
226801 (2005).]
\label{QSHE}}
\end{center}
\end{figure}

QHE, anomalous Hall effect and spin Hall effect are related 
in that they are manifestations of the phase of the wavefunctions 
in magneto-transport properties. 
An intimate relation between QHE and the spin Hall effect has been 
brought home by a more recent ``quantum spin Hall effect" 
(QSHE)\cite{konig,murakami} 
which was predicted originally for graphene.  
The idea starts as follows.  Graphene 
is described, as we shall describe in Section `QHE 
in graphene', by a Dirac equation for 
two spatial dimensions 
around each of K and K' valleys in the Brillouin zone, 
where the two valleys 
are distinguished by a pseudospin $\tau_z$.  
In the presence of a spin-orbit interaction, 
we have an extra term that couples the real spin and the pseudospin, 
and this gives rise to an energy gap (a mass gap in the language of 
the Dirac theory), as proposed by Kane and Mele,\cite{KaneMele}.  
The coupling between the real and pseudospins is related, 
as mentioned by Kane and Mele, 
to a model introduced by Haldane in 1988\cite{haldane88} for 
the quantum anomalous Hall effect (QHE in zero external 
field), for spinless electrons but with spatially 
structured magnetic fluxes.  
The gap is a topological gap, where 
the gapped state cannot indeed be reached by an adiabatic 
change of the system, so that the insulator is called a topological 
insulator.  
Since the bulk is topological, the edge states 
have to exist and called topological edge states.   
In this respect the QSHE is distinct from the spin Hall effect, and 
the whole situation is rather  similar to the IQHE, 
where an essential difference is that the edge states carry charges 
in IQHE while edge states carry spins in the QSHE.  
The energy dispersion  (Fig. \ref{QSHE}) of the edge states in a QSHE system, 
which may be thought of as arising from Kramers' doublets, 
is gapless with the two dispersion crossing at a 
point in the Brillouin zone.  
The conductivity calculated with the linear-response theory for 
each of the spin-up and spin-down electrons gives a 
quantised (Chern number) conductivity, where the 
directions of the current are opposite between the up and down 
spins, with a quantised spin Hall 
conductivity shown to be $\sigma_{xy}^{\rm spin} = e/2\pi$.   
QSHE belongs to Class AII (symplectic) in 
Figs.1,2.  
We shall come back to Haldane's model(Fig.\ref{fig:HaldaneVsFTI}) in Section `Floquet 
topological insulator'.

\begin{figure}[ht]
\begin{center}
\includegraphics[width=7cm,clip]{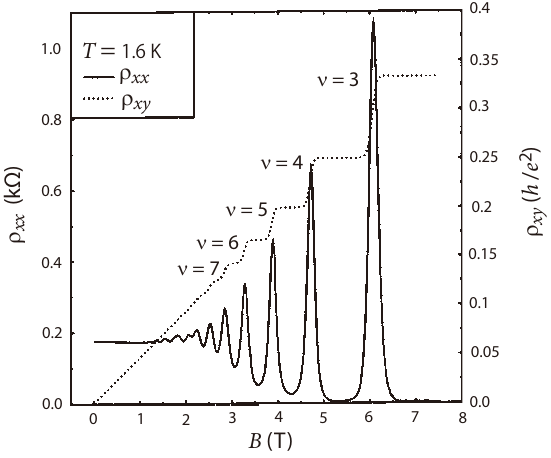}
\caption{QHE in a 
HgTe quantum well 
[after A. Pfeuffer-Jeschke et al, Physica B {\bf 256-258}, 486 (1998).]
\label{HgCdTe}}
\end{center}
\end{figure}

While the original proposal due to Kane and Mele\cite{KaneMele} was made 
for graphene, the material has too small a 
spin-orbit interaction.  Subsequently, the effect was shown to be realised 
in quantum wells with narrow-gap semiconductors, HgTe/CdTe.\cite{Bernevig06}  
While both of HgTe and CdTe have zinc-blende crystal structure, 
their band structures are affected by the spin-orbit interactions 
in significant ways.  
Namely, the bulk HgTe has a band structure 
called ``inverted", 
since a band that usually forms the valence band 
is pushed above another band that usually forms the conduction 
band, where the former consists of two (heavy- and light-mass) 
bands that touch with each other at a point in the Brillouin zone 
with opposite curvatures (i.e., the bulk system is a semimetal).   
While IQHE (Fig.\ref{HgCdTe}) was observed in 
HgTe/CdTe quantum wells and superlattices 
grown by MBE,  the QSHE itself was then detected.  
The band structure behind this is, 
when HgTe is sandwiched between CdTe to make a quantum well, 
the bands in HgTe well remain inverted for wide wells, while 
the bands become normal for thin enough wells.  So the well thickness 
can act to control the presence or otherwise of a mass gap, 
and the system should be a quantum spin Hall insulator above a 
critical thickness.  
With the spin Hall effects we can electronically 
manipulate the spin degrees of freedom, so that 
applications are explored, which is 
called the spintronics.

\clearpage

\section{Integer vs fractional quantum Hall effects}

The integer QHE has subsequently been 
developed into the fractional QHE (FQHE).  
In FQHE the Hall conductivity is quantised into 
$\sigma_{xy} = \nu e^2/h$ for fractional Landau level fillings 
$\nu = 1/3, 2/3, 3/5, ...$ as opposed to the IQHE for $\nu=$ integer.  
 Let us compare the IQHE with FQEH in this section.  
Physically, the integer QHE is primarily understood in terms of 
a one-body problem as described in this chapter, while FQHE is inherently a  many-body effect, 
and this is a customary way to distinguish the integer quantum Hall effect 
and FQHE.   
It is natural that the IQHE was originally discovered 
in Si-MOSFET, while the FQHE effect in GaAs-AlGaAs 
heterostructures, where the latter system is atomically much cleaner 
with typical mobility exceeding $\mu \simeq 10^6\;{\rm cm}^2/{\rm V\cdot s}$ against 
the former's $\sim 10^4\;{\rm cm}^2/{\rm V\cdot s}$.  
The FQHE has also observed in clean enough SiMOSFET's.  

\begin{figure}[ht]
\begin{center}
\includegraphics[width=7cm,clip]{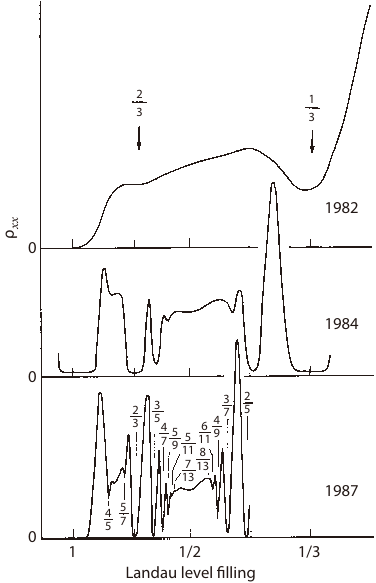}
\caption{Developments of the fractional structure 
in samples of progressively higher quality  
[after D. C. Tsui et al, Phys. Rev. Lett. {\bf 48}, 1559 (1982), A. M. Chang et al, ibid {\bf 53}, 997 (1984), R. Willett et al, 
ibid {\bf 59}, 1776 (1987)].
\label{ItoFQHE}}
\end{center}
\end{figure}

One clear way to realise that the FQHE, a many-body effect, 
emerges as the degree of disorder is lowered is to look at 
the historical developments from IQHE to FQHE 
in Fig.\ref{ItoFQHE}, which shows how the fractional effect appears 
as the sample quality (as characterised by the carrier 
mobility) becomes higher.  In 2002 
$\mu = 10\times 10^6 \;{\rm cm}^2/{\rm V\cdot s}$ is 
attained, where fractions observed include 6/25.\cite{pan02} 
In the clean limit, the system is indeed in the limit of strong 
electron correlation in that the kinetic energy is quenched 
due to Landau's quantisation so that the ratio of 
the interaction energy to the kinetic energy is infinite 
(although more rigorously we have to consider the 
inter-Landau level matrix elements).  
While the FQHE is a new class of many-body states, 
the electron-electron interaction exists for integer 
fillings as well.   
In fact,  an integer is a kind of fraction in that 
Laughlin's wavefunction for the FQHE liquid 
for clean systems, allowed for $\nu = 1/m$ with $m$ an odd integer, 
also accommodates 
$m=1$.  So the real question is the nature of 
the energy gap: in the integer QHE the excitation gap is primarily the 
one-body gap between the adjacent Landau levels (or 
between the mobility edges in disordered systems), while 
 the excitation gap in the FQHE has a many-body origin.

\begin{figure}[ht]
\begin{center}
\includegraphics[width=7cm,clip]{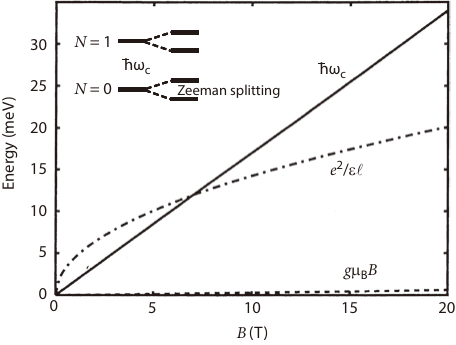}
\caption{Various energy scales (cyclotron, Coulom and Zeeman) against 
magnetic field, here plotted for GaAs.  Inset schematically 
shows the Landau levels with 
Zeeman splitting. 
\label{Energyscales}}
\end{center}
\end{figure}

The relative weight of the one-body and many-body 
natures depends on the degree of disorder (and 
the g-factor in the case in which the adjacent Landau 
levels are Zeeman-split ones).  
To quantify this, we can 
examine the relevant energy scales.  Figure \ref{Energyscales} plots 

(i) $\hbar \omega_c$ (the cyclotron energy), 

(ii) $e^2/\varepsilon \ell$ (typical size of the electron-electron Coulomb interaction), 

\noindent where $e$ is the elementary charge, 
$\varepsilon (=13$ for GaAs) the dielectric constant of the material 
and $\ell$ the magnetic length, and 

(iii) $g\mu_BB$ (the Zeeman energy), 

\noindent where $g$ is Land\'{e}'s g-factor,  $\mu_B 
= \hbar e/(2mc)$ the Bohr magneton.  We can see that 
for $B \sim$ few tesla the Zeeman energy is relatively negligible, 
while the cyclotron energy and the Coulomb energy are 
comparable.  
For disordered systems we have to compare these with the 
Landau level broadening $\Gamma$, which is roughly 
estimated in the self-consistent Born approximation 
as $\Gamma/(\hbar\omega_c) \sim 1/(\omega_c\tau_0)^{1/2}$, 
where $\tau_0$ is the scattering relaxation time in zero magnetic 
field.   So the energy scale of disorder (Landau level broadening) 
becomes comparable with  the cyclotron energy ($\sim$ 
the Coulomb energy for $B \sim 10$ T) for $\omega_c\tau_0 >\sim 1$.

While the two-dimensionality in IQHE appears in the fact that 
the Hall conductivity represents a topological quantum 
number (Chern number) in terms of the Berry's curvature, 
in FQHE the spatial dimensionality of two 
is essential in allowing, and indeed accommodating, such novel concepts as 
anyon quasi-particles with fractional statistics, 
and a description in terms of 
the composite fermion picture 
(i.e., a Chern-Simons gauge field theoretic treatment) 
of the many-body quantum liquid.

The relevant chapter  should be referred to for the FQHE, 
so suffice it to mention here that FQHE can be regarded as an IQHE of composite fermions in the composite fermion picture.  
The Landau level filling is expressed as $\nu = N_e/N_{\phi}$, 
for $N_e$ electrons in $N_{\phi}$ flux quanta, so that 
an odd fraction, say, $\nu=1/3$ implies that there are three flux quanta 
per each electron on average.  In the composite fermion picture 
we attach two flux quanta to each electron (in a 
kind of gauge transformation), and we are left with one flux quantum.  
So the $\nu=1/3$ state can be mapped to a $\nu=1$ state of 
composite fermions in 
a mean-field sense.    
One application of this correspondence is the ``global phase diagram" 
in Fig.\ref{global}.\cite{kivelson92} We have explained the $\sigma_{xx}-\sigma_{xy}$ diagram above for 
the IQHE.  If we combine this with the composite particle 
transformation, we can, again in a mean-field sense, 
a phase diagram for the integer and fractional QHE phases against 
the Landau level filling and the degree of disorder.

\clearpage

  \section{QHE in graphene}

One fascinating aspect of the condensed-matter physics is that 
we can have various field theories effectively realised on 
low-energy scales.   Recent emergence of the physics of 
massless Dirac 
particles (or Weyl particles in the language of field theoretic 
texbooks) in graphene is a prime example.   
While the three-dimensional graphite has long been studied 
extensively, experimental fabrication of graphene had to wait 
for the accomplishment by Geim's group around 2004.  
IQHE then received a strong impetus when 
seminal series of works on graphene were 
launched for graphene after around 2005.\cite{NovoselovReview05,GeimNovoselof07,CastronetoRMP09}   

\subsection{Monolayer graphene}

\begin{figure}[ht]
\begin{center}
\includegraphics[width=13cm,clip]{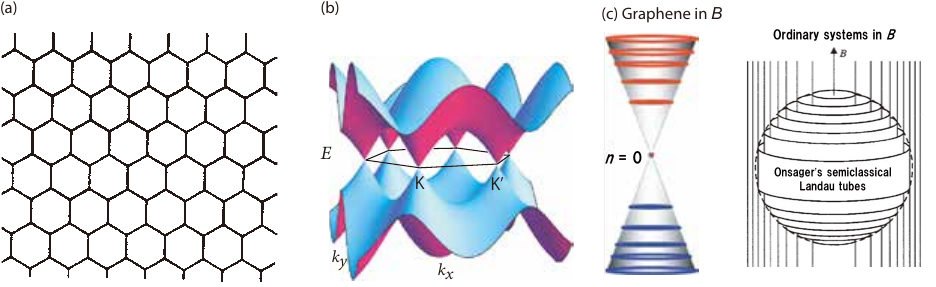}
\caption{Crystal structure (a) and the band dispersion (b) of graphene.  
(c) Landau levels around each Dirac cone in a magnetic field $B$, 
with $n=0$ Landau level highlighted.  
An attached panel on the right schematically shows 
Onsager's semiclassical Landau quantisation in terms of Landau tubes 
on the Fermi surface for ordinary systems.
\label{graphene}}
\end{center}
\end{figure}

Graphene is a monolayer graphite with a honeycomb array of carbon 
atoms (Fig.\ref{graphene}(a)), while graphite is a stack of graphene sheets (in a staggered 
manner called Bernal stacking).  
Electrons on a single layer of honeycomb lattice, despite its 
simplicity, provide rich problems in 
condensed matter physics.   
Specifically, it has long been known 
that its band dispersion (for carbon's $\pi$ orbitals) 
is composed of a pair of $k$-linear 
conical electron and hole dispersions that touch with each other at $E=0$ 
(Fig.\ref{graphene}(b)),  
so that the graphene is a
condensed-matter realisation of massless Dirac fermions 
in two spatial dimensions 
around $E=0$ (at which $E_F$ usually resides).   
This was noted by Wallece 
as early as in 1947.  Subsequently the reason why honeycomb symmetry 
implies the massless Dirac 
dispersion was revealed group-theoretically 
by Lomer and  by Coulson in the 1950's.  Band analysis in terms of 
the $k\cdot p$ perturbation was also done by Slonczewski and Weiss.  
Three-dimensional graphite, by contrast, has a band 
structure of a semimetal with a Fermi surface 
comprising small electron and hole pockets.

An inherent aspect of graphene is that it has two, inequivalent Dirac cones at K and K' points in the 
first Brillouin zone (Fig.\ref{graphene}(b)).  The region around 
each of K and K' is called a ``valley".  
To see this we should go back to the 
honeycomb lattice, which is a non-Bravais lattice that contains two (A and B) 
sublattice sites in a unit cell.   Such a lattice is called bipartite, 
where the 
energy spectrum is shown to be electron-hole symmetric about $E=0$ if we 
only consider nearest-neighbour hopping.  
The tight-binding model (for $\pi$ electrons that is relevant 
to conduction) on a honeycomb lattice 
reads, in k-space, 
\begin{eqnarray}
{\cal H} &=&  t \sum_{{\Vec k}} 
\left(
\begin{array}{cc}
0 & D({\Vec k} ) \\
D^*({\Vec k} ) & 0
\end{array}
\right), \nonumber \\  
D({\Vec k})
&=& 1 + e^{-ik_1}+e^{-ik_2}, 
\label{Dirachamiltonian}
\end{eqnarray}
where the 2$\times$2 matrix is spanned by the AB sublattices, 
$k_1, k_2$ are the wavevectors along the primitive vectors in 
the reciprocal space for the honeycomb lattice.  
The eigenenergies are $\pm |D({\Vec k})|$, giving us 
the Dirac-cone dispersion.

In the vicinity of K and K' in the Brillouin zone, we can 
use the $k\cdot p$ perturbation to 
show that the Hamiltonian is linearised in $p$ as
\begin{equation}
{\cal H} = v_F (\sigma_x \tau_z p_x + \sigma_yp_y),
\end{equation}
where $v_F \simeq 10^8$ cm/s is the Fermi velocity, and $\sigma_{\mu}$ 
is the Pauli matrix for the 2$\times$2 Hamiltonian, 
while  $\tau_z$ is another Pauli matrix representing 
the valley (K and K') degrees of freedom with 
$\tau_z = 1 (-1)$ for K (K').  
So the equation (for each value of $\tau_z$) has the same form as 
the Dirac equation for a massless particle in two spatial dimensions, 
although the equation here does not refer to a relativistic one.   
The Hamiltonian anticommutes with $\sigma_z$ (which plays the role of 
a Dirac operator $\gamma_5$ in (3+1) dimensional Dirac field), and this symmetry is 
called the chiral symmetry, again borrowing from the Dirac-electron 
nomenclature.

In a more general framework, we can generically express the Hamiltonian for 
two-band systems as 
\begin{eqnarray*}
{\cal H}({\Vec k} )= {\Vec R}({\Vec k} ) \cdot {\Vec \sigma } 
= 
\left( 
\begin{array}{cc}
R_3 & R_1-i R_2 \\
R_1+i R_2 & -R_3
\end{array}
\right)
,
\label{chiralrep}
\end{eqnarray*} 
where ${\Vec R} =\; ^t(R_1, R_2, R_3)$ is a 
three-dimensional real vector with 
$R_1({\Vec k})={\rm Re\,} D({\Vec k}) $ and 
$R_2({\Vec k})=-{\rm Im\,} D({\Vec k})$, 
while ${\Vec \sigma }=(\sigma _1,\sigma _2,\sigma _3) $ is again the 
Pauli matrix.  In this representation, the eigenenergies are 
given as $E({\Vec k}  ) =  \pm |{\Vec R}({\Vec k} )|$.  
From the $k\cdot p$ perturbation, we have 
\begin{eqnarray} 
{\cal H}({\Vec k} )  &\approx 
[(\partial _{k_x}{\Vec R}) \cdot {\Vec \sigma }]  k_x
 +
[(\partial _{k_y}{\Vec R}) \cdot {\Vec \sigma }] k_y
\label{XYrep}
\end{eqnarray} 
around each valley.

This formalism answers the question: Is the Dirac cone an accident 
for the honeycomb lattice? and when and how Dirac cones actually appear? 
The answer is that we have Dirac cones when the chiral symmetry 
exists.   When there exists a (Hermitian) operator ${\Vec \gamma }$ 
(with ${\Vec \gamma }^2 = 1$) that anticommutes with the Hamiltonian as 
$\{{\cal H}, {\Vec \gamma }\}=0,$ the Hamiltonian is 
called ``chiral-symmetric".  Equation (\ref{Dirachamiltonian}) 
has the chiral symmetry with $\gamma = \sigma_z$. In more general 
terms with Eq(\ref{XYrep}), the Hamiltonian is chiral-symmetric 
if ${\Vec R} ({\Vec k} ) \perp {\Vec n} _\gamma$ everywhere in k-space 
with a certain vector $ {\Vec n}_\gamma$, for which the 
chiral operator is given as $\gamma = {\Vec n}_\gamma \cdot {\Vec \sigma}$. 
Then we can designate the value of the chirality $\chi$ 
as $\chi=+1$ when 
the three vectors 
$(\partial _{k_x}{\Vec R}, \partial _{k_y}{\Vec R}, {\Vec n}_\gamma)$ are right-handed, 
or $\chi=-1$ when left-handed.  
See Ref\cite{aoki_hatsugai_PhysOfGraphene} for details.

\begin{figure}[ht]
\begin{center}
\includegraphics[width=11cm,clip]{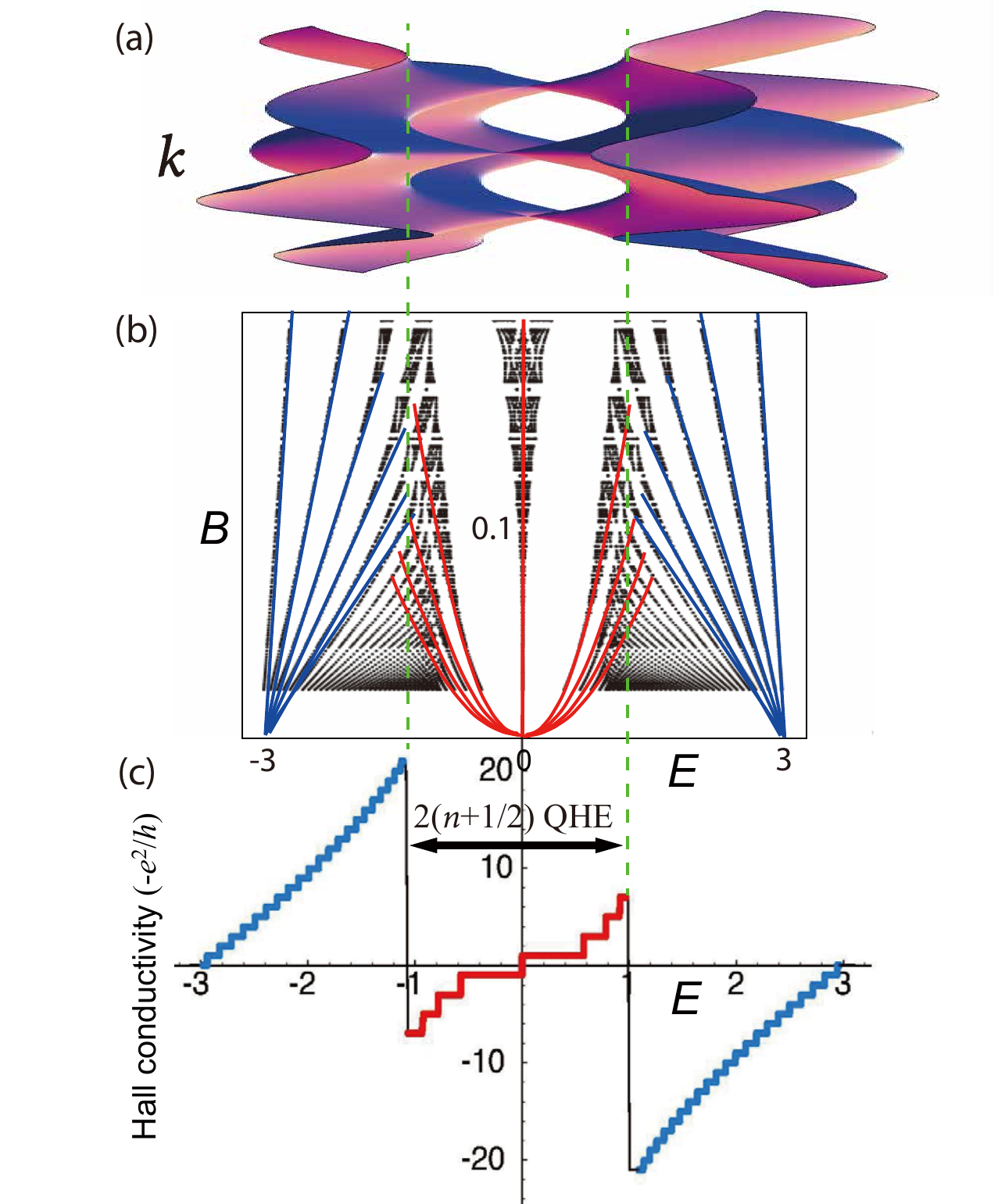}
\caption{For the honeycomb tight-binding model for graphene, we show 
(a) the full band dispersion over the entire energy range $E$, 
(b) the energy spectrum (black) against $E$ and external magnetic field $B$, 
and (c) the quantum Hall conductivity in units of $-e^2/h$.  
The unit of energy is the absolute value of the nearest-neighbour 
hopping in the tight-binding model, and 
vertical green lines mark the position of the van Hove singularities.  
In (b) Dirac-electron Landau levels are superposed in red, while 
the usual massive-fermion Landau levels, starting from the 
band edges, are superposed in blue.  The magnetic field $B$ is 
measured by the magnetic flux $\phi \equiv BS/(2\pi)$ 
penetrating each unit hexagon of area $S$ in the honeycomb lattice.  In (c) 
the Chern number is plotted 
for a magnetic field $\phi=1/31$ here. 
The red lines indicate the graphene QHE with 
$\sigma_{xy}=\pm2[n+(1/2)]e^2/h, n$: integer and factor of 
2 coming from the valley degeneracy), 
while the blue lines the usual QHE with step of one ($\sigma_{xy}=\pm ne^2/h$).
} 
\label{LandauFan}
\end{center}
\end{figure}

In the presence of a magnetic field, we can replace ${\Vec p}$ 
with ${\Vec \pi} = {\Vec p}+(e/c){\Vec A}$ in the 
$k\cdot p$ Hamiltonian.  
Then the Landau levels become, as shown by 
McClure in 1956, $E_N = \sqrt{N}\hbar \omega_c$ 
with $\omega_c = (\sqrt{2}/\ell)v_F = (2ev_F/c\hbar)\sqrt{B}$, 
$\ell$ the magnetic length as usually defined (Eqn(\ref {eq:maglength})), and 
$N=1,2,...$ ($N=-1,-2,...$) correspond to electron (hole) Landau levels.  
The Landau 
levels are $\propto \sqrt{N}$ and not uniformly spaced in sharp contrast to 
the usual, uniformly-spaced Landau levels.
  In particular, 
there is the $N=0$ Landau level right at the Dirac point $E=0$.  
The cyclotron energy $\propto \sqrt{B}$ is also unusual.  
The $N=0$ Landau level is quite peculiar, which is brought home by 
noting that the level is completely outside the Onsager's semiclassical 
quantisation scheme in magnetic fields (because the Landau tube, which 
is the set of cylinders of varying radii for the semiclassical 
quantisation, cannot be defined around the Dirac point).   
In fact the $N=0$ Landau level is an outcome of a topological property of the 
massless Dirac cone, so that its presence is `topologically protected'.  
If we go back to the original honeycomb lattice in magnetic fields, the problem becomes a 
Hofstadter butterfly for the honeycomb lattice (Fig.\ref{LandauFan}(b)), which was first obtained by Rammal in 1985.

Incidentally, it is heuristic to look at the Landau's quantisation in 
the (massive) relativistic particles.  From the Dirac equation 
for a Dirac particle in a magnetic field, 
the energy levels are 
$E_N = [m^2c^4+mc^2(\hbar\omega_c(2N+1\pm1))]^{1/2},$\cite{macdonaldRel}  
which contains the non-relativistic limit, 
$E_N = mc^2+ \hbar\omega_c(N+1\pm1/2)$, as 
the leading term 
in $\hbar\omega_c/mc^2$ expansion.  
The massless limit, $m \rightarrow 0$, can be taken 
by taking care of $\omega_c = eH/mc$ that also contains the mass if we want 
to recover the graphene Landau levels.

Soon after the fabrication of graphene samples, an anomalous IQHE 
was observed by Geim's group, and by Kim's group.\cite{NovoselovReview05,GeimNovoselof07,CastronetoRMP09}  
The IQHE in graphene has
\begin{equation}
\sigma_{xy} = -2\frac{e^2}{h}(2N+1)
\end{equation}
 as contrasted with the usual $\sigma_{xy} = -2 \frac{e^2}{h}N$ 
when the carriers are filled up to the $N$-th Landau level.  
The prefactor of 2 in these equations is for the spin degeneracy.  
The peculiarity of the Dirac cone appears as the factor $(2N+1)$ 
replacing the ordinary $N$ as 
shown by \cite{ZhengAndo,gusynin}.  There are K and K' 
valleys that contribute equally to the Hall conduction, so that 
if we factor out the valley degeneracy of 2 along with the spin degeneracy, 
we can write 
\[
\sigma_{xy} = -4\frac{e^2}{h}(N+1/2),
\] 
where, remarkably, a fraction (1/2) appears in the valley-resolved 
contribution.  

\begin{figure}[h]
\begin{center}
\includegraphics[width=14.5cm,clip]{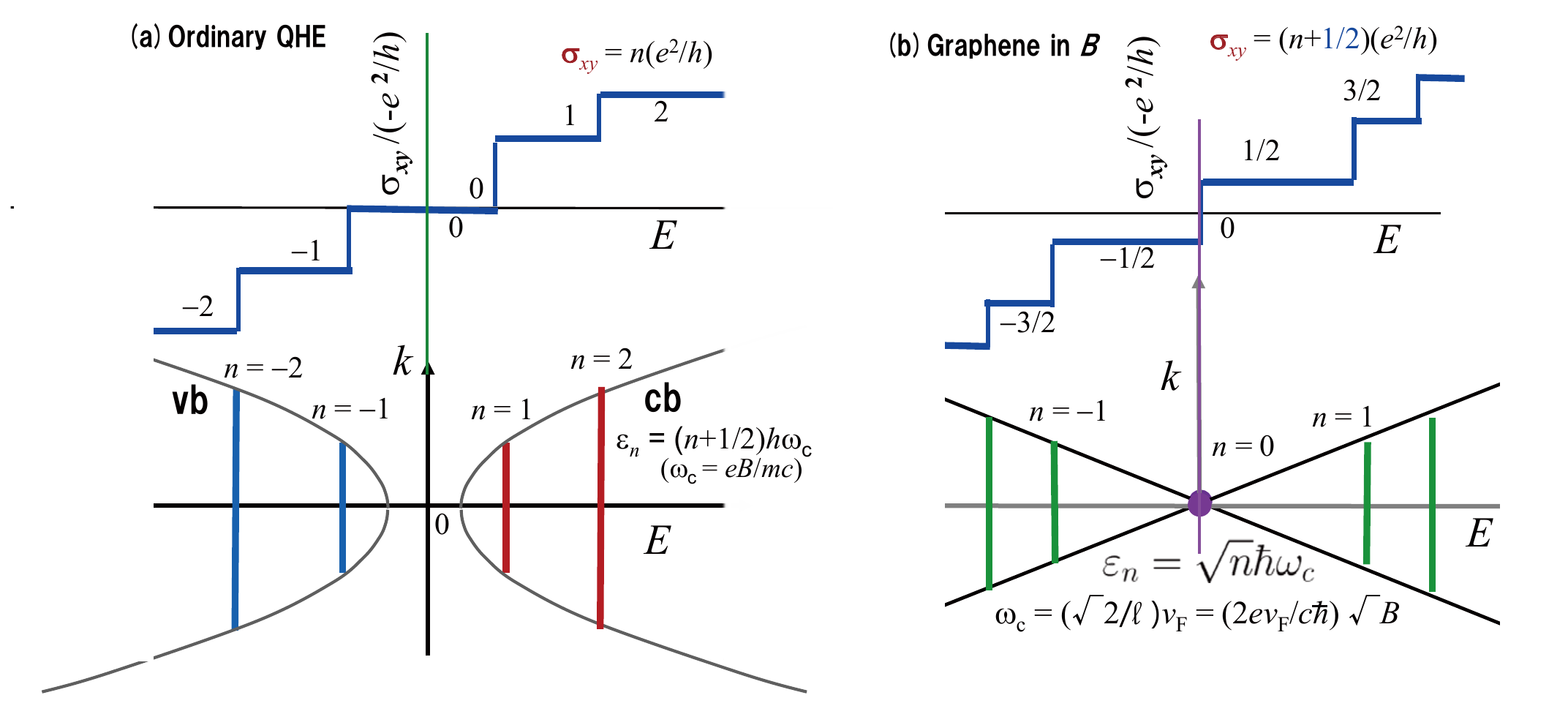}
\caption{Band dispersion and Landau levels (lower panels) and 
the QHE steps (upper) are compared between the ordinary 
2D system with valence and conduction bands (a) and 
graphene with $k$-linear, Dirac-cone dispersion (b).}
\label{graphene2}
\end{center}
\end{figure}

The situation around the $N=0$ Landau level is 
indeed unusual.  Let us compare the situation with the ordinary one where 
we have a (massive) conduction band and a (massive) valence band 
in Fig.\ref{graphene2}.  
In the latter case, we have an ordinary IQHE sequence, 
$0\rightarrow 1 \rightarrow 2 \rightarrow ...$, for electrons (in the conduction 
band)  and another, 
$0\rightarrow -1 \rightarrow -2 \rightarrow ...$, for 
holes (in the valence band).  For a Dirac cone, 
the IQHE step across $N=0$ at the Dirac point (i.e., an electron-hole symmetric point) cannot be $0\rightarrow +1$ nor $-1\rightarrow 0$, which would be incompatible 
with the electron-hole symmetry.  Instead, the step 
changes from -1/2 to +1/2 at $N=0$.  
The quantum Hall step are observed to remain robust even at 
room temperature, which is distinct 
from the usual QHE.

\begin{figure}[h]
\begin{center}
\includegraphics[width=13cm,clip]{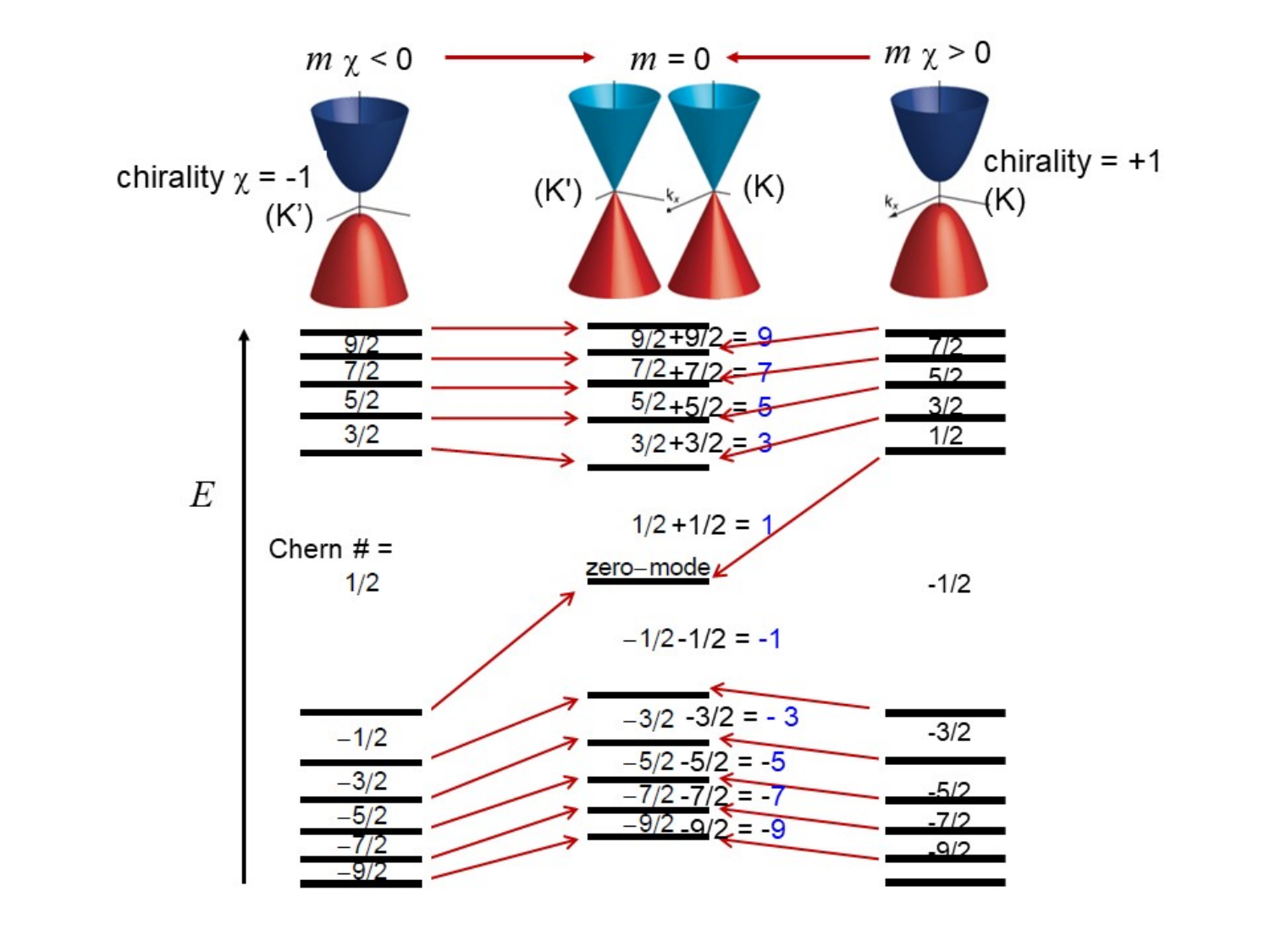}
\caption{The way in which the Landau levels of a massive Dirac fermion 
with chirality $\chi = -1$ (valley K'; left column) or +1 (K; right) cross over 
(red arrows) to 
those for the massless 
Dirac cones.  The numbers in black represent 
the Chern numbers for each valley, while the numbers in blue 
the sum of K and K' contributions.  Top insets depict the dispersions. 
After Y. Hatsugai and H. Aoki in H. Aoki and M.S. Dresselhaus (eds.): 
{\it Physics of Graphene} (Springer, 2014), Ch.7.
}
\label{masslessLimit}
\end{center}
\end{figure}

Topologically, the appearance of $1/2$ in the graphene QHE 
is actually natural, since, in the 
representation Eq.(\ref{chiralrep}), 
we can express the TKNN formula adapted for the two-band 
model that has a chiral symmetry as 
\begin{eqnarray}
C &=
- \frac {1}{8\pi}
\int \hat{{\Vec R} } \cdot 
(d \hat{{\Vec R} }
\times 
d \hat{{\Vec R} }) 
= - \frac {1}{2}\; {\rm sgn}(m\chi),
\label{grapheneTKNN}
\end{eqnarray}
where $\hat{{\Vec R} }\equiv {\Vec R}/R$, $m$ is the mass of the Dirac electron, and 
$\chi$ is the chirality.  The Hall conductivity is then 
given by $\sigma_{xy} = (e^2/h)C$.  
This is the most concise 
expression for the graphene quantum Hall number.  
The massless limit $m \rightarrow 0$ to resume the Dirac cone 
has to be taken carefully as shown in Fig.\ref{masslessLimit}.  
See Ref.\cite{HatsugaiFukuiAoki,aoki_hatsugai_PhysOfGraphene} for details.


So far we have mainly dealt with the Dirac field formalism, 
where the honeycomb lattice structure is only considered 
in a $k\cdot p$ formalism in the very vicinity of the 
Dirac points.  Then an interesting question is: 
what if we fully take account of the lattice structure?  
Will the Dirac-field result, particularly the 
QHE, be washed away soon after 
we go away from the Dirac point?
Hatsugai et al \cite{HatsugaiFukuiAoki} have looked into this problem, 
and found that the massless Dirac particle behaviour 
with the QHE $\sigma_{xy} \propto \pm(n+1/2)$ 
persists, surprisingly, up to a significanly high 
energies, at which the usual finite-mass fermion behaviour 
abruptly takes over. The boundary energies are 
just the van Hove singularities, seen in the full 
band dispersion as the positions at which 
the Dirac cones are rounded off. A technique developed in
the lattice gauge theory enabled them to calculate the behaviour
of the topological number.  
This result indicates a robustness of the topological
quantum number, and should be observable if the
chemical potential can be varied over a wide range in
graphene.  The result is summarised in Fig.\ref{LandauFan}.

As for the effect of disorder and its effect on localisation, graphene samples 
themselves are atomically clean (although there are extrinsic source of 
disorder such as charged impurities), so that graphene can have a very high mobility.  One intrinsic disorder is corrugations of the graphene sheet, called ripples.  As for the Landau quantisation in magnetic fields, ripples as represented as random hopping energies or random compontents in the magnetic 
field, both of which respect the  chiral symmetry, shown to exert anomalously small effect on the broadening of the $N=0$ 
Landau level.\cite{kawarabayashi11}    The spatial correlation 
length of the disorder, on the 
other hand, dominates the scattering between K and K' points, where 
the longer-ranged the disorder the K-K' scattering becomes less 
effective.

There are other materials that accommodate 
Dirac cones.  A typical example is an organic metal 
$\alpha$-(BEDT-TTF)$_2$I$_3$ [BEDT-TTF stands for bis(ethylenedithio)
tetrathiafulvalene], which comprises a stack of 2D layers and 
the Dirac cones are tilted and do not sit 
at the corner of the Brillouin zone.  
Subsequently, Landau quantisation characteristic of 2D and 
QHE are reported for this material.\cite{tajima06}  
Theoretically, the notion of the chiral symmetry, and the 
associated QHE are constructed for the tilted Dirac 
cones.\cite{kawarabayashi11}  In this case, we can generalise the chiral symmetry by introducing a 
generalised chiral operator $\gamma$ (which is non-Hermitian in 
general) with a chiral commutation relation 
$\gamma^{\dagger} H + H\gamma =0$ with the Hamiltonian $H$.  
The generalised chiral operator is definable as far as the 
Hamiltonian is elliptic as a differential operator.   
We can thus realise that this 
is a topological effect, related with Atiyah-Singer's 
index theorem for elliptic differential operators.   
In magnetic fields, 
we have QHE, where the peculiarity of the tilted Dirac 
cone appears as a chirality-protected ``zero-mode", i.e., $n=0$ 
Landau level at $E=0$ which remains sharp even with 
disorder, again as far as the disorder is not too rapidly-varying 
in real space.  
The zero-mode Landau level has experimentally been 
detected in the $\alpha$-(BEDT-TTF) system.\cite{tajima06}

\subsection{Optical properties in graphene}
\begin{figure}[ht]
\begin{center}
\includegraphics[width=14cm]{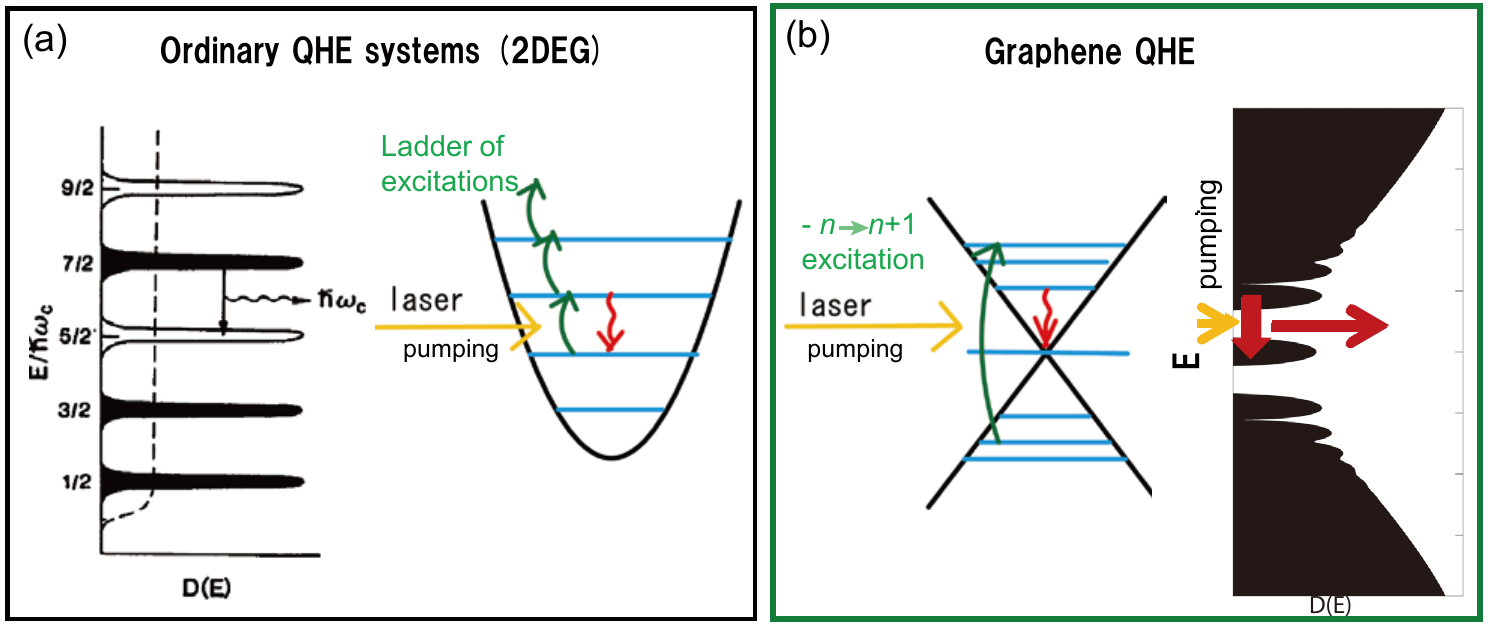}
\caption{(a) Landau levels  and 
optical pumping  are schematically shown.  
After H. Aoki, Appl. Phys. Lett. {\bf 49}, 559 (1986). 
(b) Graphene Landau levels and 
optical pumping.  
After T. Morimoto et al, Phys. Rev. B {\bf 78}, 073406 (2008). 
}
\label{LLL}
\end{center}
\end{figure}

Graphene QHE offers ample opportunities in photonics as well.  
Let us touch upon them.  An example is a proposal of 
Landau-level laser.  For the ordinary 2DEG in magnetic fields, 
Aoki proposed that, if we can pump the electrons in lower 
Landau levels to higher ones, then a cyclotron emission may realise 
a laser, where the emitted frequency (Landau level spacing 
$\hbar \omega_C \propto B$) can be tuned with the magnetic field.\cite{LLL}  
However, a difficulty is that the Landau levels in 2DEGs are 
equally spaced, so that a photon energy tuned to the spacing 
would incur a ladder of excitations rather than 
a simple population inversion.  
Then, Morimoto et al\cite{morimotoPRB08} 
suggested that, if we go over to graphene, 
the Landau levels are not equally spaced, so that 
this problem is evaded as in Fig.\ref{LLL}.  
Still,  dynamics of electrons, which involve phonons and photons 
for relaxation processes such as Auger processes, 
have to be examined.  For recent references, see \cite{wendler15}. 
A zinc-blende crystal HgCdTe, which has 
a cone and an almost flat band in the band structure 
\cite{orlita14} may also be interesting.  

\begin{figure}[ht]
\begin{center}
\includegraphics[width=15cm]{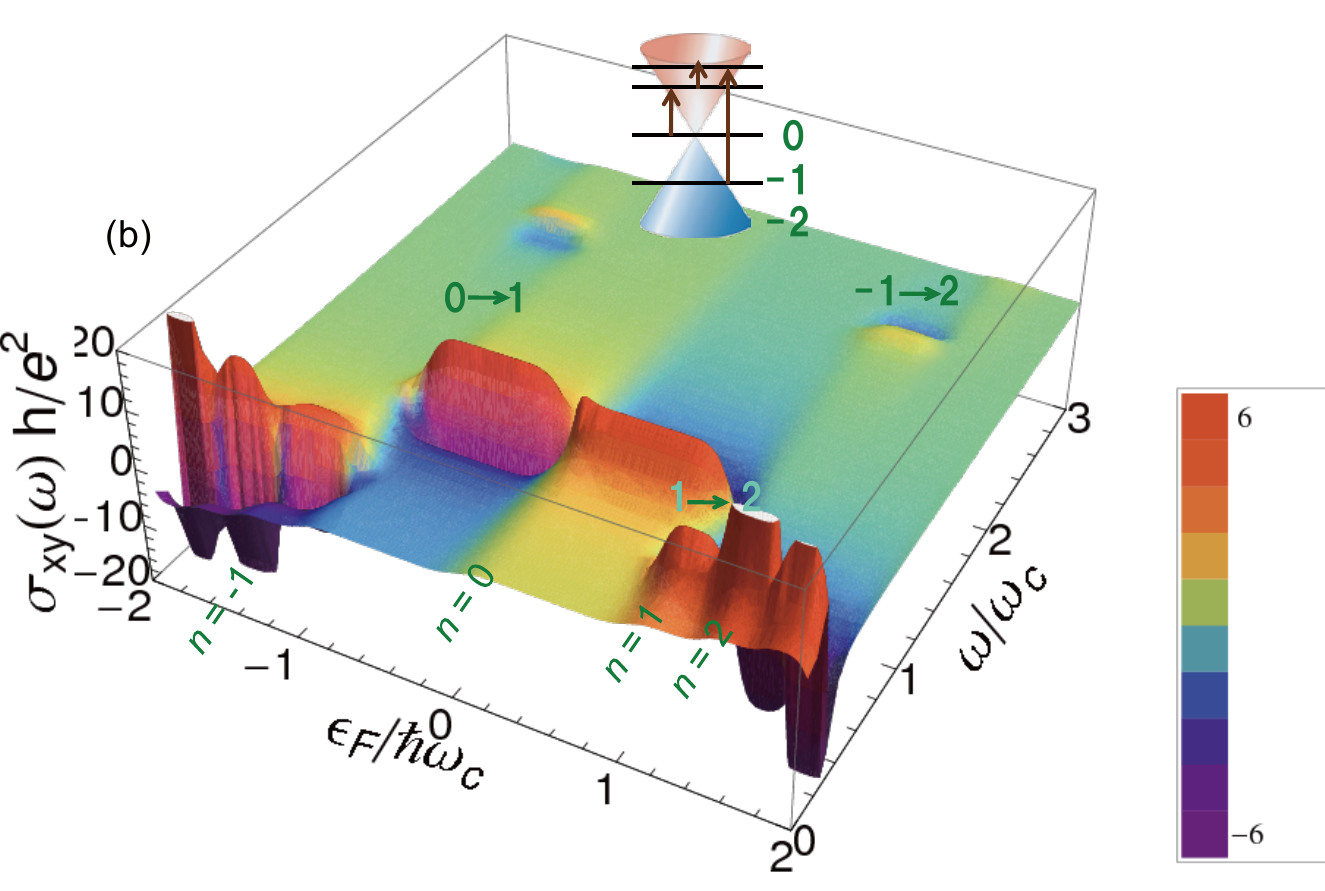}
\caption{Theoretical result for the optical Hall conductivity $\sigma_{xy}(\varepsilon_F,\omega)$ plotted against photon frequency $\omega$ and 
the Fermi energy $\varepsilon_F$ 
for the graphene QHE system.  Associated Landau level index $n$ is 
indicated, with the inset depicting the levels and 
optical transitions on a Dirac cone. 
After T. Morimoto et al, Phys. Rev. Lett. {\bf 103}, 116803 (2009).
}
\label{opticalQHE}
\end{center}
\end{figure}

One conceptially interesting question is: 
can we observe the QHE without attaching electrodes 
to the sample?  If we can accomplish this, this will 
provide a new avenue for manipulating optical properties 
through QHE physics.  
Morimoto et al\cite{morimoto_PRL09} have proposed that 
when a (linearly-polarised) light is illuminated to 
a QHE system, 
the optical Hall conductivity $\sigma_{xy}(\omega)$ against 
photon frequency $\omega$ should exhibit a characteristic feature 
typically in THz regime, as shown in Fig.\ref{opticalQHE}.  Namely, the optical Hall conductivity, 
measured through the Faraday rotation, should have 
plateaux, both in the ordinary 2DEG and in graphene in the quantum Hall regime, although the plateau height is no longer
rigorously quantised in ac. In graphene,  the optical 
conductivity $\sigma_{xy}(\omega)$ 
reflects the unusual Landau level structure. The optical 
QHE remains 
robust against the significant strength of disorder 
according to a theoretical result, which 
is attributed to an effect of the mobility gap due to 
localisation.  The estimated Hall angle in ac regime 
is of the order of the fine-structure constant $\alpha$ 
 (i.e. $\sim 7$ mrad).  
This has subsequently been experimentally detected by 
Shimano's group for a 2DEG\cite{shimanoPRL10} 
and for graphene.\cite{shimanoNatCom13}

\clearpage

\subsection{QHE and superconductivity}

As reviewed in Section 1, there is a whole spectrum of 
topological systems, which includes the topological 
superconductivity (BdG class in the classification 
table, Fig.\ref{topologicalPeriodicTable}).  So it may be heuristic to have a brief look at the 
analogy between QHE and superconductivity (SC), because 
the $2\times 2$ Hamiltonian for graphene 
has a similarity with  $2\times 2$ Hamiltonian 
for superconductors in the Nambu's spinor and Bogoliubov-de Gennes 
representations.  
Naively, the correspondence between QHE and SC would be \par
\begin{tabular}{l|c|c}
 & IQHE & SC \\ \hline
bulk & gapped (Landau levels) & gapped (SC gap) \\ \hline
edge & chiral edge states & Andreev bound states \\
\end{tabular}
\par
\begin{figure}[ht]
\begin{center}
\includegraphics[width=10cm]{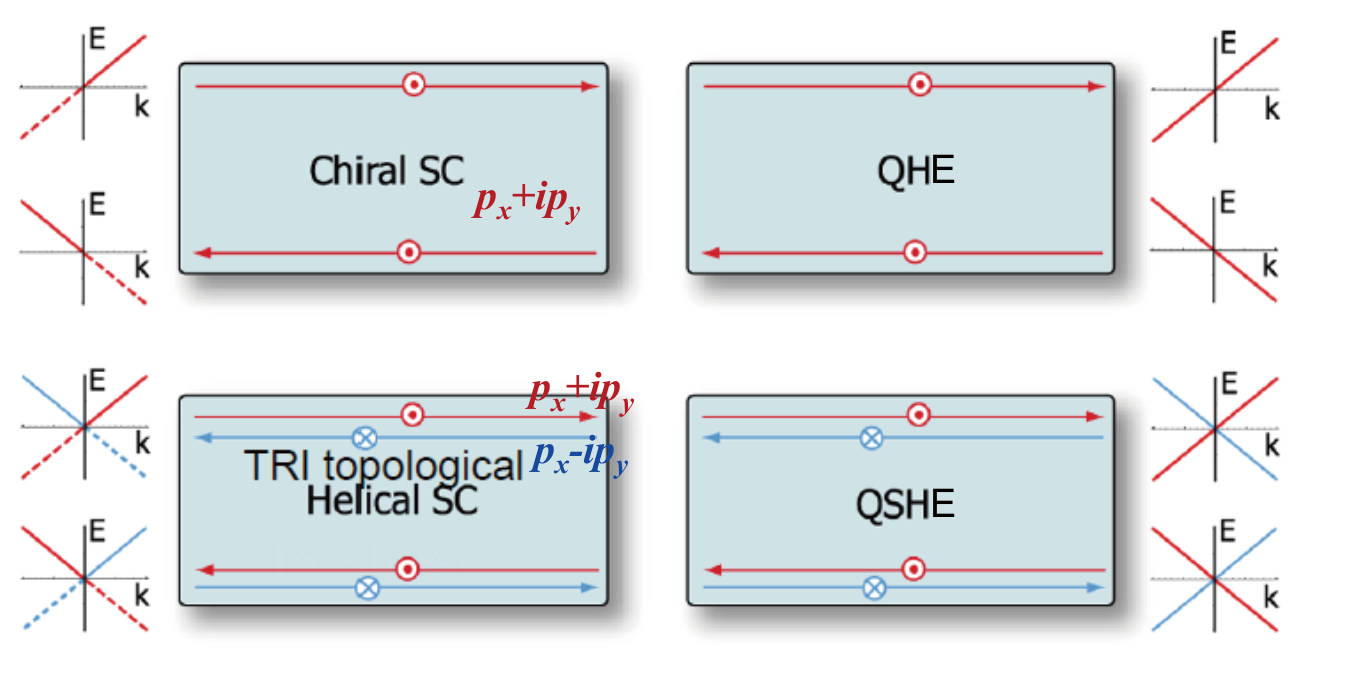}
\caption{Edge currents (arrows) in various QHE and topological SC states.  
Top row: A comparison of a 2D chiral superconductor with 
QHE, where, in both cases, the time-reversal (TR)  symmetry is broken and chiral edge states appear. 
Bottom row: A comparison of a 2D TR-invariant topological
superconductor with QSHE insulator, where, in both cases, 
TR symmetry is preserved and helical pairs of edge states appear 
for up (red, $\bigodot$) and down (blue, $\bigotimes$) spins.  Band dispersions are attached.  
After X.L. Qi et al, Phys. Rev. Lett. {\bf 102}, 187001 (2009).	
}
\label{topSCedgeStates}
\end{center}
\end{figure}

However, the analogy is not so simple: 
Only for the topological SC (time-reversal broken SC such as p+ip, d+id 
that belong to  
class D and C with spatial dimension $d=2$ 
 in the classification 
table, Fig.\ref{topologicalPeriodicTable}))  
  does the analogy hold.  Those SCs are described by a $2\times 2$ 
Hamiltonian, 
\begin{eqnarray}
{\cal H} &=&  t \sum_{{\Vec k}} 
\left(
\begin{array}{cc}
\epsilon_{{\Vec k}}-\mu & \Delta({\Vec k} ) \\
\Delta^*({\Vec k} ) & -\epsilon_{{\Vec k}}+\mu
\end{array}
\right) = {\Vec R}({\Vec k} ) \cdot {\Vec \sigma } ,
\label{topSC}
\end{eqnarray}
in the BdG formalism, where ${\Vec R} =\; ^t({\rm Re\,} \Delta({\Vec k}), 
-{\rm Im\,} \Delta({\Vec k}), \epsilon_{{\Vec k}}-\mu)$.  
The off-diagonal element, 
$\Delta$, represents the gap function, while $\epsilon_{{\Vec k}}$ 
is the band dispersion, and 
$mu$ is the chemical potential.  
The two-dimensional SC with broken 
time-reversal symmetry such as p+ip pairing is 
sometimes called a chiral SC.   We have e.g. 
$\Delta({\Vec k}) \sim k_x+ik_y$ for p+ip, which is complex, 
signifying the broken time-reversal.  
Then the topological number for the topological SC is 
given by 

\begin{eqnarray}
C &=
\frac {1}{4\pi}
\int \hat{{\Vec R} } \cdot 
\left(\frac{\partial \hat{{\Vec R}}}{\partial k_x} 
\times 
\frac{\partial \hat{{\Vec R}}}{\partial k_y}  \right) \;  d{\Vec k},
\end{eqnarray}
where $\hat{{\Vec R} }\equiv {\Vec R}/R$.  This formula 
corresponds to the TKNN expression for the QHE Chern 
number, namely Eq.(\ref{ChernCharacter}) for 2DEG, or more directly 
to Eq.(\ref{grapheneTKNN}) for graphene which is a two-band system 
as in the Nambu representation for SC.  
As for the edge currents, we can also compare them between 
the QHE and topological SC states as shown in Fig.\ref{topSCedgeStates}.  
See also Fig.\Ref{boundaryStates}.

We can also mention that, if we go over to Floquet physics, which is 
described in Section ``QHE in light-matter coupled systems - Floquet topological
insulator" below, we can realise a topological SC (d+id pairing) 
when we illuminate a d-wave SC with a circularly-polarised light.  
This is not as straighforward as in the realisation of Floquet topological
insulator described in that section, since the light field does not directly couple with 
the gap function (where a pair is electrically neutral).  
We can evade this difficulty, however, by going to 
the strong electron correlation, where the Floquet physics 
generates photon-induced many-body interactions 
(multi-site pairing interactions) rather than a 
modification of one-band structure, 
as shown by Kitamura et al.\cite{kitamuraAoki_CommPhys22} 
This provides a realisation of class C in the classification table.

  \subsection{QHE in bilayer graphene}

There has been an upsurge of developments in 
bilayer and multilayer graphene systems.  In particular, this has opened a 
seminal avenue towards twisted bilayer graphene, which has 
turned out to accommodate interesting physics and phenomena, 
encompassing QHE and superconductivity.  So let us  
look at them in this section, starting with the 
(untwisted) 
bilayer graphene.\par
\ \\

\subsubsection{Untwisted bilayer graphene}  

\begin{figure}[ht]
\begin{center}
\includegraphics[width=5cm]{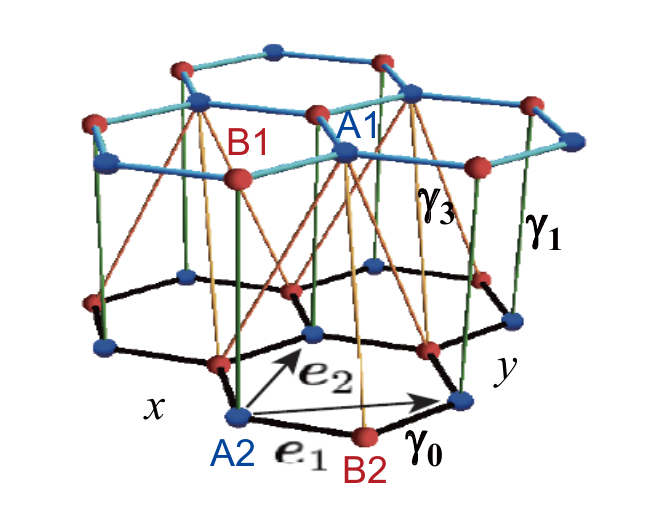}
\caption{The lattice structure of an (untwisted) bilayer graphene. 
A lattice site on B sublattice (red) in the top sheet is located just above an A sublattice site (blue) of the 
bottom sheet, and a unit cell contains four carbon atoms (A1, B1 in the top layer; 
A2, B2 in the bottom).  Main hoppings are 
the intralayer nearest-neighbour hopping ($\gamma_0$), 
the vertical interlayer hopping ($\gamma_1$), and 
an oblique interlayer hopping ($\gamma_3$).
}
\label{bilayer-lattice}
\end{center}
\end{figure}

An untwisted bilayer graphene consists of two graphene sheets 
coupled by interlayer hoppings as shown in Fig.\ref{bilayer-lattice}, where 
a B sublattice in the top sheet is located just above an A sublattice of the 
bottom sheet (called AB stacking or Bernal stacking).  
A unit cell thus contains four carbon atoms (A1, B1 in the top layer; 
A2, B2 in the bottom).  
Main hoppings are 
the intralayer nearest-neighbour hopping ($\gamma_0$), 
and the vertical interlayer hopping ($\gamma_1$).  
An oblique interlayer hopping ($\gamma_3$) also exists, which 
causes a three-fold (trigonal) warping of the Dirac 
cone around the Brillouin zone corners, $K_\pm$.  
The magnitudes of these
are $\gamma_0 \simeq 3$ eV,  $\gamma_1, \gamma_3 \simeq 0.3$ eV.

In a basis with components ($\psi_{A1} , \psi_{B1},\psi_{A2} , \psi_{B2}$), 
the Hamiltonian for the bilayer graphene is given as
\begin{equation}
H_{\rm AB}= 
\left(
\begin{array}{llll}
0 &v \pi^\dagger & 0& v_3 \pi \\
v \pi  &0& \gamma_1&0\\ 
0& \gamma_1&0 & v \pi^\dagger \\ 
v_3 \pi^\dagger &0& v \pi  &0 \\ 
\end{array}
\right)
,
\label{H-bi2}
\end{equation}
where $\pi=\xi \pi_x+i \pi_y$, 
$\Vec{\pi} = \Vec{p}+e\Vec{A}$, with 
$\Vec{A}$ being the vector potential arising from the 
applied magnetic field,
and the valley index $\xi=\pm 1$ for $K_\pm$ points.
Here $(v, v_3) \equiv (\sqrt{3} a /2 \hbar) (\gamma_0, \gamma_3)$ 
are the hopping elements expressed in terms of 
velocities with $a \simeq 0.23$ nm being 
the distance between the nearest A sites.  

The low-energy physics of bilayer graphene, 
for an energy region with $\varepsilon \ll \gamma_1$, 
is captured  in the leading order in 
$\varepsilon/\gamma_1$  by a 2 $\times$ 2 Hamiltonian
in a basis of (A1,B2) carbon sites as 
\begin{equation}
H^{\rm (eff)}_{\rm AB}
=\frac{1}{2m} 
\left(
\begin{array}{ll}
0 &(\pi^\dagger)^2 \\
\pi^2  &0 
\end{array}
\right)
+
v_3
\left(
\begin{array}{ll}
0 &\pi \\
\pi^\dagger &0 
\end{array}
\right)
,
\label{H-bilayer}
\end{equation}
with an effective mass $m=\gamma_1/(2 v^2)$.  
In the absence of magnetic fields,
the first term on the right-hand side 
gives a pair of parabolic bands, $E = \pm p^2/(2m)$, 
touching with each other.  
The second term coming from $\gamma_3$
causes the trigonal warping, and makes 
each Dirac cone 
reshaped into four small cones,
where the Lifshitz transition (change of 
the topology of the Fermi surface from four pockets 
to a single one as the Fermi energy is varied) occurs at 
$E_{\rm{Lifshitz}}= (1/2) m v_3^2  \sim 1$\, \mbox{meV}.

The Landau level spectrum in a uniform magnetic field 
$\Vec{B} = {\rm rot}\Vec{A}$ may be obtained with 
a standard procedure of introducing boson creation and 
annihilation operators,  $a^\dagger$ and $a$, as 
$(\pi,\pi^\dagger) = (\sqrt{2}\hbar/\ell) (a^\dagger,a)$
for valley $K_+$, or $(\pi,\pi^\dagger) = (\sqrt{2}\hbar/\ell) (a,a^\dagger)$
for $K_-$.  
The eigenenergies are then given by\cite{mccann-falko} 
\begin{eqnarray}
&& \varepsilon_{n,s}= s \hbar \omega_c \sqrt{n(n+1)},
\end{eqnarray}
where states are labeled by the Landau index $n = \cdots, -1,0,1,\cdots$,
and the band index $s = \pm$ 
labeling the conduction ($s=+$) and valence  ($s=-$) bands, 
and the cyclotron energy is 
$\hbar\omega_c = (\hbar eB/m) 
\simeq 3.5 (B/1 {\rm Tesla})$ meV.
Two zero-energy Landau levels appear ($n=0,-1$) 
in each valley.
\par
\ \\

\subsubsection{Twisted bilayer graphene}

Bilayer graphene physics witnessed an unexpected new 
impetus in what is called the ``magic-angle twisted bilayer graphene", 
which has turned out to accommodate rich physics.  
A technical breakthrough is that it became possible 
to put a second graphene layer on top of another with 
a designated twist angle, namely, the crystallographic 
axis of the top layer is rotated with some angle, $\theta$ from that 
of the bottom one.  The top view of the bilayer honeycomb 
lattices forms what is generally known as ``Moir\'{e} pattern", 
and is quite sensitive to $\theta$.  

For the honeycomb bilayer, this strongly affects the 
electronic structure including the formation 
of Dirac cones.  
For small $\theta$s, the 
period of the Moir\'{e} pattern is large in real space, 
which implies there are a lot of band foldings with the 
enlarged unit cell, resulting in a lot of Moir\'{e} subbands.  
Specifically, at the magic angle, $\theta \sim 1^\circ$, 
the lowest subband exhibits a dispersion that has 
almost completely flat portions.  This hosts various 
quantum phases, such as QHE as well as superconductivity.\cite{caoHerrero_SC_nature18}

If we denote the reciprocal lattice vectors
of the bottom layer as  $(\Vec{b}_1, \Vec{b}_2)$, 
the top layer has reciprocal vectors $\hat{R}\, \Vec{b}_i \quad (i=1,2)$ 
with $\hat{R}$ being the rotation matrix.  
Then the primitive lattice vectors, $\Vec{A}_i \quad (i=1,2)$, of the twisted 
bilayer graphene (TBG) satisfy 
$\Vec{A}_i\cdot \Vec{B}_j = 2\pi \delta_{ij}$, 
where 
$\Vec{B}_i = (1-\hat{R})\Vec{b}_i$ 
is the reciprocal lattice vectors for the Moir\'{e} pattern. 
We can readily see that 
\begin{equation}
|\Vec{A}_i| =  \frac{a}{2\sin(\theta/2)} \simeq a/\theta,
\label{eq_L_M}
\end{equation}
where $a \simeq 0.246\,\mathrm{nm}$ is the graphene lattice constant.
Thus the size of the unit cell in the 
Moir\'{e} system for decreasing $\theta$ blows up like $1/\theta$, which 
extends to 14 nm at the magic angle $(\theta \sim 1^\circ)$.  
Not only does the structure have a long period, but 
the double honeycomb system's top view 
comprises a periodic array of AA-, AB-, and BA-stacked 
patches. 
The Dirac points of a single-layer graphene 
reside at the Brillouin zone corners $K_+$ and $K_-$ 
separated from the $\Gamma$ point by $K = 4\pi/(3a)$ as 
described in sections above, while 
the TBG has the valleys displaced by 
\begin{equation}
\Delta K =  2K\sin(\theta/2) \approx K \theta.
\end{equation}
Now, $K_+$ and $K_-$ do not in general fall upon 
the $K_+$ and $K_-$ points after the band folding.  
In other words, the bilayer system is an incommensuate 
system for general rotation angles. 
So we have to take care of the band hybridisation 
caused by the folding for each of $K_+$ and $K_-$ valleys, 
while we can ignore the hybridisation across 
$K_+$ and $K_-$ when $K \gg \Delta K$.   

\begin{figure}[ht]
\begin{center}
\includegraphics[width=13cm,clip]{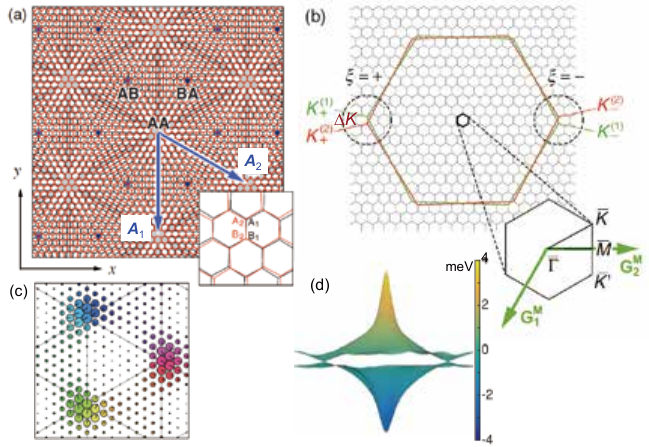}
\caption{(a) Atomic structure of the twisted bilayer graphene (TBG) with twist 
angle $\theta (= 3.89$ degrees here). AA, AB and BA 
stackings are marked, and inset is a blowup.  The primitive lattice vectors of 
the Moir\'{e} structure are denoted as $\Vec{A}_i \quad (i=1,2)$.  
(b) Brillouin-zone (Bz) folding with green (red) large hexagon 
representing the first Bz of layer 1 (2), and black small hexagons 
the Moir\'{e}  Bz.  (c) An example of maximally localised Wannier wavefunction 
in a flat band of valley $\xi = +$ of TBG with $\theta = 1.05$ degrees here.  The wavefunction amplitude is indicated by the radius of each circle, while its phase by arrows and colours.  
After M. Koshino et al, Phys. Rev. X {\bf 8}, 031087 (2018).  (d)  An example of theoretical band dispersion in the first 
Bz with $\theta = 1.05$ degrees here.  
After H.C. Po et al, Phys Rev. B {\bf 99}, 195455 (2019).
}
\label{Koshino_PRX2018}
\end{center}
\end{figure}

Figure \Ref{Koshino_PRX2018}(a) shows an 
atomic structure of TBG with AA, AB and BA 
stackings indicated. Panel (b) displays the 
Brillouin zone folding from the original ones (coloured large hexagons) 
for layers 1 and 2 into the Moir\'{e} Brillouin zone 
(black small hexagons).  
Wannier orbitals then reflect the Moir\'{e} structure, as 
typically displayed in panel (c).\cite{Koshino_PRX2018}  
Panel (d) shows a theoretical band dispersion which 
exhibits flat parts.\cite{Po_PRB2019}  
The narrow bands , especially flat bands, 
accommodate a Mott-insulator behaviour\cite{caoHerrero_MottI_nature18} 
arising from strong electron correlation in the flat band.

\begin{figure}[ht]
\begin{center}
\includegraphics[width=13cm,clip]{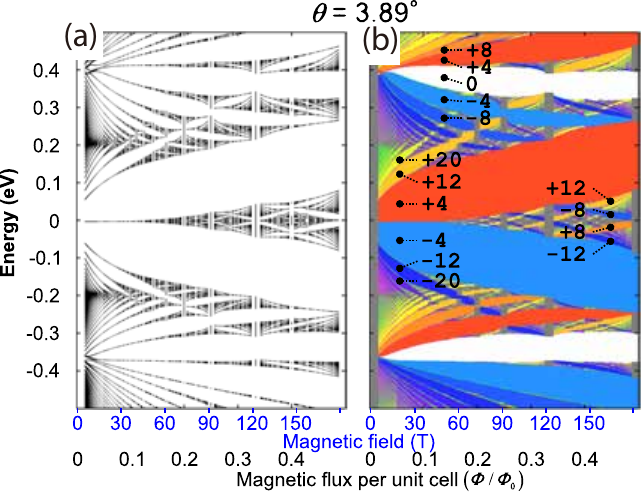}
\caption{Energy spectrum (a) and the quantum Hall effect (b) against magnetic field in the twisted bilayer graphene with a twist angle of 3.89 degrees here. In (b), the quantised values of the Hall conductivity in individual energy gaps are indicated by numbers and colours.  After P. Moon and M. Koshino,  Phys. Rev. B {\bf 85}, 195458 (2012).
}
\label{TBG_Hofstadter}
\end{center}
\end{figure}

Since the  AA, AB, and BA patches, hence the Wannier functions, 
form a long-period hexagonal lattice, 
this realises a Hofstadter problem (described in Section on 
that) when we apply an external magnetic field to 
a TBG.  Koshino and coworkers have examined this 
problem to theoretically obtain the Hofstadter butterfuly 
for TBG, Fig.\ref{TBG_Hofstadter}.\cite{moonKoshino12}  
They have obtained the Chern numbers, including the case where the bands are partially flat.

This reminds us of the Hofstadter problem for 
flat-band systems that accommodate flat bands (that are 
flat over the entire Brillouin zone), as considered in 
Ref.\cite{aoki_Hofstadter}.  An interesting point is 
the flat band either remain flat or prolifelates into 
a butterfly according as the mechanism which produces the flat band: 
flat bands that are dictated by space group, as in Lieb's model, 
remain flat in magnetic fields, while the flat bands that arise 
from quantum interference of wavefunctions, as in Mielke's and Tasaki's 
models, split into butterflies.

The TBG has turned out to harbour versatile quantum phases 
that include superconductivity.  In a typical 
example of experimentally obtained phase diagram against carrier density $n$ 
and temperature $T$,\cite{Lu_Nature19}  we can see superconducting states, 
along with metal, band insulator, and 
correlated insulating states that occur around integer values of the 
Moir\'{e} band filling.

Since the electronic state in TBG is extremely sensitive to the twist angle $\theta$, 
it is desirable to explore the inhomogeniety and disorder in the 
angle in a sample.  Uri et al\cite{Uri_Nat20} 
have employed a 
nanoscale on-tip scanning
superconducting quantum interference device (SQUID-on-tip) to obtain
tomographic images of the Landau levels in the quantum Hall state 
along with the 
local variations in $\theta$.  
Landau levels, superconducting state and correlated states 
are shown to be significantly affected 
by local variations in $\theta$ and its gradients.  
One finding is that 
the gradients of $\theta$ generate large in-plane
electric fields, which alter the
quantum Hall state by forming edge channels in the bulk.  The 
authors suggest that this may be 
important in device applications.


The bilayer graphene\cite{bilayergrapheneReview} 
is also extended to multilayer (e.g. trilayer) 
graphene systems.  
Another avenue of interest in materials is 3D porous graphene.  
Namely, a class of graphene structures shaped into three-dimensional periodic curved surfaces (``graphitic zeolites") has been considered.  
For experiments, refer to Ref.\cite{ito_PCCP17}; 
For theories, refer to Ref.\cite{koshino_3Dgraphite16} 

\clearpage

  \subsection{Edge states in graphene QHE}

\begin{figure}[ht]
\begin{center}
\includegraphics[width=3.5cm,clip]{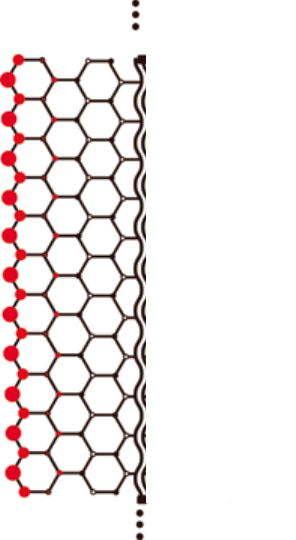}
\caption{A typical edge states in graphene in a strong magnetic field 
(here for the flux $\phi=1/5$) obtained 
numerically, with the charge density represented by 
the radius of each circle [after H. Aoki et al,  Int. J. Modern Phys. B {\bf 21}, 1133 (2007)].}
\label{graphene5}
\end{center}
\end{figure}

\begin{figure}[ht]
\begin{center}
\includegraphics[width=8cm,clip]{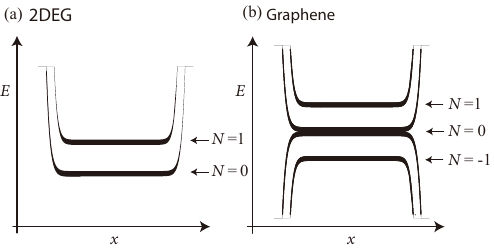}
\caption{Energy spectrum against the real-space position $x$ 
along the width for a finite sample is schematically compared between 
the ordinary QHE system (a) and graphene QHE system (b).}
\label{Fig40}
\end{center}
\end{figure}

We have described above how the graphene QHE is 
viewed topologically.  The 
topological analysis also leads to another topological character, 
namely edge states  (Fig.\ref{graphene5}) that appear 
for a finite graphene having edges in strong mangetic fields 
(which should not be confused with the graphene edges states in 
zero magnetic field, an interesting issue in its own right).  
We have also mentioned earlier that  
the bulk and edge topological numbers are equal in usual QHE, i.e., 
$
\sigma_{xy}^{\rm edge} =\sigma_{xy}^{\rm bulk}
$, which can be identified by connecting the topological integers
for the bulk and for the edge states.  
The same applies to the graphene IQHE.   
Energy spectrum against real-space position also 
differs in graphene from the usual QHE system as 
depicted in Fig.\ref{Fig40}.

Edge states in graphene in magnetic fields are beginning to be 
observed with STM as in Ref.\cite{Uri_NatPhys20}, 
which evokes a nanoscale magnetometer 
(superconducting quantum interference device; SQUID-on-tip) 
for the imaging the edge currents, where the topological component 
of the current is resolved.


\clearpage

\subsection{Electric polarisation and topological numbers}

Let us mention an aspect of topological properties 
in QHE systems that is general enough but can be 
clearly conceived in terms of graphene QHE.  
As we have seen, 
topological states, of which the QHE is a canonical 
example, carry topological numbers, in place of 
the order parameters in systems having 
spontaneously broken symmetries. 
Chern numbers characterising QHE systems 
become versatile if we consider 
lattice structures (or periodic potentials) 
as we have described in Section ``Hofstadter spectrum".\cite{Hofs}
   
The topological numbers in this problem are determined 
by a Diophantine equation for integers,\cite{TKNN} 
\begin{eqnarray}
r = t_r p + s_r q \equiv t_r p \quad ({\rm mod}\, q),
\label{diophantine}
\end{eqnarray}
where $r$ labels the energy gaps from below, 
$t_r$ is the QHE topological (Chern) number. 
Curiously, in addition to the usual QHE topological 
number, we can note that there exists  
a second topological number, 
$s_r$, appearing in the Diophantine 
equation.  This has long been known, 
but its physical meaning was revealed only recently 
by St$\check{{\rm r}}$eda and coworkers, 
where the second topological number is shown to 
represent an electric polarisation\cite{Streda06}.  

\begin{figure}[ht]
\begin{center}
\includegraphics[width=12cm,clip]{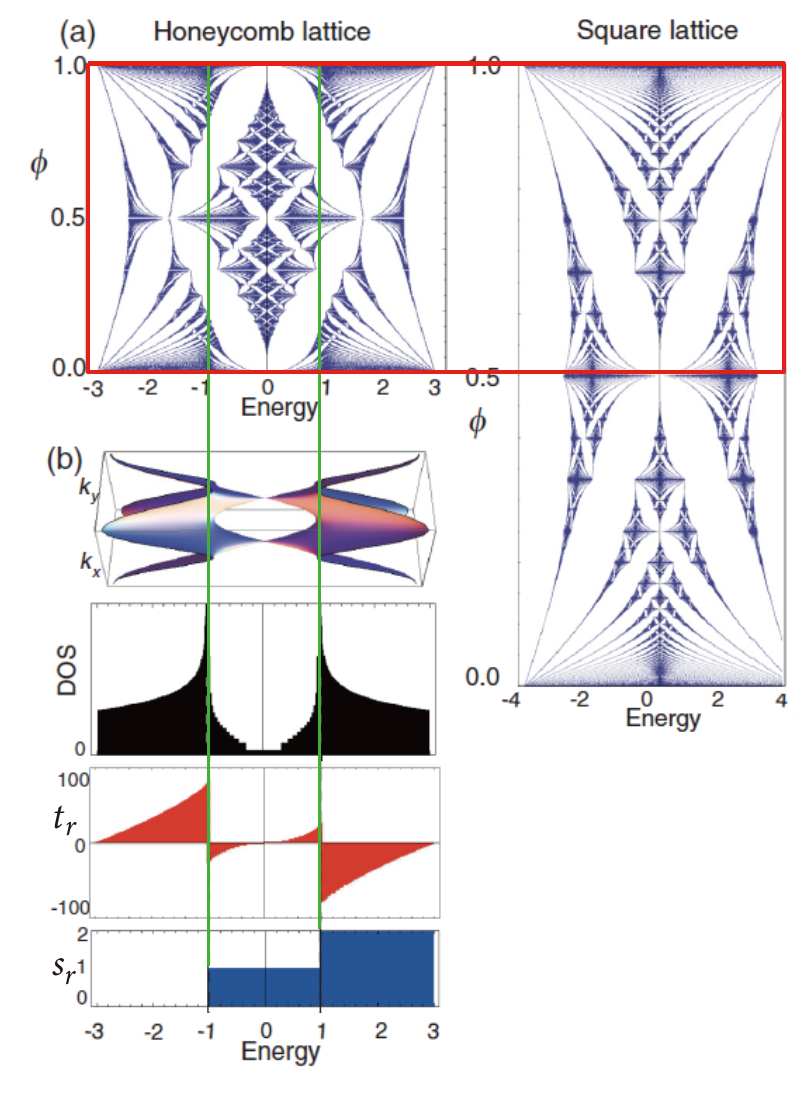}
\caption{(a) Hofstadter butterfly (one-particle energy spectrum vs magnetic field) for the tight-binding model on honeycomb (left panel) or square (right) lattices. 
The magnetic field is represented by the magnetic flux 
$\phi$ per unit cell in units of the flux quantum $\Phi_0\equiv h/e$.
To indicate the correspondence, $1/q \leftrightarrow 1/2+1/(2q)$,  between the two lattices, the energy scale is expanded and shifted on the right 
as marked with a red rectangle. 
(b) For the honeycomb lattice, the band dispersion (with the 
energy as a horizontal axis), density of states $D(E)$, 
the TKNN Chern number $t$, and the polarisation topological number 
$s$ are plotted against energy, here 
for a weak $\phi=\delta (= 1/107$).  
Vertical lines mark the van-Hove singularities.  
After H. Aoki and Y. Hatsugai, Phys. Rev. B {\bf 90}, 045206 (2014). 
}
\label{Fig_polarisation}
\end{center}
\end{figure}

Lattice structure, which exerts drastic effects on topological numbers, 
as in honeycomb lattice in graphene, 
also affects the second topological number. In the case of graphene, 
we can show that the polarisation 
quantum number behaves in a characteristic way in graphene.\cite{aokiHatugai14} To derive this we can exploit an adiabatic continuity  
between the topological numbers for 
square and honeycomb lattices as Hatsugai and coworkers 
have earlier shown\cite{HatsugaiFukuiAoki}, with which we can 
establish a correspondence between the topological numbers for 
the two lattices.  
With this, graphene is shown to be 
a ``half-flux simulator" (an adiabatic realisation of square lattice with 
half flux quantum per unit cell).  If we now apply this to the polarisation topological number in graphene in magnetic fields, 
we can show that $s_r=1$ for a wide energy region (that encompasses the two 
van Hove singularities), while $s_r=0$ or $2$ outside the vH energies, 
as shown in Fig.\ref{Fig_polarisation}.  
The second quantum number can be observable if the density 
of electrons or the polarisation itself can be measured.

In another context other than QHE, the electric polarisation for 
crystals can be, and actually should be, formulated in quantum 
mechanics 
in terms of Berry's curvature as pointed out by 
King-Smith and Vanderbild.\cite{kingsmith93}  We first note that, 
in defining the polarisation, the shift of the centre of gravity of the total 
charge can be ill-defined for an infinite crystal.  Instead, 
the polarisation (or its change, $\Delta {\Vec P}$, when a 
parameter that characterises the system is adiabatically varied) 
is expressed as an integral of a Berry's curvature as
\begin{eqnarray}
\Delta P_{\alpha} = -\frac{ie}{(2\pi)^3} \sum_n 
\int_{\rm BZ} d\bfk \int_0^1 d\lambda 
\left( \langle \frac{\partial u_{\bfk n}}
{\partial k_{\alpha}} |
\frac{\partial u_{\bfk n}}
{\partial \lambda} \rangle 
-
\langle \frac{\partial u_{\bfk n}}
{\partial \lambda} |
\frac{\partial u_{\bfk n}}
{\partial k_{\alpha}} \rangle
\right), 
\end{eqnarray}
where $u_{\bfk n}$ is a Bloch wavefunction with $n$ the band index, 
and the bookkeeping 
$\lambda$ describes 
the adiabatic change ($\lambda = 0 \rightarrow 1$).  
For a 1D system, we can cast this, with Stokes's theorem, into
\begin{eqnarray}
\Delta P = -\frac{e}{2\pi} \sum_n \left(i\oint_{\rm C} 
\langle u_{\bfk n} | 
\frac{\partial}{\partial {\Vec w}} |  u_{\bfk n}\rangle \right) 
\cdot d{\Vec w},
\end{eqnarray}
where we have denoted ${\Vec w} \equiv (\lambda, k_{\alpha})$ 
with  $k_{\alpha}$ being the wavenumber along the real-space 
axis for the polarisation measurment.  
The curvature (the quantity in the brackets above) 
is defined in a two-dimensional 
parameter space of $(\lambda, k_{\alpha})$.  
The line integral along a path C is for ${\Vec w} = (0,\pi) \rightarrow 
 (1,-\pi) \rightarrow  (0,-\pi) \rightarrow  (0,\pi)$. 
Thus the polarisation Chern number $s_r$ in the TKNN formula and the 
King-Smith and Vanderbild formula for polarisation share the 
property that both involve Berry's connection expressed as 
wavenumber-derivatives of Bloch wavefunctions, although 
the former has to do with 2D systems while the latter 
has to do with a physical quantity along a specific spatial 
direction.

\clearpage

\section{QHE in light-matter coupled systems --- Floquet topological insulator}

\subsection{Floquet formalism in general}

Nonequilibrium physics has 
recently flourished into a rich and unique field  in condensed-matter 
physics.  This not only enabled us to explore the regimes 
which are not attainable in equilibrium, but has also 
provided versatile physical concepts.  One prime example, 
in topological physics, is the 
Floquet topological insulator as initiated 
by Oka and Aoki.\cite{OkaPHE09}  
This evokes a time-periodic modulation (typically 
laser illumination) to 
put a system into nonequilibrium.  
When one deals with quantum states in a time-periodic 
modulation (such as a laser light), 
the theoretical starting point is the Floquet formalism for time-periodic 
external fields.  This is based on Floquet's theorem,\cite{Floquet1883} which is 
actually much older than Bloch's theorem for spatially periodic 
potentials --- the former dates back to 1883, 
while the latter conceived in 1928 is much more recent.

The Floquet theory is a theoretical approach to treat periodically driven 
systems, which can encompass strong eternal fields such as intense laser.  
It originates from Floquet's theorem, 
also known as 
Hill's theorem in the context of differential equations \cite{Hill1886}.  
As mentioned, the well-known Bloch's theorem  for a spatially periodic system 
is a spatial analogue of Floqeut's theorem, and there is a nice 
parallelism between the two.   
Due to the periodicity 
of external fields, the time-dependent problem can be mapped onto a {\it time-independent}
eigenvalue problem.  
Floquet topological insulator\cite{OkaPHE09} 
is indeed a remarkable ground for 
utilising the Floquet formalism, where 
a control of the topology of quantum systems by 
external time-periodic modifications, i.e., 
an application of a circularly-polarised light 
to graphene (and other multi-band systems), 
to make the system into a topological 
quantum Hall insulator, thus providing a prime example 
of engineering of topological properties.


Floquet's theorem is a general statement
about the solution of an ordinary differential equation 
with  a time-periodic potential.  In terms of 
the time-dependent Schr\"{o}dinger equation, 
\begin{eqnarray}
i\frac{d}{dt} \Psi(t)
&=
H(t) \Psi(t),
\label{schrodinger}
\end{eqnarray}
where $\Psi(t)$ is the wave function, and the Hamiltonian $H(t)$ is assumed to be periodic in time $t$ with period ${\cal T}$,
$H(t+{\cal T})=H(t)$, the Floquet theorem 
dictates that the solutions should have a form 
\begin{eqnarray}
\Psi_\alpha(t)
&=
e^{-i\varepsilon_\alpha t} \Phi_{\alpha}(t).
\label{floquet_state}
\end{eqnarray}
Here $\alpha$ labels the eigenfunction, 
$\Phi_{\alpha}(t)=\Phi_{\alpha}(t+{\cal T})$ is a periodic function of $t$, 
and the real quantity $\varepsilon_\alpha$ is called the quasienergy, which is
unique up to multiples of the frequency $\Omega \equiv 2\pi/{\cal T}$.  
So we can immediately see that $\Psi$ and $\varepsilon$ are temporal 
counterparts of Bloch's wavefunction 
and the crystal momentum which is
unique up to multiples of $2\pi/a$, the size of the reciprocal 
vector.

Since the modification is time-periodic, we can 
readily Fourier-expand the Schr\"{o}dinger equation on the 
time axis as
\begin{eqnarray}
  \sum_n (H_{mn}-n\Omega\delta_{mn}) \Phi_{\alpha}^n
    &=
      \varepsilon_\alpha \Phi_{\alpha}^m.
  \label{floquet}
\end{eqnarray}
Here, we have Fourier-expanded the wavefunction as $\Phi_{\alpha}(t)=\sum_n e^{-in\Omega t}\, \Phi_{\alpha}^n$ 
with $\Phi_\alpha^n$ called the $n$th Floquet mode, and 
\begin{eqnarray}
  H_{mn}
    &\equiv 
      \frac{1}{\cal T}\int_{0}^{\cal T} dt \;
      e^{i(m-n)\Omega t} H(t)
  \label{floquet_hamiltonian}
\end{eqnarray}
is the Floquet matrix form of the Hamiltonian. 
Thus the quasienergy $\varepsilon_\alpha$ corresponds to the eigenvalues 
of the infinite dimensional Floquet matrix $H_{mn}-n\Omega\delta_{mn}$.
If $\varepsilon_\alpha$ is an eigenvalue, 
the same holds for $\varepsilon_\alpha+n\Omega$ for an arbitrary integer $n$, 
and we can impose a condition, 
$-\frac{\Omega}{2}<\varepsilon_\alpha\le\frac{\Omega}{2}$, 
and end up with a ladder of energies with a spacing $\Omega$, 
just as we have a ladder of Bloch bands within the first Brillouin zone. 

As a consequence of the Floquet theorem, 
the time-dependent Schr\"{o}dinger equation 
is mapped onto a time-independent problem, but 
a price to pay is we have now to solve an infinite-dimensional 
matrix eigenvalue problem with a new degree of freedom, the 
Floquet index $n$.
Technically, we have to deal with the problem, which is non-equilibrium and 
ac-modulated, by combining Keldysh Green's function formalism for non-equilibrium and the Floquet formalism for the ac modulation, into 
a method dubbed Floquet Green's function.  
Since the Floquet formalism has to do with nonequilibrium 
physics, we have also to note that there are both 
transient quantum states just after the external field is swiched on, 
followed by subsequent nonequilibrium steady states (sometimes called 
``NESS").  This occurs e.g. when the system
during laser excitation  is subject to dissipation due to a 
heat bath, in which case the balance between pumping and relaxation 
determines the relaxation dynamics, which can be 
treated with Floquet Green's function.  
For many-body systems, incidentally, 
we can formulate 
the Floquet-DMFT (DMFT: dynamical mean-field theory) 
by combining the nonequilibrium DMFT 
and the Floquet method, which is out of the scope here.\cite{aokiRMP}

Now, let's describe the Floquet topological insulator, 
following Refs\cite{OkaPHE09,OkaAokiHMF10}.  
With laser illumination, 
transport properties, such as the Hall conductivity, 
can still be treated with the Kubo formula if we 
evoke the Floquet states.  
You might imagine that quantum topological quantities 
like Berry's curvature that dominates the QHE, defined originally 
for systems in equilibrium, would become ill-defined in non-equilibruim.  
This, however, is not the case --- the only thing we have to 
do, for the Floquet steady states, 
is just define and calculate the curvature by replacing 
the usual wavefunction with the Floquet 
wavefunctions.  Thus 
the conductivity tensor $\sigma_{ab}$ in magnetic fields in 
an intense ac field $\Vec{A}_{ac}$ is expressed as
\begin{eqnarray}
&&\sigma_{ab}(\Vec{A}_{ac})=i\int \frac{d\Vec{k}}{(2\pi)^2}
\sum_{\alpha,\beta\ne\alpha}
\frac{[f_\beta(\Vec{k})-f_\alpha(\Vec{k})]}{\ve_\beta(\Vec{k})-\ve_\alpha(\Vec{k})}\nonumber
\\
&&\hspace{0.8cm}\times
\frac{
\langle\langle \Phi_\alpha(\Vec{k})|J_b|\Phi_{\beta}(\Vec{k})\rangle\rangle
\langle\langle \Phi_\beta(\Vec{k})|J_a|\Phi_{\alpha}(\Vec{k})\rangle\rangle
}{\ve_\beta(\Vec{k})-\ve_\alpha(\Vec{k})+i\eta},
\end{eqnarray}
where $f_\alpha(\Vec{k})$ is the distribution function 
in the nonequilibrium for the $\alpha$-th Floquet state, 
$\Vec{J}$ is the current operator, 
and $\eta$ a positive infinitesimal. 
There, the energy is replaced with the Floquet quasi-energy, and 
the inner product $\langle\langle | \rangle\rangle$ includes 
the time average.  

The Hall conductivity then becomes 
a TKNN-type formula, 
\begin{eqnarray}
\sigma_{xy}(\Vec{A}_{\rm ac})=e^2\int \frac{d\Vec{k}}{(2\pi)^d}\sum_\alpha
f_\alpha(\Vec{k})\left[\nabla_{\Vec{k}}\times{\cal A}_\alpha(\Vec{k})\right]_z ,
\end{eqnarray}
where the gauge field 
${\cal A}_\alpha(\Vec{k}) \equiv 
-i\langle\langle \Phi_\alpha(\Vec{k})|\nabla_{\Vec{k}}|\Phi_\alpha(\Vec{k})
\rangle\rangle$ 
is defined in terms of the Floquet state $\Phi$.  Note that 
the occupation $f_\alpha(\Vec{k})$, 
which differs from the equilibrium Fermi-Dirac distribution, 
depends on the original band index $i$ and the Floquet 
index $m$, and depends on e.g. how the 
heat bath is attached to the system.


\clearpage

\subsection{Floqeut formalism for graphene --- Floquet topological insulator}

\begin{figure}[ht]
\begin{center}
\includegraphics[width=8cm,clip]{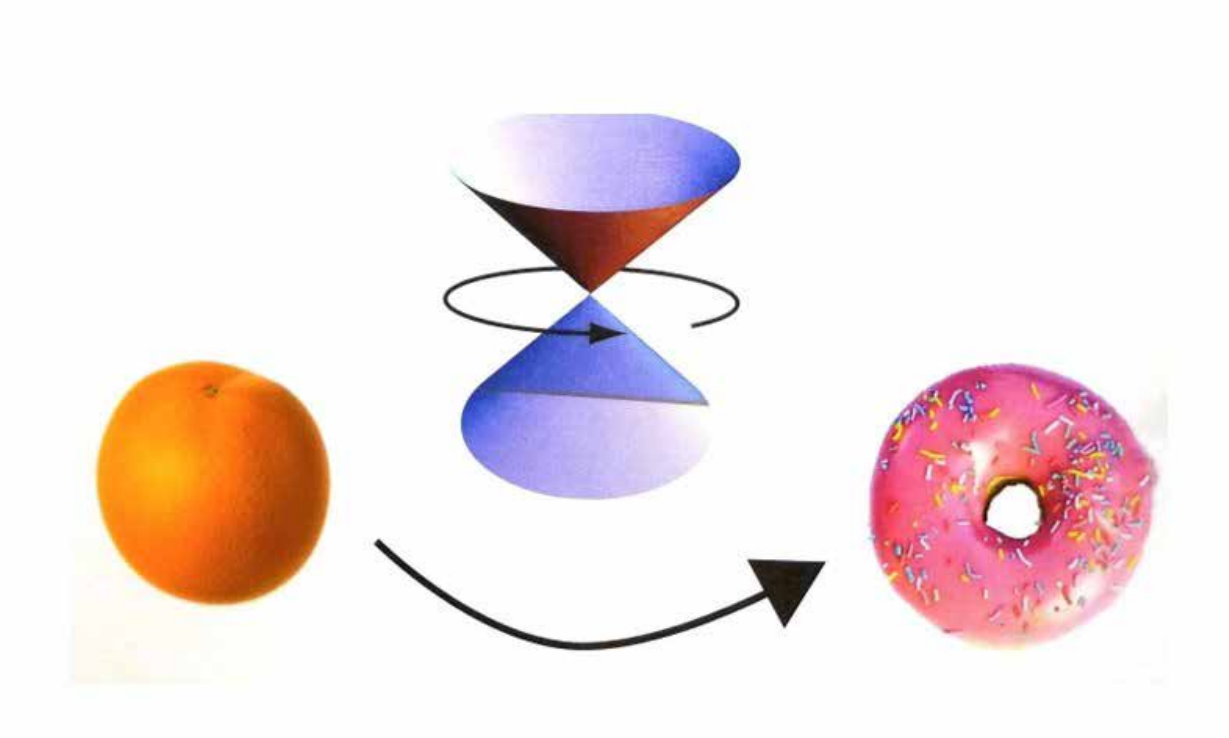}
\caption{A schematic picture for an illumination 
of a circularly-polarised light changing the topology of 
the quantum states in a Dirac system.  Top panel 
represents a Dirac cone in a circular laser, while the bottom one 
symbolises a change of the topology.  Figure courtesy of ETH Quantum Optics group.}
\label{fig:alex}
\end{center}
\end{figure}

The above is a general framework.  Now let us apply this to 
graphene, which leads us to the Floquet topological insulator (FTI),\cite{OkaPHE09}  
as symbolically depicted in Fig.\ref{fig:alex}.

For graphene, which has Dirac electrons with electron and hole 
branches, each matrix element in the Floquet Hamiltonian becomes 
a $2\times 2$ matrix.  Namely, the Hamiltonian is 
$H(t) = \tau_zv[k^x+A_{\rm ac}^x(t)]\sigma_x+v[k^y+A_{\rm ac}^y(t)]\sigma_y$,
where $\tau_z=\pm 1$ labels the valleys ($K$ and $K'$ points in graphene), 
$v$ is the velocity (set to $v=1$ here), 
and  $\sigma_i$ the Pauli matrices.
A circularly-polarised laser has the vector potential 
$(A_{\rm ac}^x, A_{\rm ac}^y) = A({\rm cos}\Omega t, {\rm sin}\Omega t),$ 
where $A \equiv F/\Omega$ with $F$ being the field strength. 
With Fourier transformed Floquet states $|\Phi(t)\rangle=
\sum_me^{-im\Omega t}|u_\alpha^m\rangle$, the Floquet equation becomes 
\begin{eqnarray}
\sum_{n}H^{mn}|u^n_\alpha\rangle = (\ve_\alpha+m\Omega)|u^m_\alpha\rangle
,
\end{eqnarray}
where, for graphene, we have 
$H^{mm}=\left(^{0\;k_x-ik_y}_{k_x+ik_y\;0}\right)$
for the diagonal elements, 
while the off-diagonal elements 
depend on $\tau_z$, i.e., $H^{mm+1}=\left(^{0A}_{0\;0} \right),
\;H^{mm-1}=\left(^{0\;0}_{A0} \right)$
for $K$ point ($\tau_z=1$), so that we have\par

\begin{figure}[h]
\begin{center}
\includegraphics[width=9cm]{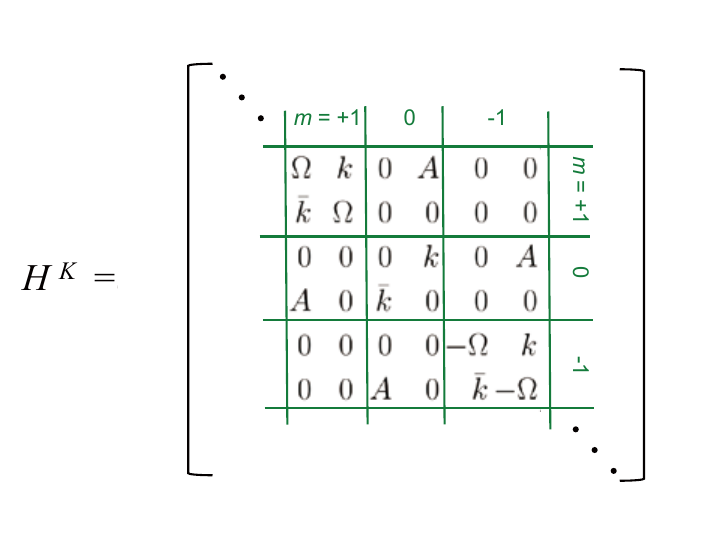}
\end{center}
\end{figure}

\noindent where $k \equiv k_x-ik_y, \bar{k} \equiv k_x+ik_y$.  For $K'$ point ($\tau_z=-1$), on the other hand, $H^{mm+1}=\left(^{0\;\;\;0}_{-A0} \right),
\;H^{mm-1}=\left(^{0-A}_{0\;\;0} \right)$.

When an ac field (such as laser) is applied to the Dirac system, 
we have thus a series of Floquet branches replicated from the 
original Dirac bands.  This is just a two-band version 
of the  Floquet bands, but 
a unique and essential effect of the circularly-polarised light 
is that this turns the system into a topological system, 
just as in QHE, where the topological nature is 
characterised by Chern number.  
Consequently, a topological gap opens at every $\varepsilon=$ integer $\times (\Omega/2)$ 
in the quasi-energy band structure (Fig. \ref{fig:dirac2G}(d)), 
as also seen in the density of states 
in  Fig. \ref{fig:dirac2G}(e).  
The gap at $\varepsilon=\pm \Omega/2$ is the largest, 
which is related to one-photon assisted transport and 
exists also in a linearly-polarised light.  
More importantly, the circularly-polarised case has a gap opening right 
at the Dirac point,  $\Vec{k}=0,\;\ve=0$. 
The quasi-energy around the point is 
$\ve_\alpha=\tilde{\ve}_i+m\Omega$
with  $\tilde{\ve}_1=
(\sqrt{4A^2+\Omega^2}+\Omega)/2,\;
\tilde{\ve}_2=
(-\sqrt{4A^2+\Omega^2}+\Omega)/2$.  
Here we label the states with $\alpha = (i, m)$, 
where $i$ is the original band index while $m$ the Floquet index.  
The $\alpha=(1,-1),\;(2,0)$ bands are descendants 
of the original Dirac bands, and the dynamical gap $2\kappa$ between 
them is 
\begin{equation}
2\kappa = \sqrt{4A^2+\Omega^2}-\Omega.
\end{equation}
The dynamical gap first grows quadratically with $A$ like 
$2\kappa\sim 2A^2/\Omega$, followed by an asymptote 
$2\kappa\sim 2A-\Omega$.  

To be precise, the quasi-energy is a sum of the dynamical phase 
and the Aharonov-Anandan (AA) phase (a nonadiabatic extension of 
Berry's phase)\cite{AA} as
\begin{eqnarray}
\ve_\alpha=\langle\langle \Phi_\alpha|H(t)|\Phi_\alpha\rangle\rangle+\gamma_\alpha^{\rm AA}/T,
\label{eq:quasienergysum}
\end{eqnarray}
where the AA phase is given by 
$
\gamma^{\rm AA}_\alpha \equiv T\langle\langle \Phi_\alpha|i\partial_t|\Phi_\alpha\rangle\rangle = \pm\pi\left\{[4(A/\Omega)^2+1]^{-1/2}-1\right\}.
$
In Berry's curvature for 
the Floquet state in Fig. \ref{fig:dirac2G}(b), 
we can see a conspicuous peak at $\Vec{k}= 0$,
\begin{eqnarray}
\left[\nabla_k\times{\cal A}_\alpha(\Vec{k})\right]_z \sim 
\pm \frac{1}{2}\kappa(k^2+\kappa^2)^{-3/2},
\label{eq:Berrycurvature}
\end{eqnarray}
where $\pm$ corresponds to $\alpha=(1,m),(2,m)$.  

\begin{figure}[h]
\begin{center}
\includegraphics[width=12cm,clip]{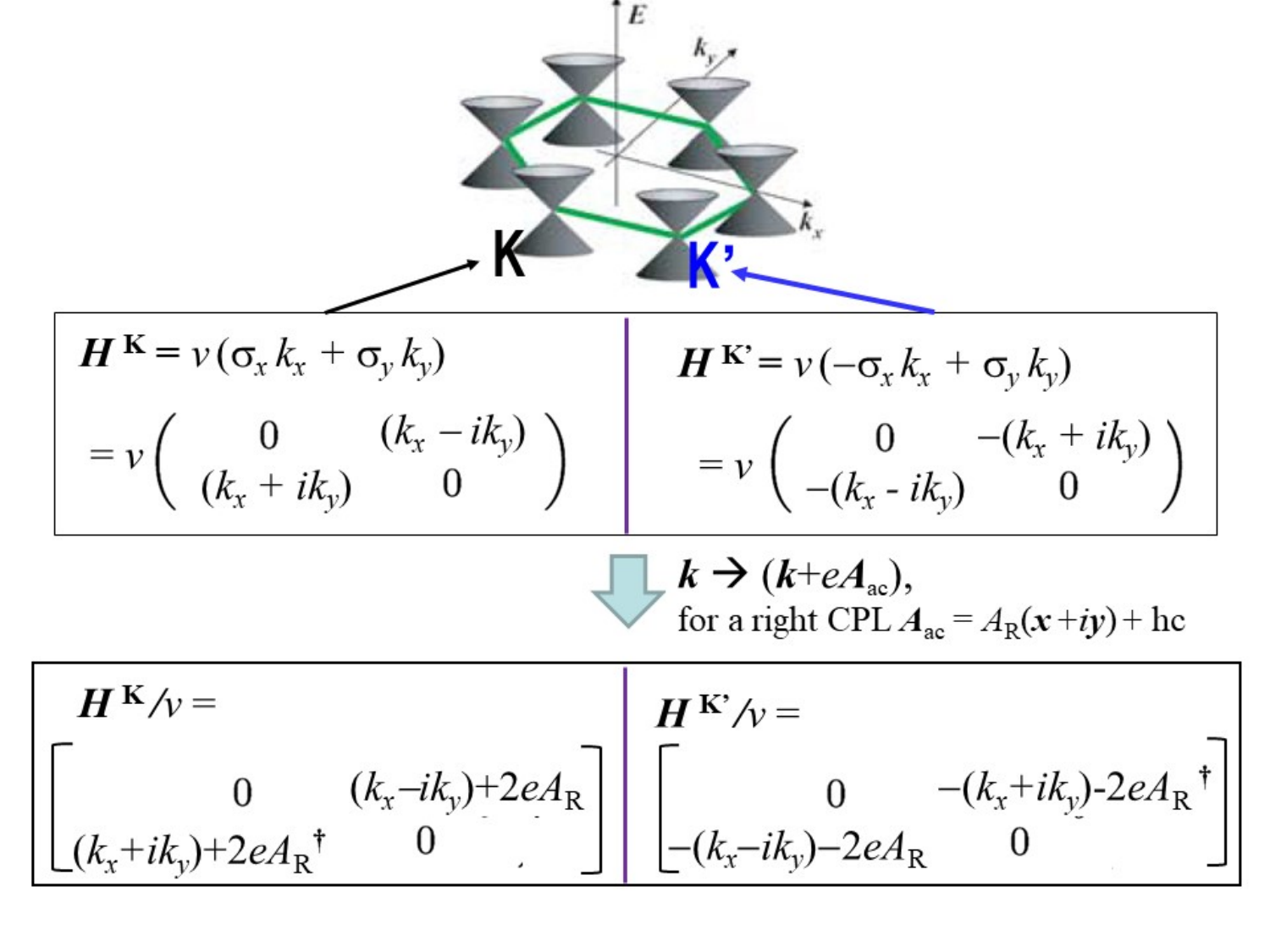}
\caption{For each of the valley K and valley K', we write down 
the Hamiltonian in $2\times 2$ forms.  When a circularly-polarised 
light is illuminated, we replace 
the momentum as $\Vec{k} \rightarrow (\Vec{k}+e{\Vec A}_{\rm ac})$, 
where ${\Vec A}_{\rm ac} = A_R({\Vec x}+i{\Vec y}) +$ h.c. (for 
a right polarisation here).  We can show that $H^{\rm K}$ and 
$H^{\rm K'}$ give the same Chern density.
}
\label{fig:whyValleyIndep}
\end{center}
\end{figure}

Now, in graphene we have two valleys, K (with $\tau_z = +1$) and K' ($\tau_z = -1$), and we can show that 
both valleys contribute to the FTI Hall conductivity with the same sign 
and add up, i.e., $\tau_z$ does not appear in the Berry's curvature 
in the above expression.  The reason for this is that, if we look at 
the structure of the Hamiltonian for each of the two valleys as in 
Fig.\ref{fig:whyValleyIndep}, the vector potential representing the circularly-polarised light 
enters into the expression in such a way that the chirality 
does not affect the Chern density.

\begin{figure}[h]
\begin{center}
\centering 
\includegraphics[width=11cm]{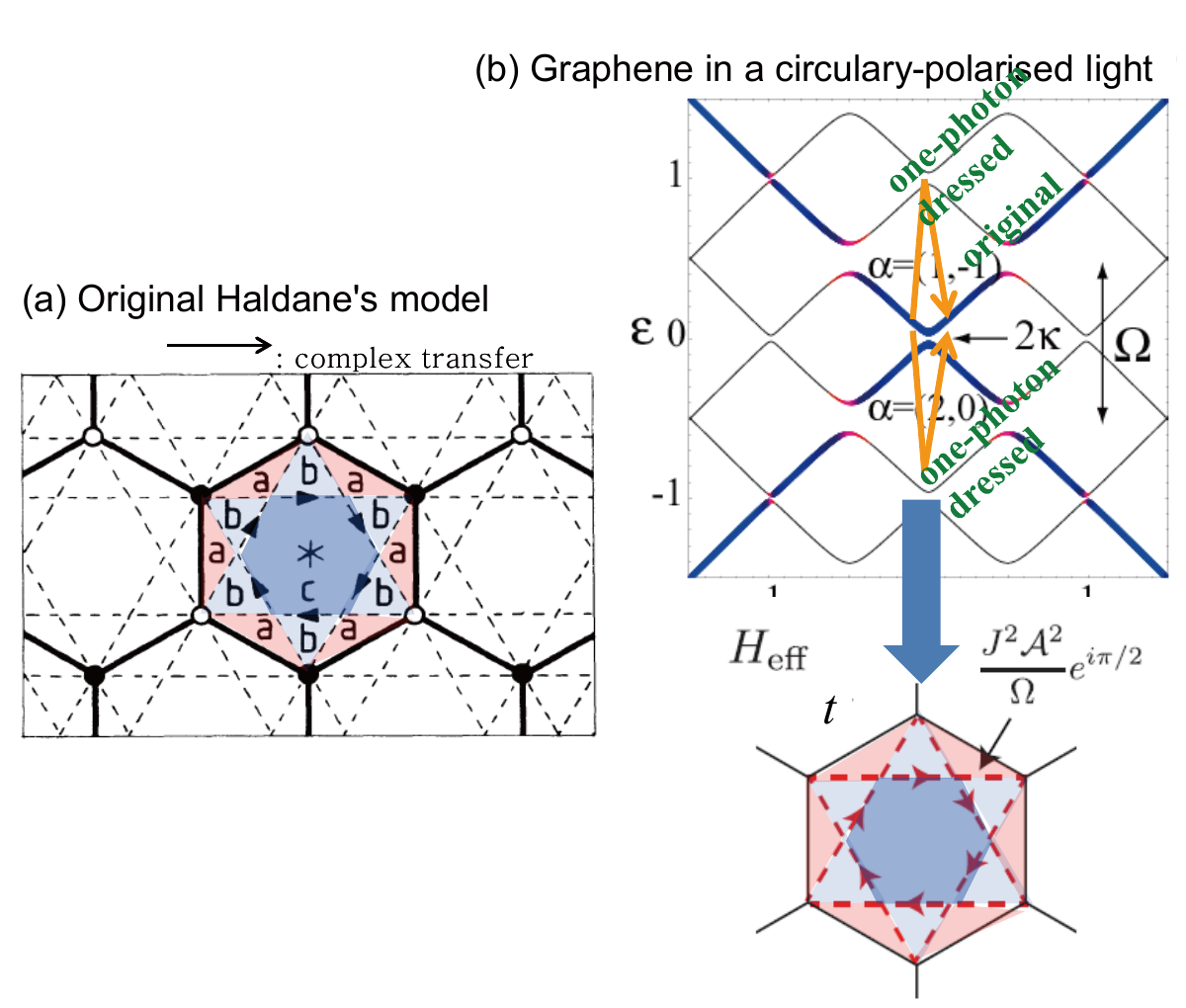}
\caption{(a) The original Haldane's model for the anomalous QHE.  
Dashed lines represent second-neighbour complex hopping (with 
a positive phase along the arrow, negative in the opposite 
direction).  This may be viewed as coming from magnetic flux 
penetrating the unit hexagon directed out of the plane (in 
the portions of the hexagon marked in red) 
or into the plane (blue), with the total mangetic flux being zero.
(b) Graphene in a circularly-polarised ligth has a series of 
Floquet subbands comprising Dirac cones separated by the 
laser frequency $\Omega$.  From the second-order processes between 
the original band and the one-photon dressed bands (double 
orange arrows) emerge the effective Floquet Hamiltonian that 
exactly coincides with the Haldane's model. 
}
\label{fig:HaldaneVsFTI}
\end{center}
\end{figure}

\begin{figure}[h]
\begin{center}
\centering 
\includegraphics[width=16cm]{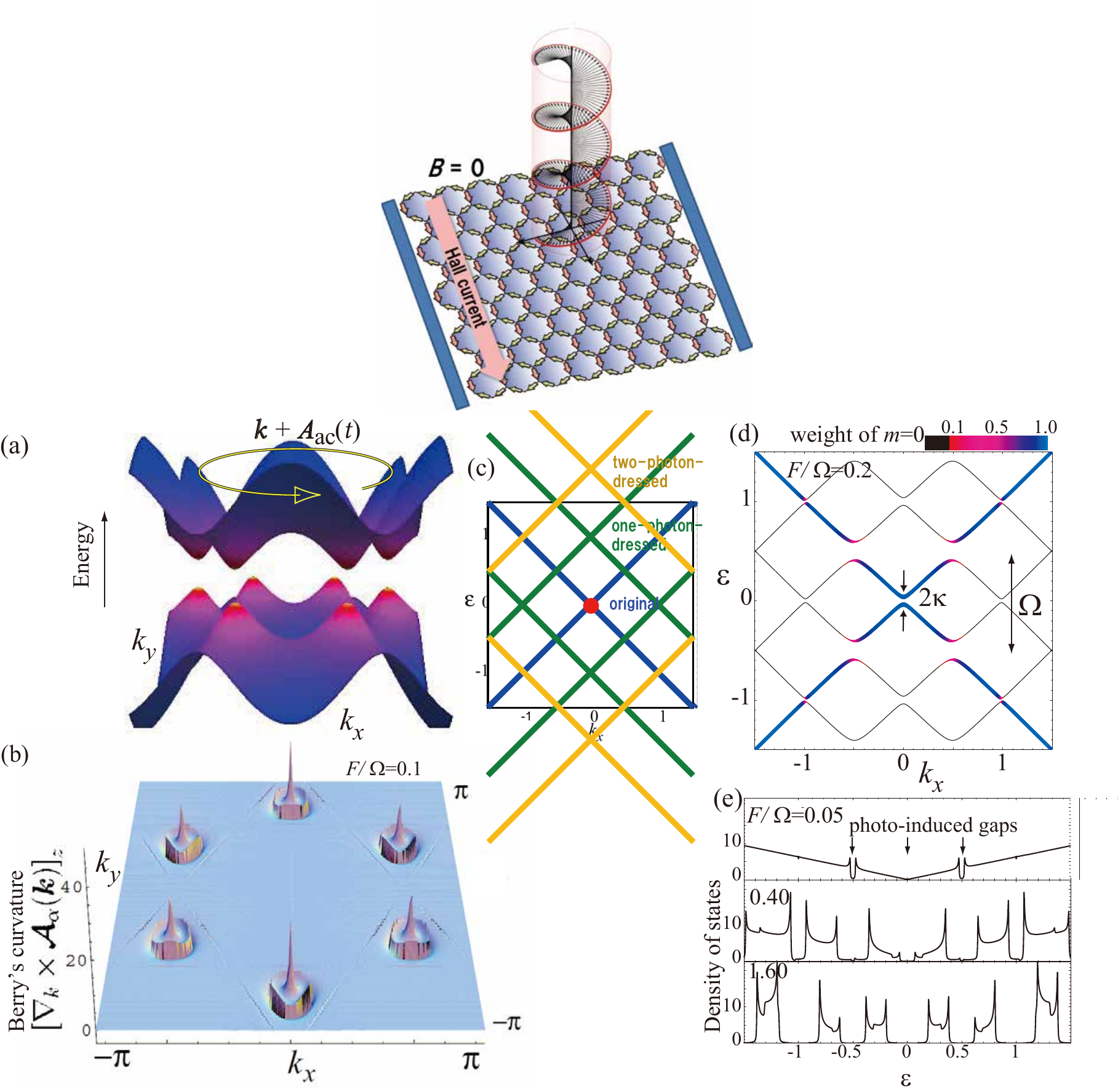}
\caption{Top inset: Floquet topological insulator, 
which 
arises when graphene is illuminated by a circularly-polarised 
laser is schematically shown.  Blue bars stand for electrodes for 
Hall effect measurement.  
(a) Band dispersion, with topological gaps arising, when each electron's momentum follows a trajectory $\Vec{k}+\Vec{A}_{\rm ac}(t)$ in a circularly-polarised light is schematically shown.  
(b) The photo-induced Berry's curvature $[\nabla_{\Vec{k}}\times{\cal A}_\alpha(\Vec{k})]_z$ for 
$\alpha=(1,m)$ 
for $F/\Omega =0.2$, $\Omega/v=1$.  
(c) A series of Floquet subbands generated from the original Dirac cone 
is schematically shown against $k_x$ (measured from each Dirac point), 
for which level anticrossing will give the next panel. 
(d) The Floquet quasi-energy (black curves) plotted against $k_x$ 
with $k_y=0$ for $F/\Omega =0.2$. The colour coding
represents the weight of the static ($m=0$) component. 
(e) Density of states for various field strengths $F$. 
After T. Oka and H. Aoki, Phys. Rev. B {\bf 79}, 081406(R) (2009).
}
\label{fig:dirac2G}
\end{center}
\end{figure}

You might have wondered: why an ac field (laser) 
can produce a dc response (dc Hall current)? 
This is exactly due to the geometric phase: 
the ac field changes the system into a Chern 
insulator, which naturally accommodates the 
Hall current ($\propto$ Chern number), despite the absence of static 
external magnetic field.  Indeed, right after the 
proposal by Oka and Aoki,\cite{OkaPHE09} Kitagawa et al\cite{kitagawa} have shown that 
the effective Floquet Hamiltonian for the honeycomb 
lattice in a circularly-polarised light is, 
to the leading (second) order in the high-frequency 
($1/\Omega$) expansion, is exactly coincides with 
Haldane's celebrated model\cite{haldane88}  
for the anomalous QHE (i.e., 
QHE in zero external magnetic field), as shown in Fig.\ref{fig:HaldaneVsFTI}.  The second-order 
processes correspond to the photon absorption/emission twice 
between the original Dirac band and the first ($n=\pm1$) 
Floquet bands.  Thus the FTI shares Class A (unitary) 
 in Fig.\ref{topologicalPeriodicTable} with 
QHE and AQHE.  
The emergence of the Floquet topological insulator explained 
above is summarised in Fig.\ref{fig:dirac2G}.  
If we change the right-circularly 
polarisation into the left-circularly 
polarisation, the Hall response changes sign.

\begin{figure}[h]
\begin{center}
\centering 
\includegraphics[width=9cm]{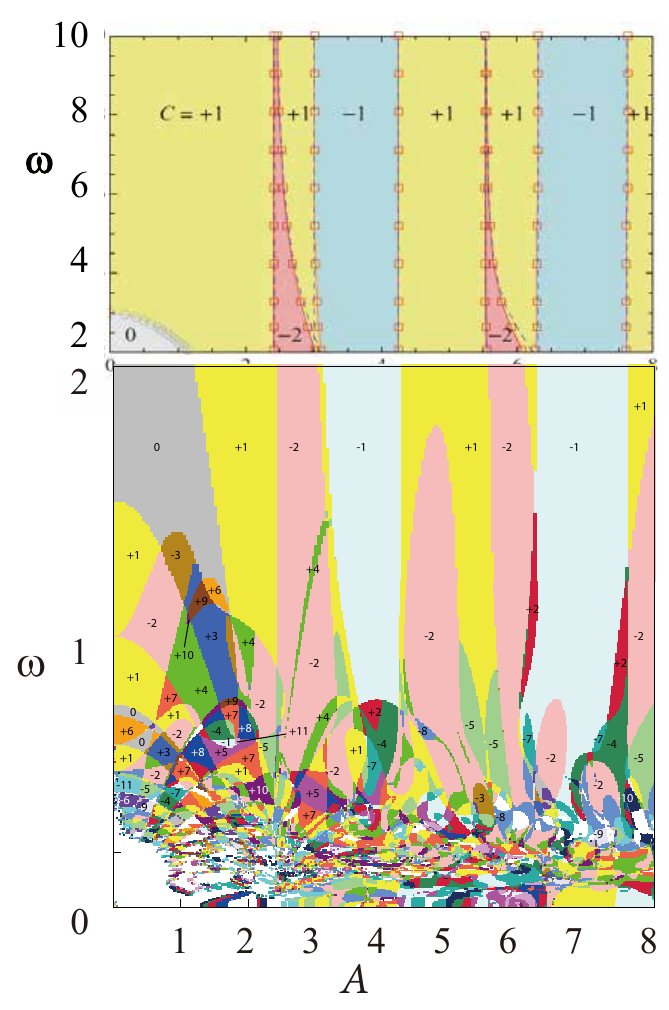}
\caption{Topological phase diagram against the frequency $\omega$ 
and the amplitude $A$ of the laser for a honeycomb lattice 
illuminated by a circularly-polarised laser.  Top (bottom) 
panel is for $\omega > 2\;(\omega<2)$.  
After T. Mikami et al, Phys. Rev. B {\bf 93}, 144307 (2016).
}
\label{fig:fractalPhaseDiag}
\end{center}
\end{figure}

The topological phases naturally depend strongly on the frequency and 
intensity of applied laser field.  For the
honeycomb lattice driven by circularly-polarised light, 
Fig.\ref{fig:fractalPhaseDiag} is a theoretical result for 
the Chern numbers against the frequency $\omega$ 
and the amplitude $A$ of the laser.\cite{mikami16}  
Various values of 
the topological number appear, with the phase diagram 
becoming more intricate as $\omega$ is decreased, 
where many Floqeut subbands become superposed with 
band anticrossing.

Since the FTI is a non-equilibrium state, 
we have to take care of the non-equilibrium (i.e., non-thermal) distributions 
which differ from the equilibrium Fermi-Dirac distribution.  
Then the FTI Hall conductivity is somewhat blurred from 
the QHE quantised values.  This depends on the situation: 
whether (i) the system is isolated (as in cold-atom systems), in which 
case the Hall conductance significantly deviates from the quantised values, 
or (ii) the system is coupled to a reservoir (in usual 
solid-state systems), in which case the Hall conductance 
is (a) closer to the quantised values for the driving frequency $\Omega \gg W$ with 
$W$ being the electronic band width, or 
(b) rather deviates for $\Omega \sim W$, where 
the situation is determined by the competition between the 
 reservoir-induced cooling and the photocarrier excitations.\cite{dehghani15} 

Experimentally, the detection of the Floquet topological insulator was 
first observed on the surface of a 3D topological insulator 
(Bi$_2$Se$_3$) by Gedik's group.\cite{gedik13} The 3D topological insulator 
has a topological surface (2D) state that has a Dirac dispersion.  
When they have looked at 
the bulk energy spectrum with ARPES, the gap opening is observed when 
a circularly-polarised light is illuminated on the surface.  
Time evolution of the gaps were even observed with 
time-resolved ARPES.  The next detection of FTI came from 
 photonics, which has an interesting 
analogy with electronic systems.  Namely, Rechtsman's group\cite{rechtsman13} 
has constructed a honeycomb array of 
spiral photonic guides, where they succeeded in 
observing a photonic analogy of the Floquet 
topological insulator.  There, the drive by 
an external circularly-polarised light is 
mimicked by the spiral structure, and the 
topological nature of the resulting photonic state 
is confirmed by detecting topological photonic edge states.  
The third experimental observation of FTI was done in the 
cold-atom physics.\cite{jotzu14} 
 Namely, Esslinger's group in ETH Z\"{u}rich 
has used a cold-atom system on a honeycomb 
optical lattice, which was shaken in a circular 
motion.  The topological state was detected as a 
Hall drift of the cold atom system, and they have 
even obtained the AQHE phase diagram of the Haldane's model 
by tuning the phase dominating the Hamiltonian with 
linear $\leftrightarrow$ circular motions.

Finally came an experimental realisation\cite{mciver20} 
of the FTI in 
graphene itself, which the original theoretical proposal 
\cite{OkaPHE09} had in mind.  
McIver and coworkers in 
MPI Hamburg had an experimental setup 
for graphene, where a graphene sample 
with four terminals is optically driven by 
an ultrafast mid-infrared circularly polarised laser in femtosecond pulse forms.  
The excited Hall current (perpendicular to an applied 
electric field in the four-terminal) is then led via a microstrip transmission line  
to a photoconductive switch for current detection.  
The obtained result shows a clear plateau structure in the 
Hall conductivity and associated gap openings.

\begin{figure}[t]
\begin{center}
\centering 
\includegraphics[width=16cm]{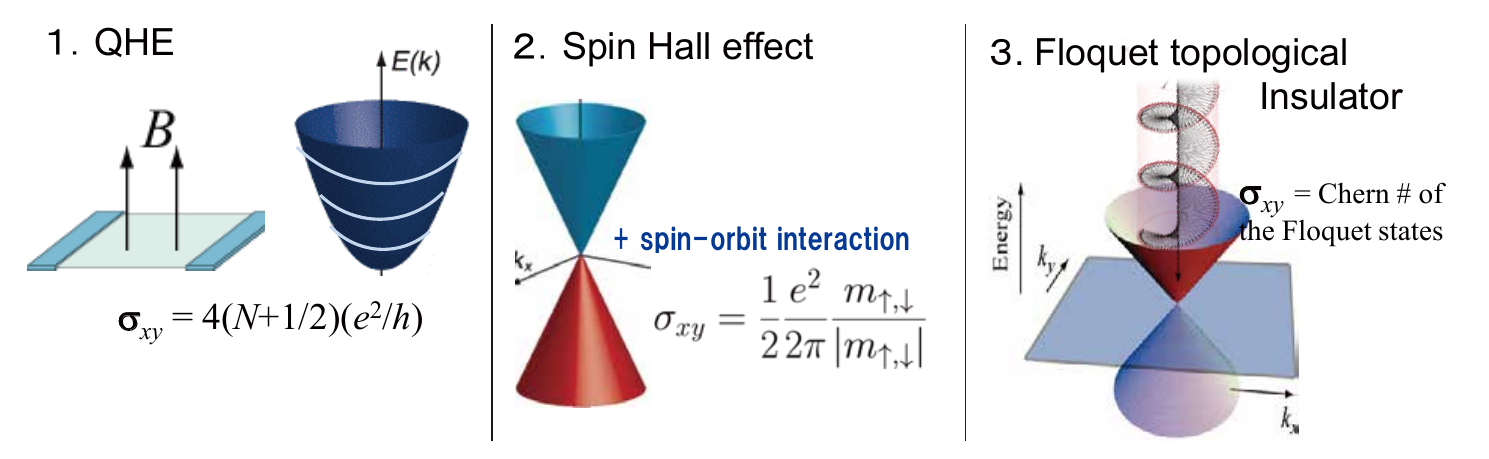}
\caption{Three examples of quantum Hall effects: (Left) the original QHE system, (middle) the spin Hall effect in topological insulators for a Dirac system with a spin-orbit coupling,  
and (right) the Floquet topological insulator.
}
\label{fig:threeQHEs}
\end{center}
\end{figure}

Thus, for the quantum Hall effects,  we have now three examples: 
(i) the original QHE system, (ii) the spin Hall effect in topological insulators for a Dirac system with a spin-orbit coupling,\cite{KaneMele} 
and (iii) the Floquet topological insulator (Fig.\ref{fig:threeQHEs}).

Talking of the photonic FTI, there is an entirely 
different avenue for electromagnetic 
waves for QHE-like states.  This was initiated 
theoretically by Haldane and coworkers,\cite{haldaneOptical} 
which was subsequently confirmed experimentally 
by Wang and coworkers.\cite{wangNature09}  
A photonic analogue of quantum spin-Hall effect (QSHE) has also been 
explored in photonic crystals.   By manipulating 
honeycomb photonic crystals, it is possible to have 
a photonic Hamiltonian from the $k\cdot p$ formalism 
in an appropriate photonic basis, which has 
the same form as the HgTe quantum
wells which is the birth place of QSHE.\cite{Bernevig06} 

\clearpage

\section{Summary}

To summarise the chapter, we have surveyed the quantum Hall effect, 
primarily in a perspective of topological systems.  So we have 
started from the classification of topological systems by generic 
symmetries to show a rich variety of topological systems 
which were kicked off by the integer quantum Hall effect. 
Then we have surveyed theoretical reasoning and 
various experimental results.  Due to the bulk-edge 
correspondence for topological systems, edge states 
are among the essential feature in QHE.  We then 
looked at how QHE contributes not only to 
the resistance standard, but to the determination 
of the fine-structure constant and definition of the new SI.  
We have also looked into QHE in periodic systems with 
fractal energy spectra and QHE in three dimensions.  
Then we delve into the anomalous quantum Hall effect and 
spin quantum Hall effect as relatives of the IQHE, 
and also touched upon integer vs fractional quantum Hall effects.  
Then we elaborated QHE in graphene as a novel arena, both 
for monolayer and bilayer graphenes.  We have included 
optical properties, and an analogy between QHE and 
superconductivity.  For the bilayer graphene, 
we have explored the twisted bilayer graphene which 
harbour rich physics.  Last but not least, 
the final section looks into QHE in light-matter coupled systems, 
where Floquet topological
insulator is described  both theoretically and 
experimentally as a prime example of nonequilibrium 
realisation of topological states.  

These perspectives give a widened horizon of the 
condensed-matter physics, and we can envision 
further developments into diverse directions.


{\bf Further Reading}

Aoki H (1987) Reports on Progress in Physics 50: 655. 

Aoki H and Dresselhaus MS (eds.) (2014): 
{\it Physics of Graphene}. Springer.

Chakraborty T and Pietil\"{a}inen P (1988) The Fractional Quantum
Hall Effect. Springer.

Das Sarma S and Pinczuk A (eds.) (1997) Perspective in Quantum Hall Effects. New York: Wiley.

Zyun Francis Ezawa (2013): 
{\it Quantum Hall Effects} 3rd Ed. World Scientific. 

Landwehr G (1986) Festk\"{o}rperprobleme 26: 17.

Prange RE and Girvin SM (eds.) (1990) The Quantum Hall Effect,
2nd edn. New York: Springer.

Yoshioka D (2002) The Quantum Hall Effect. Springer.


\end{document}